\documentclass[12pt]{article}

\usepackage{amsmath}
\usepackage{graphicx,psfrag,epsf}
\usepackage{enumerate}
\usepackage{natbib}

\usepackage{algorithmicx}
\usepackage{algorithm}
\usepackage{algpseudocode}
\usepackage{amssymb}
\usepackage{amsthm}
\usepackage{array}
\usepackage{authblk}
\usepackage[american]{babel}
\usepackage{bm}
\usepackage{booktabs}
\usepackage{caption}
\usepackage{color}
\usepackage{float}
\usepackage{framed}
\usepackage{makecell}
\usepackage{makeidx}
\usepackage{mathtools}
\usepackage[maxfloats=80]{morefloats}
\usepackage{multirow}
\usepackage{pdfsync}
\usepackage[section]{placeins}
\usepackage{physics}
\usepackage{relsize}
\usepackage{setspace}
\usepackage{siunitx}
\usepackage{subfigure}
\usepackage{url}

\usepackage{xr-hyper}
\usepackage{hyperref}

\theoremstyle{definition}

\newtheorem{remark}{Remark}

\def\I{\mathrm{I}}
\def\H{\mathrm{H}}

\def\W{\mathrm{W}}

\DeclareMathOperator{\vect}{vec}

\DeclareMathOperator{\E}{E}
\DeclareMathOperator{\Var}{Var}

\DeclareMathOperator{\N}{N}

\newcommand{\blind}{0}

\begin{document}

\def\spacingset#1{\renewcommand{\baselinestretch}%
{#1}\small\normalsize} \spacingset{1}


\if0\blind
{
  \title{\bf Bayesian Variable Selection for Gaussian copula regression models}
  \author{A. Alexopoulos \\
    Department of Statistical Science, University College London\\
    and \\
    L. Bottolo\thanks{
    The authors gratefully acknowledge the MRC grant MR/M013138/1, The BHF-Turing Cardiovascular Data Science Awards 2017 and The Alan Turing Institute under the Engineering and Physical Sciences Research Council grant EP/N510129/1. The authors are thankful to Katherine Schon and Michael Johnson for providing the data of the Ataxia-Telangiectasia disorder and Temporal Lobe Epilepsy examples and to H\'{e}l\`{e}ne Ruffieux and Verena Zuber for insightful comments. The authors are also grateful to the editor, associate editor and two anonymous referees for their valuable comments that greatly improved the presentation of the paper.}
    \hspace{.2cm} \\
    Department of Medical Genetics, University of Cambridge\\
    The Alan Turing Institute, London\\
    MRC Biostatistics Unit, University of Cambridge}
  \maketitle
} \fi

\begin{abstract}
We develop a novel Bayesian method to select important predictors in regression models with multiple responses of diverse types. A sparse Gaussian copula regression model is used to account for the multivariate dependencies between any combination of discrete and/or continuous responses and their association with a set of predictors. We utilize the parameter expansion for data augmentation strategy to construct a Markov chain Monte Carlo algorithm for the estimation of the parameters and the latent variables of the model. Based on a centered parametrization of the Gaussian latent variables, we design a fixed-dimensional proposal distribution to update jointly the latent binary vectors of important predictors and the corresponding non-zero regression coefficients. For Gaussian responses and for outcomes that can be modeled as a dependent version of a Gaussian response, this proposal leads to a Metropolis-Hastings step that allows an efficient exploration of the predictors' model space. The proposed strategy is tested on simulated data and applied to real data sets in which the responses consist of low-intensity counts, binary, ordinal and continuous variables.
\end{abstract}

\vfill

\spacingset{1.45}

\section{Introduction}

The identification of important predictors in linear and non-linear regression models is one of the most frequently studied questions in statistical theory. In Bayesian statistics, this problem is known as Bayesian variable selection (BVS) and for Gaussian responses, there is an extensive literature for an efficient detection of important predictors in both single- and multi-response regression models. In the case of a single response, a list of relevant papers includes, but is not limited to, \cite{george1997approaches}, \cite{liang2008mixtures}, \cite{guan2011bayesian} and \cite{Rockova2018}, while \cite{brown1998multivariate}, \cite{holmes2002accounting} deal with the same problem in the multi-response case. BVS in single-response non-linear regression models has also received great attention. In \cite{dellaportas2002bayesian} and \cite{forster2012reversible} the corresponding methods are reviewed and advances are proposed.

More recently, there has been an increasing interest in the joint analysis of outcomes of diverse types, for instance, continuous, binary, categorical and count data, given their availability from studies involving multivariate data, see for example \cite{hoff2007extending}, \cite{murray2013bayesian}, \cite{zhang2015bayesian} and \cite{bhadra2018inferring}. In regression analysis, the most popular model used to account for the multivariate dependencies between any combination of discrete and/or continuous responses is the Gaussian copula regression (GCR) model \citep{song2009joint} in which each response is associated with a (potentially different) set of predictors. When only Gaussian responses are considered, this is known as Seemingly Unrelated Regression (SUR) model \citep{zellner1962efficient}. Recent contributions for sparse Bayesian SUR models include, for instance, \cite{wang2010sparse} and \cite{Deshpande2019}. Bayesian methods for the estimation of the regression coefficients of the GCR model with a fixed set of predictors have also been proposed \citep{pitt2006efficient}. 
Despite the growing Bayesian literature regarding an efficient selection of important predictors, to the best of our knowledge, variable selection for the GCR model has not been attempted. We propose here the first fully Bayesian approach for model selection in regression models with multiple responses of diverse types.

The main obstacle of the application of BVS in single-response non-linear models, as well as in the SUR model and the GCR model with responses of diverse types, is the non-tractability of the marginal likelihood. To overcome this problem, Markov chain Monte Carlo (MCMC) algorithms are based either on Laplace approximation, see for example \cite{bove2011hyper}, or on a Metropolis-Hastings (M-H) step in which the dimension of the proposal distribution is not fixed at each iteration \citep{forster2012reversible}. The latter is an application of the Reversible Jump algorithm \citep{green1995reversible} which is known to experience a low acceptance rate when the trans-dimension proposal distribution is not devised carefully, resulting in MCMC samplers with poor mixing \citep{brooks2003efficient}. In addition, in current applications, the number of predictors is often very large and any MCMC algorithm for BVS in both linear and non-linear regression has to be designed carefully in order to explore successfully the ultra-high dimensional model space consisting of all the possible subsets of predictors \citep{bottolo2010evolutionary,Lamnisos2009}.

The main contribution of this paper is the development of a Bayesian approach for the joint update of the latent binary vector of important predictors and the corresponding vector of non-zero regression coefficients for each response of the GCR model. By using the proposed strategy, we perform BVS without any approximation. We also avoid the Reversible Jump algorithm by utilizing the Gaussian latent variables of the GCR model to construct a proposal distribution defined on a fixed-dimensional space. For Gaussian responses and for the outcomes that can be modeled as a dependent version of a Gaussian response, e.g., the Probit model for binary data, the designed proposal distribution allows the \lq\lq implicit marginalisation\rq\rq\ of the non-zero regression coefficients in the M-H  acceptance probability \citep{holmes2006bayesian}, favoring the exploration of the predictors model space in a more efficient manner.

Modeling the dependence amongst the responses is another key aspect in GCR. Until recently devising an efficient MCMC algorithm for a structured (constrained) covariance matrix that includes the identifiability conditions for the non-linear responses has been a difficult task. Here, we follow the solution proposed by \cite{talhouk2012efficient} and specify a conjugate prior on the correlation matrix. We utilize the parameter expansion for data augmentation \citep{liu1999parameter,van2001art} to expand the correlation into a covariance matrix. We also adopt the idea of covariance selection to obtain a parsimonious representation of the dependence amongst the responses. Based on the theory of decomposable Gaussian graphical models \citep{lauritzen1996graphical}, we use the hyper-inverse Wishart distribution as the prior density for the covariance matrix. This prior allows some of the off-diagonal elements of the inverse covariance matrix to be identical to zero and to estimate the conditional dependence pattern of the observations \citep{webb2008bayesian}.

We tested the performance of our model, Bayesian Variable Selection for Gaussian Copula Regression (BVSGCR), in a comprehensive simulation study and compared the performance of our new approach with conventional Bayesian methods for the selection of important predictors in single-response (linear and non-linear) regression models, see \cite{holmes2006bayesian}, \cite{fruhwirth2009improved} and \cite{dvorzak2016sparse}.

We also applied the proposed method to two real data sets. The first data set includes a combination of nine continuous, binary and ordered categorical responses. These responses are phenotypic traits of a rare disorder called Ataxia-Telangiectasia. We analyzed one of the largest cohorts of patients, consisting of $46$ individuals affected by the disease \citep{schon2018genotype}. Our model borrows information across the outcomes in order to identify important associations between the responses and a set of genetic and immunological predictors that have been collected in the same study. 
The second data set consists of four counts and one ordered categorical response which are measured in $122$ individuals suffering from Temporal Lobe Epilepsy \citep{johnson2016systems}. Our interest lies in the identification of associations between the responses and $162$ correlated genes that have been identified in a recent gene-network analysis related to cognition abilities and epilepsy \citep{johnson2016systems}. 
Finally, in both real data sets, we compared the predictive ability of the BVSGCR model with widely-used single-response linear and non-linear sparse regression models.

The rest of the paper is organized as follows. In Section \ref{sec:GCR} we provide a brief presentation of the GCR model and the prior distributions on the regression coefficients and the correlation structure. In Section \ref{sec:MCMCsec} we describe the novel MCMC algorithm that we propose for BVS when a combination of discrete and/or continuous responses are considered. Section \ref{sec:simstudy} presents the results of the simulation study and in Section \ref{sec:real} we apply the proposed model on two real data sets with missing values in the outcome variables which led to a straightforward modification of the designed MCMC algorithm. Finally, in Section \ref{sec:discussion} we conclude with a short discussion.

\section{Gaussian copula regression model}
\label{sec:GCR}

%

In the following, all vectors, in bold font, are understood as column vectors and the superscript \lq\lq $T$\rq\rq is used to denote the transpose of a vector or a matrix. Matrices are also indicated in bold font. The lower-case notation will be reserved for the observations with the corresponding random variables in capital letters.

\subsection{Gaussian copulas}
\label{sec:copula}

An $m$-variate function $C (u_1,\ldots,u_m)$, where
$
C :[0,1]^m \rightarrow [0,1],
$
is called a copula if it is a continuous distribution function and each marginal is a uniform distribution function on $[0,1]$. \cite{sklar1959fonctions} proved that any joint cumulative distribution function (cdf) of continuous random variables can be completely specified by its marginal distributions and a unique copula $C$. 
If $F_1(\cdot),\ldots,F_m(\cdot)$ are the marginal cdfs of a combination of $m$ continuous and discrete random variables $Y_1,\ldots,Y_m$, their joint cdf can be specified through a specific copula function $C$ as
\begin{equation*}
F_{Y_{1},\ldots ,Y_{m}}(y_{1},\ldots ,y_{m})=C\{F_{Y_{1}}(y_{1}),\ldots
,F_{Y_{m}}(y_{m})\}.
\end{equation*}
A copula function which is commonly used for modeling the dependence structure of any combinations of continuous and discrete variables is the Gaussian copula, see for example \cite{hoff2007extending} and \cite{murray2013bayesian}.
The Gaussian copula $C$ is specified through the function
\begin{equation}
C(u_{1},\ldots ,u_{m};\bm{R})=\Phi _{m}\{\Phi ^{-1}(u_{1}),\ldots ,\Phi
^{-1}(u_{m});\bm{R}\},
\label{eq:gausscop}
\end{equation}
where $\Phi_m( \cdot ; \bm{R})$ is the cdf of an $m$-variate Gaussian distribution with zero mean vector and correlation matrix $\bm{R}$ and $\Phi^{-1}( \cdot )$ is the inverse of the univariate standard normal cdf. Thus, taking in eq. \eqref{eq:gausscop} $u_{k} = F_{Y_{k}}(y_{k})$, for each $k=1,\ldots,m$, we specify the cdf of $\bm{Y}=(Y_1,\ldots,Y_m)$ to be the Gaussian copula function. \cite{xue2000multivariate} proves that the density of the Gaussian copula is
\begin{equation}
c(u_{1},\ldots ,u_{m};\bm{R})=|\bm{R}|^{-\frac{1}{2}}\exp \left\{ -\frac{1}{2}(
\tilde{\bm{z}}^{T}\bm{R}^{-1}\tilde{\bm{z}}-\tilde{\bm{z}}^{T}\tilde{\bm{z}}
)\right\},
\label{eq:gausscopdens}
\end{equation}
where $\tilde{\bm{z}}$ is an $m$-dimensional vector and $\tilde{z}_{k} = \Phi^{-1}(u_{k})$, $k=1,\ldots,m$, is known as the normal score which follows marginally the standard Gaussian distribution. Then, $\widetilde{\bm{Z}}$ is an $m$-variate Gaussian distribution with zero mean vector and correlation matrix $\bm{R}$, i.e., $\widetilde{\bm{Z}} \sim \N_{m}(\bm{0},\bm{R})$. If all the marginal distributions are continuous, the matrix $\bm{R}$ can be interpreted as the correlation matrix of the elements of $\bm{Y}$ and zeros in its inverse imply the conditional independence among the corresponding elements of $\bm{Y}$. However, in the presence of discrete random variables, the notion of conditional independence has to be interpreted with care; zeros in $\bm{R}^{-1}$ imply that the observed variables are independent conditionally only on the latent variables \citep{webb2008bayesian}. Note also that if $\bm{R}$ is the identity matrix the elements of $\bm{Y}$ can be considered to be independent despite the presence of discrete variables, see \cite{xue2000multivariate}, \cite{song2009joint} and \cite{talhouk2012efficient} for a detailed discussion.

\subsection{Regression model}

Let $\bm{Y} =(\bm{y}_1^{T},\ldots,\bm{y}_n^{T})$ be the $(n\times m)$-dimensional matrix of observed data where each $\bm{y}_i=(y_{i1},\ldots,y_{im})$ consists of any combination of $m$ discrete and/or continuous responses. Moreover, let $\bm{x}_{ik}$ be the $p_{k}$-dimensional vector of predictors for the $i$th sample and the $k$th response. To model the marginal distribution of each $y_{ik}$, we specify its cdf $F_{k}(y_{ik};\bm{x}_{ik},\bm{\beta}_{k},\bm{\theta}_{k})$ to be the cdf of any parametric distribution. Our notation emphasizes the dependence of each $y_{ik}$ on the $p_{k}$-dimensional response-specific vector of predictors $\bm{x}_{ik}$, the associated regression coefficients $\bm{\beta}_{k}$ and the response-specific parameters $\bm{\theta}_{k}$.

The GCR model is described by the transformation
\begin{equation}
y_{ik}=h_{ik}^{-1}(\tilde{z}_{ik}),\,\,\ \widetilde{\bm{Z}}_{i}\sim \mathrm{N}_{m}(\bm{0},\bm{R}),
\label{eq:yh}
\end{equation}
where $\tilde{z}_{i1},\ldots,\tilde{z}_{im}$ are realizations from $\widetilde{\bm{Z}}_{i}$ and $h_{ik}(\cdot )=\Phi^{-1} \{F_{k}(\cdot; \bm{x}_{ik},\bm{\beta}_{k},\bm{\theta}_{k})\}$ for each $i=1,\ldots,n$ and $k=1,\ldots,m$.

By assuming that each $F_{k}(\cdot)$ is a member of the exponential family, we obtain the multivariate Generalized Linear Model presented in \cite{song2009joint} as the multivariate extension of the well-known single-response Generalized Linear Model \citep{mccullagh1989generalized} with $\bm{\theta}_{k}$ the specific vector of parameters for the $k$th response. The SUR model is a special case of eq. \eqref{eq:yh} when all margins are univariate Gaussian with mean $\bm{x}_{ik}^{T}\bm{\beta}_{k}$ and variance $\theta_{k}$ and $\bm{R}$ is the correlation matrix. The multi-response Probit regression model of \cite{chib1998analysis} is obtained from eq. \eqref{eq:yh} by specifying each margin to be the cdf of a Bernoulli random variable with probability of success $\Phi(\bm{x}_{ik}^{T}\bm{\beta}_{k} )$. Finally, setting $F_{k}(y_{ik};\bm{x}_{ik},\bm{\beta}_{k},\bm{\theta}_{k}) = \Phi(\theta_{kc}-\bm{x}_{ik}^{T}\bm{\beta}_{k} )$, $c=1,\ldots ,C_{k}-1$, with $C_{k}$ the number of categories, eq. \eqref{eq:yh} becomes a regression model for a combination of binary and ordinal observations and $\bm{\theta}_{k}=(\theta_{k1},\ldots\theta_{k,C_{k}-1})$ consists of the cut-points for the $k$th ordinal observation \citep{mccullagh1980regression}. In both multi-response Probit and ordinal regression models the matrix $\bm{R}$ is in the correlation form for identifiability conditions \citep{chib1998analysis}.

Eq. \eqref{eq:yh} implies that the joint likelihood function of the observations $\bm{Y}$ conditional on the correlation matrix $\bm{R}$, the regression coefficients $\bm{B}=\vect(\bm{\beta}_{1},\ldots,\bm{\beta}_{m})$ and the parameters $\bm{\Theta}=\vect(\bm{\theta}_{1},\ldots,\bm{\theta}_{m})$ is an intractable function of non-computable high-dimensional integrals \citep{song2009joint}.

\subsection{Prior distributions}
\label{sec:priors}

In this section, we specify the prior distributions on the regression coefficients $\bm{B}$, the associated sparsity prior on the inclusion probability and the correlation matrix $\bm{R}$ of the model. By noting that for different choices of the marginal cdfs $F_{k}(\cdot)$, $k=1,\ldots,m$, we have different vectors of parameters $\bm{\theta}_{k}$, we will assign their prior distributions differently in each simulated and real data example. See Supp. Mat. Section S.1 for details.

\subsubsection{Variable selection}
\label{sec:BVS}

We use a hierarchical non-conjugate model to assign a prior distribution on the regression coefficients of the GCR model defined in eq. \eqref{eq:yh}. A point mass at zero is specified on the regression coefficients of the unimportant predictors, whereas a Gaussian distribution is assigned to the non-zero regression coefficients \citep{george1997approaches}. By utilizing the binary latent vector $\bm{\gamma}_{k} =(\gamma_{k1},\ldots,\gamma_{kp_{k}})$, $k=1,\ldots,m$, where $\gamma_{kj}$ is $1$ if, for the $j$th predictor and the $k$th response, the regression coefficient is different from zero and $0$ otherwise, we assume that for each $k$
\begin{equation}
\beta _{kj}|\gamma _{kj}\overset{\text{iid}}{\sim }(1-\gamma _{kj})\delta
_{0}+\gamma _{kj}\mathrm{N}(0,v),\,\,\ j=1,\ldots,p_{k},  \label{eq:Bprior}
\end{equation}
\begin{equation*}
\gamma _{kj}|\pi _{k}\overset{\text{iid}}{\sim }\mathrm{Ber}(\pi _{k}),\,\,\ j=1,\ldots ,p_{k}
\end{equation*}
\begin{equation}
\pi _{k}\sim \mathrm{Beta}(a_{k},b_{k}),
\label{eq:Betaprior}
\end{equation}
where $\delta_{0}$ denotes a point mass at zero and $v$ is a fixed value. It is common practice to standardize the predictor variables, taking $v= 1$ in order to place appropriate prior mass on reasonable values of the non-zero regression coefficients \citep{hans2007shotgun}. Integrating out $\pi _{k}$, it is readily shown that marginally
\begin{equation}
p(\bm{\gamma}_{k}) = \frac{\mathrm{B}\left(a_{k} + |\bm{\gamma}_{k}| ,b_{k} + p_{k}-|\bm{\gamma}_{k}|  \right)}{\mathrm{B}(a_{k},b_{k})},
\label{eq:Gmarg}
\end{equation}
where $|\bm{\gamma}_{k}| = \sum_{j=1}^{p_{k}}\gamma_{kj}$. The hyper-parameters $a_{k}$ and $b_{k}$ can be chosen using prior information about the number of important covariates associated with the $k$th response and its variance. See \citep{kohn2001nonparametric} for further details on the elicitation of the hyper-parameters $a_{k}$ and $b_{k}$ in eq. \eqref{eq:Betaprior}.

The sparsity prior \eqref{eq:Betaprior} is a common choice in single-response BVS as well as in the sparse SUR model \citep{wang2010sparse}. 
When the predictors are common to all responses, the primary inferential question may shift to the identification of key predictors that exert their influence to a large fraction of responses at the same time. When a large number of responses are regressed independently on the same set of predictors, \cite{Richardson2010} proposed to decompose the \emph{a priori} \lq\lq cell\rq\rq\ inclusion probability into its marginal effects, i.e., $\pi _{kj} = \pi_{k} \times \rho_{j}$, $\pi _{kj} \in [0, 1]$. The idea behind this decomposition is to control the level of sparsity for each response $k$ through a suitable choice of the hyperparameters $a_{k}$ and $b_{k}$ as in \eqref{eq:Betaprior}, while $\rho_{j} \sim \mathrm{Ga}(c,d)$ captures the relative \lq\lq propensity\rq\rq\ of predictor $j$ to influence several responses at the same time. For a detailed discussion regarding the effect of the prior $p(\rho_{j})$ on BVS in multiple-response regression models, see \cite{Ruffieux2020}.

In the following, for ease of notation, we denote by $\bm{\beta}_{\bm{\gamma}_{k}}$ all non-zero elements of $\bm{\beta}_{k}$ and analogously, we indicate by $\bm{x}_{i,\bm{\gamma}_{k}}$, the elements of $\bm{x}_{ik}$ corresponding to those elements of ${\bm{\gamma}_{k}}$ equal to 1. Let $\bm{X}_{k}$ be the $n \times p_{k}$ matrix of all available predictors for the $k$th response. Similarly to the vector case, $\bm{X}_{\bm{\gamma}_{k}}$ denotes the $n \times |\bm{\gamma}_{k}|$ matrix in which the columns are selected according to the latent binary vector $\bm{\gamma}_{k}$. Finally, we refer to  $\bm{\Gamma}= (\bm{\gamma}_{1},\ldots,\bm{\gamma}_{m})$ as the latent binary matrix.

\subsubsection{Correlation matrix \textbf{\emph{R}} and adjacency matrix \textbf{\emph{G}}}
\label{sec:covsel}

To specify a prior distribution on the correlation matrix $\bm{R}$, we follow \cite{talhouk2012efficient} and use the parameter expansion for data augmentation strategy \citep{liu1999parameter} to expand the correlation matrix into a covariance matrix. This choice allows us to specify conjugate prior distributions on the resulting covariance matrix. In particular, we utilize the hyper-inverse Wishart \citep{dawid1993hyper} to allow for zero entries in the inverse of the covariance matrix. Details can be summarized as follows.

First, to expand $\bm{R}$ into a covariance matrix, we define the transformation $\widetilde{\bm{W}}=\widetilde{\bm{Z}}\bm{D}$, where $\widetilde{\bm{Z}}$ is the $n \times m$ matrix of the Gaussian latent variables and $\bm{D}$ is an $m \times m$ diagonal matrix with elements $\delta_{k}$, $k=1,\ldots,m$. Then,
\begin{equation}
\vect(\widetilde{\bm{W}})\sim \mathrm{N}_{mn}(\bm{0}, \bm{\Sigma} \otimes \bm{I}_{n}),
\label{eq:W}
\end{equation}
where $\bm{\Sigma} = \bm{D}\bm{R}\bm{D}$ and $\bm{I}_{n}$ is a diagonal matrix of dimension $n$ that encodes the independence assumption amongst the observations. Then, a conjugate prior distribution can be assigned on $\bm{\Sigma}$ and updated at each iteration of the MCMC algorithm before it is projected back to $\bm{R}$ using the inverse transformation $\bm{R} = \bm{D}^{-1}\bm{\Sigma} \bm{D}^{-1}$. We also utilize the theory of decomposable models to perform a conjugate analysis of the covariance structure of the model since the hyper-inverse Wishart distribution is a conjugate prior distribution for the covariance matrix $\bm{\Sigma}$ with respect to the adjacency matrix $\bm{G}$ of a decomposable graph $\mathcal{G}$. The diagonal elements in the adjacency matrix are always restricted to be 1 to ensure the positive definiteness of $\bm{G}$. 

Second, we assign the following prior structure on $\bm{G}$ and $\bm{D}$. Let $g_{\ell}$, $\ell=1,\ldots,m(m-1)/2$, be the binary indicator for the presence of the $\ell$th off-diagonal edge in the lower triangular part of the adjacency matrix $\bm{G}$ of the decomposable graph $\mathcal{G}$. We assume that
\begin{equation}
g_{\ell}  \sim  \mathrm{Ber}(\pi_{\bm{G}}),\,\,\, \pi_{\bm{G}} \sim \mathrm{Unif}(0,1),\,\,\, \ell=1,\ldots,m(m-1)/2,\,\,\, \bm{G} \in \mathcal{G}.
\label{eq:Gprior}
\end{equation}
For a detailed discussion regarding compatible priors on $\bm{G}$ for decomposable graphs, see \cite{Bornn2011}.

We denote by $p(\bm{G})$ the induced marginal prior on the adjacency matrix and define the joint distribution of $\bm{D}$, $\bm{R}$ and $\bm{G}$ as
\begin{equation*}
p(\bm{D},\bm{R},\bm{G}) = p(\bm{D}|\bm{R})p(\bm{R}|\bm{G})p(\bm{G}),
\label{eq:DRG}
\end{equation*}
where
\begin{equation}
\delta _{k}^{2}|r^{kk}\overset{\text{iid}}{\sim }\mathrm{IGam}
((m+1)/2,r^{kk}/2),\,\,\, k=1,\ldots,m
\label{eq:delta}
\end{equation}
with $r^{kk}$ the $k$th diagonal element of $\bm{R}^{-1}$. Assuming a uniform prior for $\bm{R}|\bm{G}$, it can be shown that, for a decomposable graph $\mathcal{G}$, $\bm{\Sigma} |\bm{G} \sim \H\I\W_{\mathcal{G}}(2,\bm{I}_m)$ \citep{talhouk2012efficient}. 



\section{MCMC sampling strategy}
\label{sec:MCMCsec}

We are interested in sampling from the joint posterior distribution $p(\bm{B},\bm{\Gamma},\bm{\Theta},\widetilde{\bm{Z}},\bm{D},\bm{R},\bm{G}|\bm{Y})$. To draw samples from the specified model, we design a novel MCMC algorithm which proceeds as follows. We first update the regression coefficients $\bm{B}$, the latent binary matrix $\bm{\Gamma}$, the parameters $\bm{\Theta}$ and the Gaussian latent variables $\widetilde{\bm{Z}}$ in $m$ blocks. In particular, for each $k=1,\ldots,m$, we sample jointly $(\bm{\beta}_{k},\bm{\gamma}_{k})$, then we update $\bm{\theta}_{k}$ and, finally, we draw realizations from the Gaussian latent variables $\widetilde{\bm{Z}}_{k}=(\widetilde{Z}_{1k},\ldots ,\widetilde{Z}_{nk})$. In the last step, we draw the correlation matrix $\bm{R}$ and the adjacency matrix $\bm{G}$ from their full conditional distributions.

Algorithm \ref{algo:MCMC} presents the pseudo code of the designed MCMC algorithm where the subscript \lq\lq $-k$\rq\rq\ implies that the corresponding matrix (or vector) consists of all the elements except those that are related to the $k$th response.

\begin{algorithm}[!h]
\caption{MCMC algorithm for sampling from the joint distribution $p(\bm{B},\bm{\Gamma},\bm{\Theta},\widetilde{\bm{Z}},\bm{D},\bm{R},\bm{G}|\bm{Y})$}
\label{algo:MCMC}
\begin{algorithmic}[1]
 \State Set the number of iterations $S$.
 \For{\texttt{$s = 1,\ldots,S$ }}
\For{\texttt{$k = 1,\ldots,m$ }}
\State Sample from  $p(\bm{\beta}_{k},\bm{\gamma}_{k}  |\bm{y}_{k},\bm{\theta}_{k}, \widetilde{\bm{Z}}_{-k},\bm{D},\bm{R})$
 \State Sample from  $p(\bm{\theta}_{k}|\bm{y}_{k},\bm{\beta}_{k},\bm{\gamma}_{k},
\widetilde{\bm{Z}}_{-k},\bm{D},\bm{R})$
\State Sample from $p(\tilde{\bm{z}}_{k}|\bm{y}_{k},\bm{\beta}_{k},\bm{\gamma}_{k},\bm{\theta}_{k}, \widetilde{\bm{Z}}_{-k},\bm{D},\bm{R})$
 \EndFor
 \State Sample from $p(\bm{D},\bm{R}|\bm{Y},\widetilde{\bm{Z}},\bm{G})$
 \State Sample from $p(\bm{G}|\bm{Y},\widetilde{\bm{Z}},\bm{D},\bm{R})$
\EndFor
\end{algorithmic}
\end{algorithm}

For the large majority of responses, drawing $(\bm{\beta}_{k},\bm{\gamma}_{k})$ in step 4 is complicated due to the unavailability of the full conditional distributions in closed-form expression and a M-H step is therefore required. However, because of the uncertainty associated with the latent binary vector $\bm{\gamma}_{k}$, the proposal distribution requires a different dimension at each MCMC iteration. In this framework, a commonly used tool is the Reversible Jump algorithm \citep{green1995reversible}, although it may experience a low acceptance rate, resulting in a MCMC sampler with poor mixing \citep{brooks2003efficient,Lamnisos2009}.

To avoid the Reversible Jump algorithm, we design a proposal distribution defined on a fixed-dimensional space that can be used for the joint update of $\bm{\beta}_{k}$ and $\bm{\gamma}_{k}$. It is worth noticing that we conduct the update of $(\bm{\beta}_{k},\bm{\gamma}_{k})$ by first integrating out the latent variables $\widetilde{\bm{Z}}_{k}$. This accelerates the convergence of the proposed MCMC algorithm since conditioning on $\widetilde{\bm{Z}}_{k}$ induces many restrictions on the admissible values of $(\bm{\beta}_{k},\bm{\gamma}_{k})$, see \cite{pitt2006efficient}. Drawing samples from the posterior distributions of the remaining parameters and the Gaussian latent variables of the model can be conducted by using standard MCMC algorithms which we also describe briefly in this section.


\subsection{Proposal distribution for variable selection}
\label{sec:sampleB}

To sample jointly the regression coefficients $\bm{\beta}_{k}$ and the latent binary vector $\bm{\gamma}_{k}$ for each $k=1,\ldots,m$ (step 4 Algorithm \ref{algo:MCMC}), we design a M-H step that targets the distribution with density
\begin{equation}
p(\bm{\beta}_{k},\bm{\gamma}_{k}|\bm{y}_{k},\bm{\theta}_{k},\widetilde{\bm{Z}}_{-k},\bm{D},\bm{R})
\propto p(\bm{y}_{k}|\bm{\beta}_{k},\bm{\gamma}_{k},\bm{\theta}
_{k},\widetilde{\bm{Z}}_{-k},\bm{D},\bm{R})p(\bm{\beta}_{k}|\bm{\gamma}_{k})p(\bm{\gamma}_{k}),
\label{eq:targetBG}
\end{equation}
where $p(\bm{\beta}_{k}|\bm{\gamma}_{k})$ and $p(\bm{\gamma}_{k})$ are the prior densities on $\bm{\beta}_{k}$ and $\bm{\gamma}_{k}$ as defined in eq. \eqref{eq:Bprior} and eq. \eqref{eq:Gmarg}, respectively.

In the case of a continuous response $\bm{y}_{k}$, we have from eq. \eqref{eq:gausscopdens} that
\begin{eqnarray}
p(\bm{y}_{k}|\bm{\beta}_{k},\bm{\gamma}_{k},\bm{\theta}_{k},\widetilde{\bm{Z}}
_{-k},\bm{D},\bm{R}) &\propto &\exp \left\{\frac{1}{2}(1-r^{kk}) \sum_{i=1}^{n}\tilde{z}_{ik}^{2}-\sum_{i=1}^{n}\sum_{\ell =1,\ell \neq k}^{m}r^{k\ell }\tilde{z}_{ik}\tilde{z}_{i\ell }+\right.   \notag \\
&&+\left. \sum_{i=1}^{n}\log f_{k}(y_{ik};\bm{x}_{i,\bm{\gamma}_{k}},
\bm{\beta}_{\bm{\gamma}_{k}},\bm{\theta}_{k})\right\},  \label{eq:cont_lik}
\end{eqnarray}
where we condition on $\widetilde{\bm{Z}}_{-k}$ with an abuse of notation \citep{pitt2006efficient} since $\tilde{z}_{ik}=\Phi ^{-1}\{F_{k}(y_{ik};\bm{x}_{i,\bm{\gamma}_{k}}, \bm{\beta}_{\bm{\gamma}_{k}},\bm{\theta}_{k})\}$ is a shorter expression of the transformation in eq. \eqref{eq:yh}, $r^{k\ell}=(\bm{R}^{-1})_{k\ell}$ and $f_{k}(y_{ik}; \bm{x}_{i,\bm{\gamma}_{k}},\bm{\beta}_{\bm{\gamma}_{k}},\bm{\theta}_{k})$ denotes the probability density function of $F_{k}(y_{ik}; \bm{x}_{i,\bm{\gamma}_{k}},\bm{\beta}_{\bm{\gamma}_{k}},\bm{\theta}_{k})$.

If $\bm{y}_{k}$ is a discrete response, we have that
\begin{equation}
p(\bm{y}_{k}|\bm{\beta}_{k},\bm{\gamma}_{k},\bm{\theta}_{k},\widetilde{\bm{Z}}
_{-k},\bm{D},\bm{R})=\prod_{i=1}^{n}\left\{ \Phi \left( \dfrac{u_{ik}-\tilde{\mu} _{ik|-k}}{\tilde{\sigma} _{k|-k}}\right) -\Phi \left( \dfrac{\ell _{ik}-\tilde{\mu} _{ik|-k}}{\tilde{\sigma}_{k|-k}}\right) \right\},
\label{eq:disc_lik}
\end{equation}
where $\ell_{ik} = \Phi^{-1}\{F_{k}(y_{ik}-1;\bm{x}_{i,\bm{\gamma}_{k}},\bm{\beta}_{\bm{\gamma}_{k}},
\bm{\theta}_{k})\}$ and $u_{ik} =\Phi^{-1}\{F_{k}(y_{ik};\bm{x}_{i,\bm{\gamma}_{k}},\bm{\beta}_{\bm{\gamma}_{k}},
\bm{\theta}_{k})\}$ with $\tilde{\mu} _{ik|-k}=\bm{R}_{k,-k}\bm{R}_{-k,-k}^{-1}\tilde{\bm{z}}_{i,-k}$ and $\tilde{\sigma}
_{k|-k}^{2}=1-\bm{R}_{k,-k}\bm{R}_{-k,-k}^{-1}\bm{R}_{-k,k}$. Thus, the conditional density in eq. \eqref{eq:targetBG} will be intractable for the majority of the continuous and for all the discrete distributions that can be used for the marginal modeling of the $k$th response.

Instead of relying on the Reversible Jump algorithm or the Laplace approximation to sample from eq. \eqref{eq:targetBG}, we utilize a M-H step with a fixed-dimensional proposal distribution obtained by re-parameterizing the Gaussian latent variables $\widetilde{\bm{Z}}_{k}$. More precisely, for each $i=1,\ldots,n$ and $k=1,\ldots,m$, we set
\begin{equation}
\begin{cases}
      Z_{ik} = \bm{x}_{i,\bm{\gamma}_{k}}^{T}\bm{\beta}_{\bm{\gamma}_{k}} + \delta_{k}\widetilde{Z}_{ik} & \text{if} \,\, \bm{y}_{k} \,\, \text{is continuous,} \\
      Z_{ik} = \bm{x}_{i,\bm{\gamma}_{k}}^{T}\bm{\beta}_{\bm{\gamma}_{k}} + \widetilde{Z}_{ik} & \text{if} \,\, \bm{y}_{k} \,\, \text{is discrete},
\end{cases}
\label{eq:transZ}
\end{equation}
where $Z_{ik}$ is marginally normal distributed with mean $\mu_{ik}=\bm{x}_{i,\bm{\gamma}_{k}}^{T}\bm{\beta}_{\bm{\gamma}_{k}}$ and variance either $\sigma^2_{k}=\delta_{k}^{2}$ (if continuous) or $\sigma^2_{k}=1$ (if discrete). Thus, the vector of realizations $\bm{z}_{k}$ is a sufficient statistics for $(\bm{\beta}_{k},\bm{\gamma}_{k})$, whereas $\tilde{\bm{z}}_{k}$ is ancillary \citep{Yu2011}. Irrespectively of the type of the response, eq. \eqref{eq:transZ} creates a link between the Gaussian Copula model, where the Gaussian latent variables are non-centered, and BVS that requires a centered parametrization. With this transformation, we also aim to keep the advantages of using (data augmentation) centered auxiliary variables when performing BVS \citep{holmes2006bayesian}. In particular, the fact that the vector $\bm{z}_{k}$ retains information about the likelihood is key since, regardless of the response's type, it allows an effective update of $\bm{\beta}_{k}$ given a change in the predictors set.

To construct a proposal distribution for a M-H step that targets eq. \eqref{eq:targetBG}, we replace the intractable density $p(\bm{y}_{k}|\bm{\beta}_{k},\bm{\gamma}_{k},\bm{\theta}_{k},\widetilde{\bm{Z}}_{-k},\bm{R})$ with the Gaussian density of the latent variable $\bm{Z}_{k}$ conditioned on $\widetilde{\bm{Z}}_{-k}$. Exploiting the fact that $\bm{\beta}_{k}$ is quadratic in
$p(\bm{z}_{k}|\bm{\beta}_{k},\bm{\gamma}_{k},\widetilde{\bm{Z}}_{-k},\bm{D},\bm{R})$, given a candidate value of the latent binary vector $\bm{\gamma}_{k}^{\ast}$, we propose the non-zero regression coefficients $\bm{\beta}_{k}^{\ast}$ from the distribution with density
\begin{align}
q(\bm{\beta}_{k}^{\ast }|\bm{\gamma}_{k}^{\ast},\bm{z}_{k},\widetilde{\bm{Z}}_{-k},\bm{D},\bm{R})& \propto p(\bm{z}_{k}|\bm{\beta}_{k}^{\ast},\bm{\gamma}_{k}^{\ast},\widetilde{\bm{Z}}_{-k},\bm{D},\bm{R})
p(\bm{\beta}_{k}^{\ast }|\bm{\gamma}_{k}^{\ast })  \notag \\
& \propto \mathcal{N}_{|\bm{\gamma}_{k}^{\ast }|}(\bm{\beta}_{k}^{\ast };\bm{V}_{\bm{\gamma}_{k}^{\ast }}\bm{X}_{\bm{\gamma}_{k}^{\ast }}^{T}\sigma_{k|-k}^{-2}\{\bm{z}_{k}-\sigma_{k}\tilde{\bm{\mu}}_{k|-k}\},
\bm{V}_{\bm{\gamma}_{k}^{\ast }}),
\label{eq:Bprop}
\end{align}
where $\sigma_{k|-k}^{2}$ is the diagonal element of the conditional covariance matrix of $\bm{Z}_{k}|\widetilde{\bm{Z}}_{-k}$, $\bm{V}_{\bm{\gamma}_{k}^{\ast}}^{-1}=\sigma_{k|-k}^{-2}\bm{X}_{\bm{\gamma}^{\ast }_{k}}^{T}\bm{X}_{\bm{\gamma}^{\ast }_{k}}+v^{-1}\bm{I}_{\left\vert \bm{\gamma}^{\ast }_{k}\right\vert }$ and $\tilde{\bm{\mu}}_{k|-k}$ is the conditional mean of $\widetilde{\bm{Z}}_{k}|\widetilde{\bm{Z}}_{-k}$. The proposal distribution in eq. \eqref{eq:Bprop} shows two important features: (i) in the covariance matrix $\bm{V}_{\bm{\gamma}^{\ast }_{k}}$, the term $\bm{X}_{\bm{\gamma}^{\ast }_{k}}^{T}\bm{X}^{\ast }_{\bm{\gamma}_{k}}$ is rescaled by the conditional variance $\sigma _{k|-k}^{-2}$ and (ii) the proposal mean is shifted by the rescaled conditional mean $\sigma_{k}\tilde{\bm{\mu}}_{k|-k}$, where the factor $\sigma_{k}$ brings $\bm{z}_{k}$ and $\tilde{\bm{\mu}}_{k|-k}$ on the same scale. Thus, the realizations $\bm{z}_{k}$ of the Gaussian latent variables $\bm{Z}_{k}|\widetilde{\bm{Z}}_{-k}$ in eq. \eqref{eq:Bprop} are recentered to remove the effect of conditioning on $\widetilde{\bm{Z}}_{-k}$, i.e., $\E(\bm{Z}_{k}|\widetilde{\bm{Z}}_{-k}-\sigma_{k}\tilde{\bm{\mu}}_{k|-k})=
\E(\bm{Z}_{k})=\bm{X}_{\bm{\gamma}_{k}}\bm{\beta} _{\bm{\gamma}_{k}}$, see Supp. Mat. Section S.3 for details.

\remark{If $\bm{R}$ is the identity matrix, then the proposal distribution \eqref{eq:Bprop} becomes
\begin{equation}
q(\bm{\beta}_{k}^{\ast }|\bm{\gamma}_{k}^{\ast },\bm{z}
_{k},\bm{D})\propto \mathcal{N}_{|\bm{\gamma}_{k}^{\ast }|}(\bm{\beta}_{k}^{\ast};\bm{V}_{\bm{\gamma}_{k}^{\ast }}\bm{X}_{\bm{\gamma}_{k}^{\ast }}^{T}\sigma
_{k}^{-2}\bm{z}_{k},\bm{V}_{\bm{\gamma}_{k}^{\ast }}),
\label{eq:noR}
\end{equation}
since $\bm{V}_{\bm{\gamma}^{\ast }_{k}}^{-1}= \sigma_{k}^{-2}\bm{X}_{\bm{\gamma}^{\ast}_{k}}^{T}\bm{X}_{\bm{\gamma}^{\ast }_{k}} + v^{-1}\bm{I}_{\bm{\gamma}^{\ast }_{k}}$ and the conditional mean $\tilde{\bm{\mu}}_{k|-k}$ is equal to zero. When no information is shared across responses, the proposal distribution in eq. \eqref{eq:noR} coincides with the full conditional distribution of the non-zero regression coefficients used in single-response Probit ($\sigma^2_{k}=1$) regression model \citep{albert1993bayesian,holmes2006bayesian}. It is worth noticing that, regardless of the similarities with the Probit model, this proposal density has been obtained as a special case of eq. \eqref{eq:Bprop}.

\remark{For some continuous and discrete data including the Gaussian, binary, ordered and nominal categorical responses, the proposal density in eq. \eqref{eq:Bprop} allows the \lq\lq implicit marginalisation\rq\rq\ of the regression coefficients when the joint update of $(\bm{\beta}_{k},\bm{\gamma}_{k})$ in the M-H step is performed. For this type of responses, the acceptance probability $\alpha$ of the joint move is
\begin{equation}
\alpha =1\wedge \frac{\left\vert \bm{V}_{\bm{\gamma}_{k}^{\ast }}\right\vert^{1/2}v^{|\bm{\gamma}_{k}|/2}\exp \left(\frac{1}{2}\bm{m}_{\bm{\gamma}
_{k}^{\ast }}^{T}\bm{V}_{\bm{\gamma}_{k}^{\ast }}^{-1}\bm{m}_{\bm{\gamma}
_{k}^{\ast }}\right) p(\bm{\gamma}_{k}^{\ast })q(\bm{\gamma}_{k}|\bm{\gamma}
_{k}^{\ast })}{\left\vert \bm{V}_{\bm{\gamma}_{k}}\right\vert ^{1/2}v^{|
\bm{\gamma}_{k}^{\ast }|/2}\exp \left( \frac{1}{2}\bm{m}_{\bm{\gamma}
_{k}}^{T}\bm{V}_{\bm{\gamma}_{k}}^{-1}\bm{m}_{\bm{\gamma}_{k}}\right) p(
\bm{\gamma}_{k})q(\bm{\gamma}_{k}^{\ast }|\bm{\gamma}_{k})},  \label{eq:MH}
\end{equation}
where
$\bm{m}_{\bm{\gamma}^{\ast }_{k}}=\bm{V}_{\bm{\gamma}^{\ast }_{k}}\bm{X}_{\bm{\gamma}^{\ast }_{k}}^{T}
\sigma_{k|-k}^{-2}(\bm{z}_{k}-\sigma_{k}\tilde{\bm{\mu}}_{k|-k})$. Similarly to \cite{holmes2006bayesian}, in eq. \eqref{eq:MH} the current and proposed value of the regression coefficients $(\bm{\beta}_{k},\bm{\beta}_{k}^{\ast })$ do not appear. 
For details, see Supp. Mat. Section S.4.}

Finally, since the main focus of this paper is to provide a proposal distribution for the regression coefficients given the latent binary vector, any proposal distribution for $\bm{\gamma}_{k}^{\ast}$ can be used. 
Here, we use a modified version of the proposal distribution designed by \cite{guan2011bayesian}. This proposal makes efficient use of the inexpensive evaluation of the marginal association between each response and the predictors in order to propose a new latent vector $\bm{\gamma}_{k}^{\ast}$. See Supp. Mat. Section S.2 for a detailed description.

\subsection{Sampling the Gaussian latent variables}

Next, we describe the sampling strategy for the Gaussian latent variables (step 6 Algorithm \ref{algo:MCMC}). If the $k$th response is continuous then the transformation \eqref{eq:yh} is a one-to-one transformation. In this case $\tilde{\bm{z}}_{k}$ is updated deterministically by setting $\tilde{z}_{ik}=\Phi ^{-1}\{F_{k}(y_{ik};\bm{x}_{i,\bm{\gamma}_{k}},
\bm{\beta}_{\bm{\gamma}_{k}},\bm{\theta} _{k})\}$ for each $i=1,\ldots,n$ and $k=1,\ldots,m$.
For a discrete response $\bm{y}_{k}$, we have that
\begin{equation}
p(\tilde{z}_{ik}|\bm{y}_{k},\bm{\beta}_{k},\bm{\gamma}_{k},\bm{\theta}_{k},
\widetilde{\bm{Z}}_{-k},\bm{D},\bm{R})\propto\mathcal{N}(\tilde{z}_{ik};\tilde{\mu} _{ik|-k},\tilde{\sigma}_{k|-k}^{2})\mathbb{I}\{\tilde{z}_{ik}\in (\ell _{ik},u_{ik}]\},
\label{eq:zupdate}
\end{equation}
where $\tilde{\mu}_{ik|-k}$, $\tilde{\sigma}_{k|-k}^{2}$, $\ell_{ik}$ and $u_{ik}$ are defined in eq. \eqref{eq:disc_lik}. Therefore, each $\tilde{z}_{ik}$ has to be sampled from the Gaussian distribution $\mathrm{N}(\tilde{\mu}_{ik|-k},\tilde{\sigma}_{k|-k}^{2})$ truncated on the interval $(\ell_{ik},u_{ik}]$.


\subsection{Sampling the matrix \textbf{\emph{D}}, the correlation matrix \textbf{\emph{R}} and the adjacency matrix \textbf{\emph{G}}}
\label{sec:sampleR}

To update $(\bm{D},\bm{R})$ and $\bm{G}$ in the designed MCMC Algorithm \ref{algo:MCMC} (step 8 and 9), we follow \cite{talhouk2012efficient} and work in the space of the scaled Gaussian latent variables $\widetilde{\bm{W}}=\widetilde{\bm{Z}}\bm{D}$. 
In practice, we obtain samples from $p(\bm{\Sigma}|\widetilde{\bm{W}}, \bm{G})$ and transform them using the inverse transformation $\bm{R} = \bm{D}^{-1}\bm{\Sigma} \bm{D}^{-1}$ where $\bm{D}$ is sampled from its prior distribution in eq. \eqref{eq:delta}. To sample the adjacency matrix $\bm{G}$ from $p(\bm{G}|\bm{Y},\widetilde{\bm{Z}},\bm{D},\bm{R})$, we target the distribution with density
\begin{equation}
p(\bm{G}|\widetilde{\bm{W}}) = \dfrac{p(\widetilde{\bm{W}}|\bm{G})p(\bm{G})}{\sum_{\bm{G}}p(\widetilde{\bm{W}}|\bm{G})p(\bm{G})},
\label{eq:postG}
\end{equation}
where the summation in the denominator is over all the decomposable graphs $\mathcal{\bm{G}}$, $p(\bm{G})$ is defined through eq. \eqref{eq:Gprior} and $p(\widetilde{\bm{W}}|\bm{G})$ can be computed analytically due to the tractability of the hyper-inverse Wishart distribution, see Supp. Mat. Section S.5 for further details. We sample from eq. \eqref{eq:postG} by using a M-H step in which, conditionally on the current adjacency matrix $\bm{G}$, a new graph is proposed by adding or deleting an edge between two vertices whose index has been chosen randomly between the vertices that belong to a decomposable graph. The proposed graph is then accepted or rejected using the accept/reject mechanism of the M-H step which targets the density in eq. \eqref{eq:postG}. Finally, conditionally on $\widetilde{\bm{W}}$ and the updated adjacency matrix $\bm{G}$, we sample $\bm{\Sigma}$ from its conditional distribution $\H\I\W_{\mathcal{G}}(2+n,\bm{I}_m+\widetilde{\bm{W}}^{T}\widetilde{\bm{W}})$.


\remark{The choice of a decomposable graphical model can be relaxed to include non-decomposable graphs although at a higher computational cost. Within our framework, posterior samples of the adjacency matrix can be obtained by using the \texttt{BDgraph} algorithm proposed by \cite{Mohammadi2019} directly on the space of the scaled Gaussian latent variables $\widetilde{\bm{W}}$ where the problem of the intractable normalizing constants appearing in $p(\widetilde{\bm{W}}|\bm{G})$ for non-decomposable graphs is circumvented by using the solutions proposed by \cite{Wang2012} and \cite{Lenkoski2013}. As a note of caution, by specifying a non-decomposable graphical model, (all else unchanged) the induced prior on $\bm{R}$ for the incomplete prime components of $\mathcal{\bm{G}}$ may not be uniform. See Supp. Mat. Section S.5 for a detailed discussion.}

\subsection{Sampling the response-specific parameters}

Sampling the response-specific parameters $\bm{\Theta}$ depends on the marginal cdfs of the BVSGCR model. If $F_{k}(\cdot)$ is the cdf of a normal distribution, since $\theta_{k}=\delta^{2}_{k}$, the posterior samples of $\theta _{k}$ are obtained as a by-product of procedure described above. If the margins are ordinal or negative binomial, as in the real examples considered below, the response-specific parameters are sampled using a M-H step.

\section{Simulation study}
\label{sec:simstudy}

In this section, we compare the performance of the proposed Bayesian variable selection for Gaussian Copula Regression (BVSGCR) model with widely-used methods for BVS in single-response linear and non-linear regression models. We tested our multivariate method in two simulated data sets consisting of a combination of Gaussian, binary and ordinal responses, Section \ref{sec:Ataxia-type} and Gaussian and count responses, Section \ref{sec:epilepsy-type}, respectively. 

We used the marginal posterior probability of inclusion (MPPI) \citep{george1997approaches} to assess the predictor-response association, defined as the frequency a particular predictor is
included in a model during the MCMC exploration. To illustrate the performance of the different methods, we used the Receiver Operating Characteristic (ROC) curve. For a given response, the ROC curve plots the proportion of correctly detected important predictors (true positive rate - TPR) against the proportion of misidentified predictors (false positive rate - FPR) over a range of specified thresholds for the MPPI. 
To take into account the Monte Carlo error, we reported the mean of TPR and FPR over the simulated replicates for each scenario considered along with the corresponding averaged areas under the curve and their standard deviations. We also assess the accuracy of the non-zero regression coefficients' posterior credible intervals by using a modified version of the interval score described in \cite{gneiting2007strictly}.

\subsection{Data generation}
\label{sec:simmodel}

To generate the correlated predictors, we followed \cite{rothman2010sparse} and simulated, for each $i=1,\ldots,n$ and $k=1,\ldots,m$, $\bm{x}_{ik} \sim \N_{p_{k}}(0,\bm{S})$, where $\bm{S}_{jj'} = 0.7^{|j-j'|}$ is the $(j,j')$th element of $\bm{S}$, $j,j'=1,\ldots,p_{k}$, implying the same unit marginal variance. 
We also assumed that, for all responses, we had the same set of available predictors. 

\begin{table}[t]
\caption{Values of the parameters used to simulate the predictors and response variables described in Section \ref{sec:simmodel} and \ref{sec:Ataxia-type}.}
\label{tab:sim_values}
\centering
\begin{tabular}{ccccccccc}
\hline
& Response & $n$ & {$m$} & $p_{k}$ & $\pi _{1}$ & $\pi _{2}$ & $b$ & $s^{2}$
\\ \hline
\multirow{2}{*}{Scenario I \& III} & Gaussian & \multirow{2}{*}{$50$} & \multirow{2}{*}{$6\left\{\begin{array}{c} 3 \\ 3 \end{array} \right. $ \& $4\left\{\begin{array}{c} 1 \\ 3 \end{array} \right. $} & \multirow{2}{*}{$30$} & \multirow{2}{*}{$0.15$} & \multirow{2}{*}{$0.95$} & $
1$ & $1$ \\
& Discrete &  &  &  &  &  & $0.5$ & $0.2$ \\
\rule{0pt}{15pt} \multirow{2}{*}{Scenario II \& IV} & Gaussian &
\multirow{2}{*}{$100$} & \multirow{2}{*}{$6\left\{\begin{array}{c} 3 \\ 3 \end{array} \right. $ \& $4\left\{\begin{array}{c} 1 \\ 3 \end{array} \right. $} & \multirow{2}{*}{$100$} & \multirow{2}{*}{$0.05$} & \multirow{2}{*}{$0.95$} & $1$ & $1$ \\
& Discrete &  &  &  &  &  & $0.5$ & $0.2$ \\ \hline
\end{tabular}
\end{table}

We simulated a sparse vector of regression coefficients $\bm{B}$ and a sparse inverse correlation matrix $\bm{R}^{-1}$ according to the structure described in Section \ref{sec:priors}. More precisely, we first constructed the $p \times m$ matrix $\bm{B} = \bm{\Gamma}_{1} \odot \bm{\Gamma}_{2} \odot \bm{B}_{3} $, where $p=\sum_{k=1}^mp_{k}$ and $\odot$ denotes the Hadamard matrix product, as follows. Each cell of $\bm{\Gamma}_{1}$ has independent Bernoulli entries with success probability $\pi_{1}$, $\bm{\Gamma}_{2}$ has rows that are either all ones or all zeros and $\bm{B}_{3}$ consists of independent draws from $\N(b,s^{2})$. The decision regarding the zero rows in $\bm{\Gamma}_{2}$  has to be made using $p$ independent Bernoulli variables with probability of success $\pi_{2}$. As noted by \cite{rothman2010sparse}, using this simulation scheme, $(1-\pi_{2})p$ predictors are expected to be irrelevant for all the responses and each relevant predictor will be associated on average with $\pi_{1}m$ responses. We set $\bm{\beta}_{k}$ to be the $k$th column of $\bm{B}$. The choice of the parameters $\pi_{1}$,$\pi_{2}$,$b$ and $s^{2}$ will be different for each simulated scenario and it is summarized in Table \ref{tab:sim_values}. Finally, we used the correlation matrix of the autoregressive model of order one to simulate $\widetilde{\bm{Z}}_{i} \overset{\text{iid}}{\sim }\mathrm{N}_m(0,\bm{R})$ for each $i=1,\ldots,n$ and we set $\bm{R}_{kk'} = 0.8^{|k-k'|}$ for the $(k,k')$th element of $\bm{R}$, $k,k' =1,\ldots,m$, which implies a tri-diagonal sparse inverse correlation matrix. Supp. Mat. Figure S.1 shows the graphs implied by the non-zero pattern of $\bm{R}^{-1}$ used in the simulation study.

\subsection{Mixed Gaussian, binary and ordinal responses}
\label{sec:Ataxia-type}

We tested the proposed model in two different scenarios. Both scenarios consist of $m=6$ responses with three Gaussian, one binary and two ordinal (with three and four categories) variables. In Scenario II, we generated $20$ replicates with $100$ samples and $p_{k}=100$ predictors, $k=1,\ldots,m$, whereas in Scenario I we simulated the same number of replicates with $n=50$ and $p_{k}=30$. We constructed the sparse vector of regression coefficients as described in Section \ref{sec:simmodel} by setting the values the parameters $\pi_{1}$, $\pi_{2}$, $b$ and $s^{2}$ as shown in Table \ref{tab:sim_values}. With this choice of the parameters $\pi_{1}$ and $\pi_{2}$, in each scenario we simulated on average between four and five predictors associated with each response and with a small probability that they will be the same across responses since $\pi_{1}m<1$.

To simulate realizations from the correlated responses, for $k=1,2,3$, we set $F_{k}(\cdot;\bm{x}_{ik},\bm{\beta}_{k},\theta_{k})$ to be the cdf of the Gaussian distribution with mean $\bm{x}_{ik}^{T}\bm{\beta}_{k}$ and variance $\theta_{k}=3$. For $k=4$, we set $F_{k}(\cdot;\bm{x}_{ik},\bm{\beta}_{k},\theta_{k})$ to be the cdf of the Bernoulli distribution with mean $\Phi(\bm{x}_{ik}^{T}\bm{\beta}_{k})$ and $\theta_{k}=0$ and for $k=5,6$, we set $F_{k}(\cdot;\bm{x}_{ik},\bm{\beta}_{k},\bm{\theta}_{k}) = \Phi(\theta_{kc}-\bm{x}_{ik}^{T}\bm{\beta}_{k})$ in order to simulate ordinal responses with $C_{5} =3$ and $C_{6}=4$ categories and cut-points $\theta_{kc}$, $c=1,\ldots,C_{k}-1$, which are drawn from a $\mathrm{Unif}(0,1)$ and $\mathrm{Unif}(1,2)$, respectively. 

We used the MCMC sampler presented in Algorithm \ref{algo:MCMC} to obtain posterior samples of the parameters and the latent variables of the BVSGCR model as well as to compute the MPPI for each predictor-response association. To estimate the parameters and the corresponding MPPIs for the single-response regression models, we employed widely-used MCMC algorithms for sparse linear Gaussian \citep{george1997approaches} and non-linear \citep{holmes2006bayesian} regression models with the same proposal distribution for the selection of important predictors as described in Section \ref{sec:sampleB}. In all MCMC algorithms we chose the hyper-parameters $a_{k}$ and $b_{k}$ in eq. \eqref{eq:Betaprior} following \cite{kohn2001nonparametric}, where the mean and the variance of the beta distribution are matched with the \emph{a priori} expected number of important predictors associated with each response ($\E(\bm{\gamma}_{k})=5$) and its variance ($\Var(\bm{\gamma}_{k})=9$). We also set $v=1$ in eq. \eqref{eq:Bprior} since all predictors have been simulated with the same unit marginal variance. Finally, Supp. Mat. Section S.1 presents the prior distribution on the response-specific parameters $\bm{\theta}_{k}$ in the Gaussian and ordinal case. We ran each MCMC algorithm for $30,000$ iterations using the first $10,000$ as burn-in period, storing the outcome every $20$ iterations to obtain $1,000$ posterior samples for each model.

\begin{table}[t]
\caption{Area under the ROC curves for the BVSGCR model and for independent single-response regression models in the simulated Scenario I and II. Results are averaged over $20$ replicates with standard errors in brackets. Within each response, the best performance is highlighted in bold.}
\label{tab:auc_mixed}
\centering
\begin{tabular}{ccccc}
\hline
\multirow{2}{*}{Response} & \multirow{2}{*}{Regression model} & Scenario I & Scenario II & \\
\cline{3-5}
& &$n=50$ \& $p_{k}=30$ & $n=100$ \& $p_{k}=100$ & \\
\hline
\multirow{2}{*}{Gaussian}
& BVSGCR & $\bm{0.86}$ $(0.08)$ & $\bm{0.88}$ $(0.09)$ \\
& Single-response & $0.77$ $(0.09)$ & $0.78$ $(0.10)$ \\ \rule{0pt}{15pt}
\multirow{2}{*}{Gaussian}
& BVSGCR & $\bm{0.82}$ $(0.10)$ & $\bm{0.90}$ $(0.07)$ \\
& Single-response & $0.78$ $(0.11)$ & $0.81$ $(0.12)$ \\ \rule{0pt}{15pt}
\multirow{2}{*}{Gaussian}
& BVSGCR & $\bm{0.80}$ $(0.13)$ & $\bm{0.91}$ $(0.06)$ \\
& Single-response & $0.72$ $(0.13)$ & $0.82$ $(0.10)$ \\ \rule{0pt}{15pt}
\multirow{2}{*}{Binary}
& BVSGCR & $0.83$ $(0.10)$ & $0.81$ $(0.07)$ \\
& Single-response & $0.83$ $(0.10)$ & $0.81$ $(0.08)$\\ \rule{0pt}{15pt}
\multirow{2}{*}{Ordinal (3 categories)}
& BVSGCR & $\bm{0.76}$ $(0.10)$ & $\bm{0.82}$ $(0.07)$ \\
& Single-response &$0.74$ $(0.09)$ & $0.77$ $(0.09)$ \\ \rule{0pt}{15pt}
\multirow{2}{*}{Ordinal (4 categories)}
& BVSGCR & $\bm{0.85}$ $(0.07)$ & $\bm{0.87}$ $(0.07)$ \\
& Single-response &$0.77$ $(0.09)$ & $0.82$ $(0.07)$\\
\hline
\end{tabular}
\end{table}

Table \ref{tab:auc_mixed} displays the average area under the ROC curve (standard errors in brackets) for both Scenario I and II whereas Supp. Mat. Figure S.2 presents the average (over $20$ replicates) ROC curves for each one of the $m=6$ responses simulated in Scenario II. Taken together, they indicate that for the analysis of correlated Gaussian, binary and ordinal responses, the BVSGCR model achieves a higher sensitivity in the Gaussian and the ordinal variables, and for the latter irrespectively of the number of the simulated categories. Similarly to the sparse SUR model with covariance selection, this is due to the ability of our model to account for the correlation between the responses which can induce false positive results when they are analyzed only marginally. In addition, the proposal distribution for the non-zero regression coefficients in eq. \eqref{eq:Bprop} is tailored to take advantage of the estimated sparse inverse correlation structure, resulting in a more efficient algorithm for BVS. 

For the binary response, the performance of the BVSGCR model and single-univariate regression model is almost identical. This is in keeping with \cite{chib1998analysis} and \cite{talhouk2012efficient} that the estimates of the regression coefficients in multi-response Probit regression models are robust to the specification of the correlation structure including the case $\bm{R}= \bm{I}_{m}$ which corresponds, in our framework, to the single-response Probit model. 

We also investigated the effect of the \lq\lq implicit marginalisation\rq\rq\ of the regression coefficients in eq. \eqref{eq:MH} when the joint update of $(\bm{\beta}_{k},\bm{\gamma}_{k})$ in the M-H step is performed. To do so, we used the proposal density in eq. \eqref{eq:noR} that does not account for the correlation between responses and, more important, does not allow for the marginalization of the regression coefficients in the M-H step when Gaussian, binary and ordinal marginal distributions are jointly considered. 
Supp. Mat. Figure S.4 shows that a better performance is achieved across all responses and in particular in the Gaussian and ordinal case when the proposal density in eq. \eqref{eq:Bprop} is used.

To assess the effect of the covariance selection procedure, we present in Supp. Mat. Figure S.6, the ROC curves obtained by a specialized version of the proposed algorithm that does not allow any element of the inverse correlation matrix to be identically zero. Interestingly, the displayed ROC curves suggest that the Gaussian graphical model for covariance selection is crucial for an efficient identification of the important predictors. In particular, for the subset of Gaussian responses, the BVSGCR model with full $\bm{R}^{-1}$ is preferable than single-response linear regression models but it is not better than a model with sparse $\bm{R}^{-1}$. More important, in the case of discrete data, single-response regression models perform better in variable selection than the BVSGCR model when $\bm{R}^{-1}$ is a full matrix. A closer inspection of the MCMC output reveals that the selection of important predictors is affected by the difficult estimation of the inverse correlation matrix when the sample size is small and a full $\bm{R}^{-1}$ is enforced. With a larger sample size ($n=1,000$ data not shown) results are less affected by the specification of the covariance structure.

\begin{figure}[t]
\centering
\includegraphics[height=0.4\textheight]{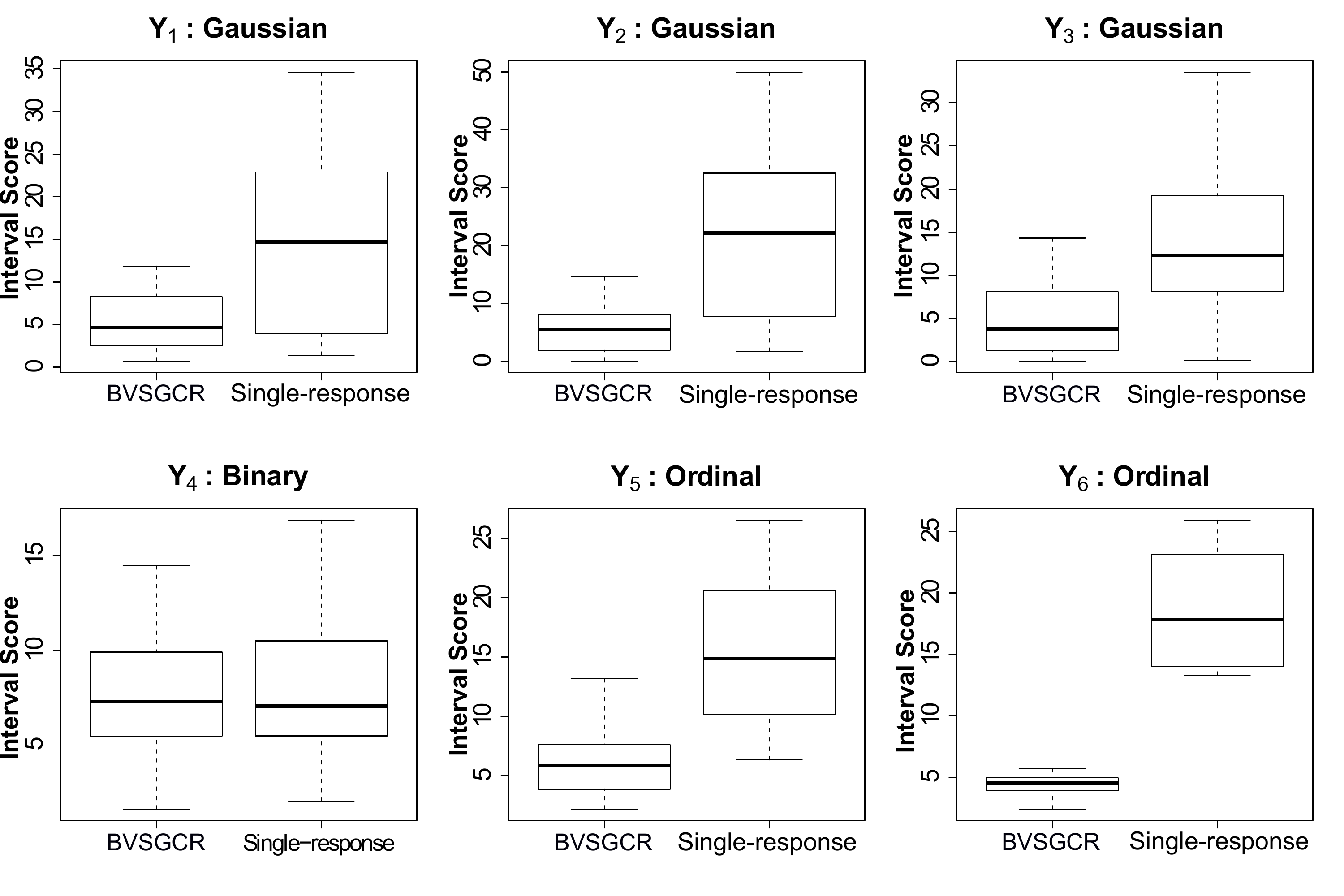}
\caption{Boxplot (over $20$ replicates) of the average interval score for the $95\%$ credible intervals of the non-zero simulated regression coefficients obtained by the BVSGCR model and by single-response regression models in the simulated Scenario II.}
\label{fig:IS_Mixed_n100}
\end{figure}

Finally, we also evaluated the estimation of the regression coefficients obtained by the BVSGCR model and compared with single-response regression models by using a modified version of the scoring rule presented in \cite{gneiting2007strictly}. In our set-up, the interval score rewards narrow posterior credible intervals and incurs a penalty proportional to the significance level of the interval if the simulated non-zero regression coefficient is not included, see also Supp. Mat. Section S.7 for its formal definition. Figure \ref{fig:IS_Mixed_n100} displays the boxplots (over $20$ replicates) of the average interval scores for the $95\%$ credible intervals of the non-zero simulated regression coefficients in Scenario II for the BVSGCR model and single-response regression models. It is apparent that by using the proposed model, we obtained a more accurate estimation of the non-zero regression coefficients for all the responses, except, unsurprisingly, for the binary case.

\subsection{Mixed Gaussian and count responses}
\label{sec:epilepsy-type}

In this section, we present the results of the application of the BVSGCR model in a simulated experiment in which the responses consist of a combination of one Gaussian and three count responses. We followed the same strategy described in Section \ref{sec:simmodel} to generate the set of correlated predictors and the sparse vector of regression coefficients by choosing the parameters $\pi_{1}$, $\pi_{2}$, $b$ and $s^{2}$ as described in Table \ref{tab:sim_values} for the Gaussian and discrete responses. We also used the same correlation matrix $\bm{R}$ as described in Section \ref{sec:simmodel}.

We considered two different scenarios. Both consist of $m=4$ responses with one Gaussian, two negative-binomial and one binomial. In Scenario IV, we generated $20$ replicates with $n=100$ samples and $p_{k}=100$ predictors, $k=1,\ldots,m$, whereas in Scenario III we simulated the same number of replicates with $n=50$ and $p_{k}=30$. Using eq. \eqref{eq:yh}, we simulated the responses by specifying $F_1(\cdot;\bm{x}_{i1},\bm{\beta}_1,\theta_1)$ to be the cdf of the Gaussian distribution with mean $\bm{x}_{i1}^{T}\bm{\beta}_1$ and variance $\theta_1=1$, for $k=2$ and $k=3$, $F_{k}(\cdot;\bm{x}_{ik},\bm{\beta}_{k},\theta_{k})$ to be the cdf of the negative binomial distribution with mean $\theta _{k}\{1+\exp (\bm{x}_{ik}^{T}\bm{\beta}_{k})\}^{-1}$ with $\theta _{2}=\theta _{3}=0.5$, and finally $F_{4}(\cdot;\bm{x}_{i4},\bm{\beta}_{4},\theta_{4})$ to be the
cdf of the binomial distribution $\mathrm{Bin}(10,\{1+\exp (\bm{x}_{i4}^{T}\bm{\beta}_{4})\}^{-1})$ with $\theta_{4}=10$.

To estimate the parameters and the Gaussian latent variables of the BVSGCR model, we utilized the MCMC presented in Algorithm \ref{algo:MCMC}. We compared the results with the same MCMC algorithm used in Section \ref{sec:Ataxia-type} for the Gaussian response. For the count responses, we used the MCMC algorithm based on the representation of the negative binomial and binomial logistic regression model as a Gaussian regression model in auxiliary variables implemented in the R-package \texttt{pogit} by \cite{dvorzak2016sparse}. After setting $\E(\bm{\gamma}_{k})=5$ and $\Var(\bm{\gamma}_{k})=9$, the parameters $a_{k}$ and $b_{k}$ of the beta prior on the probability of predictor-response association were chosen as in Section \ref{sec:Ataxia-type} and we matched the moments of \texttt{pogit} prior specification for $\bm{\gamma}_{k}$ with these values. Finally, Supp. Mat. Section S.1 presents the prior distributions on the response-specific parameters $\theta_{k}$ for the Gaussian and negative-binomial responses. We ran the MCMC algorithms for $30,000$ iterations, storing the outcome every $20$ iterations, after a burn-in period of $10,000$ iterations to obtain $1,000$ posterior samples for each model.

\begin{table}[t]
\caption{Area under the ROC curves for the BVSGCR model and for independent single-response regression models in the simulated Scenario III and IV. Results are averaged over $20$ replicates with standard errors in brackets. Within each response, the best performance is highlighted in bold.}
\label{tab:auc_epilepsy}
\centering
\begin{tabular}{c c c c c}
\hline
\multirow{2}{*}{Response} & \multirow{2}{*}{Regression model} & Scenario III & Scenario IV & \\
\cline{3-5}
& &$n=50$ \& $p_{k}=30$ & $n=100$ \& $p_{k}=100$ & \\
\hline
\multirow{2}{*}{Gaussian}
& BVSGCR & $\bm{0.82}$ $(0.13)$ & $\bm{0.86}$ $(0.10)$ \\
& Single-response & $0.80$ $(0.13)$ & $0.80$ $(0.12)$\\ \rule{0pt}{15pt}
\multirow{2}{*}{Negative Binomial}
& BVSGCR & $\bm{0.70}$ $(0.16)$ & $\bm{0.76}$ $(0.11)$\\
& Single-response & $0.68$ $(0.15)$ & $0.72$ $(0.11)$\\ \rule{0pt}{15pt}
\multirow{2}{*}{Negative Binomial}
& BVSGCR & $\bm{0.72}$ $(0.14)$ & $\bm{0.76}$ $(0.10)$\\
& Single-response & $0.70$ $(0.15)$ & $0.71$ $(0.13)$\\ \rule{0pt}{15pt}
\multirow{2}{*}{Binomial}
& BVSGCR & $0.91$ $(0.10)$ & $0.91$ $(0.07)$\\
& Single-response & $0.91$ $(0.09)$ & $0.91$ $(0.06)$\\
\hline
\end{tabular}
\end{table}

Table \ref{tab:auc_epilepsy} displays the average area under the ROC curve for both Scenario III and IV, whereas Supp. Mat. Figure S.3 presents the average (over $20$ replicates) ROC curves for each one of the $m=4$ simulated responses in Scenario IV. Overall, it is evident, especially in the simulated Scenario IV, that by employing the proposed model, we achieved a better selection of important predictors compared to single-response regression models, apart from the binomial response. Similarly to the binary case, it seems there isn't a clear advantage of the BVSGCR model over single-response non-linear regression models when the marginal distribution is only parameterized by the probability of success.

\begin{figure}[t]
\centering
\includegraphics[height=0.4\textheight]{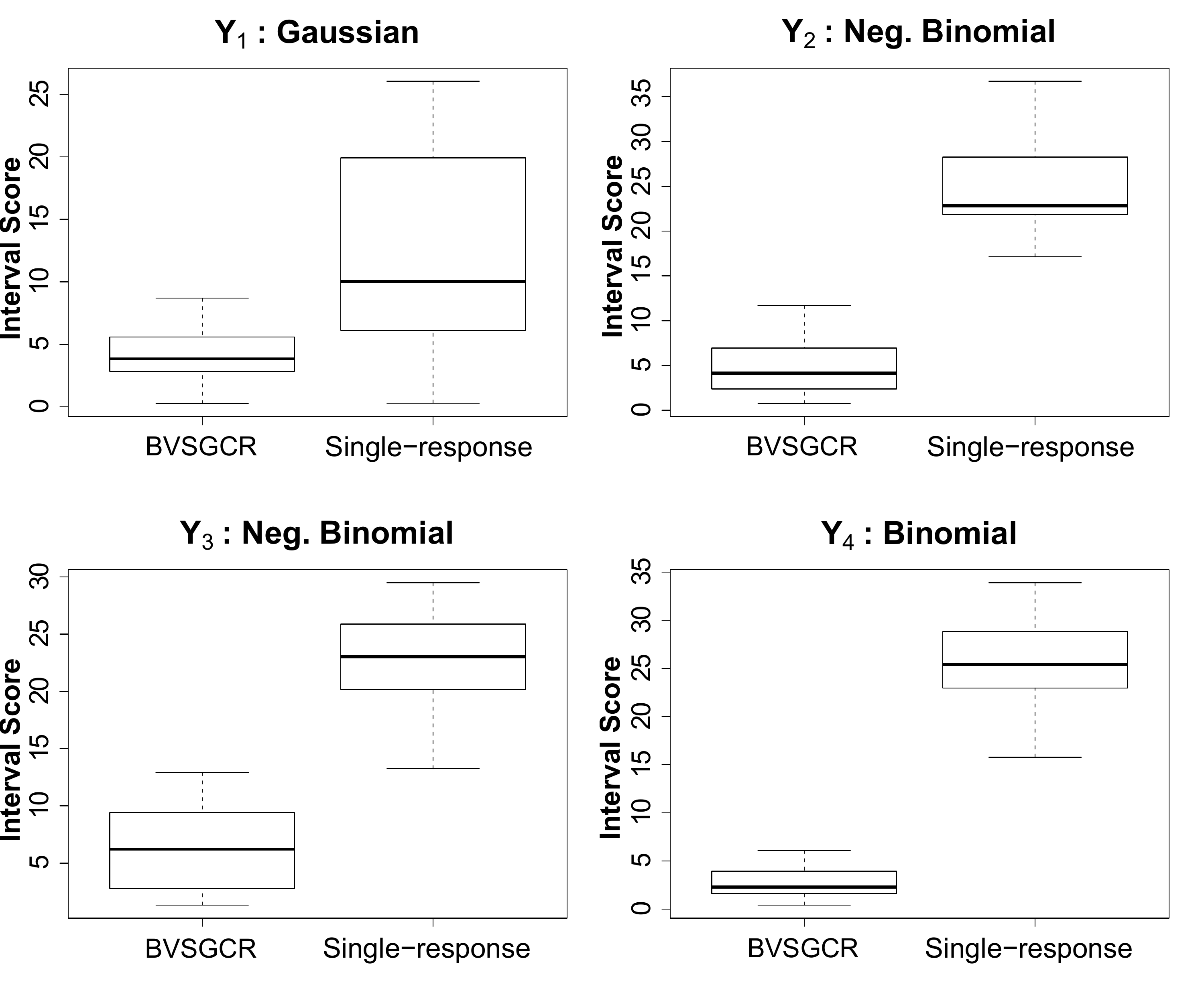}
\caption{Boxplot (over $20$ replicates) of the average interval scores for the $95\%$ credible interval of the non-zero simulated regression coefficients obtained by the BVSGCR model and by single-response regression models in the simulated Scenario IV.}
\label{fig:IS_Count_n100}
\end{figure}

Figure \ref{fig:IS_Count_n100} displays the boxplot (over 20 replicates) of the average interval scores for the $95\%$ credible intervals of the non-zero simulated regression coefficients in Scenario IV for the BVSGCR model and single-response regression models. The boxplots indicate that our model delivers more accurate estimates of the simulated regression coefficients than those obtained by using single-response regression models across all responses, including the binomial case. Thus, while the ROC curve for the binomial response shows that the ranking of the predictors based on the estimated MPPI is the same between the proposed model and the \texttt{pogit} algorithm, BVSGCR attains on average narrower 95\% credible intervals.

For the simulated Scenario IV, Supp. Mat. Figure S.7 compares the performance of the BVSGCR model with its specialized version when $\bm{R}^{-1}$ is a full matrix. Interestingly, the performance is almost identical and both are better than single-response regression models, except for the binomial case. A closer look at the MCMC output reveals that, despite a sparse simulated inverse correlation structure and a small sample size ($n=100$), the estimation of a full $\bm{R}^{-1}$ is still feasible in this scenario with four simulated responses. 

There is also another possible explanation regarding these results which is apparent in Supp. Mat. Figure S.5. With the exemption of the Gaussian outcome, for the binomial and negative-binomial responses, there is no \lq\lq implicit marginalisation\rq\rq\ of the regression coefficients in the M-H step. In this case, the proposal distribution in eq. \eqref{eq:Bprop} that takes into account the correlation between the responses performs no better than the proposal distribution in eq. \eqref{eq:noR} that does not make use of this information. It turns out that, when the \lq\lq implicit marginalisation\rq\rq\ is not possible, the acceptance probability of the joint update of $(\bm{\beta}_{k},\bm{\gamma}_{k})$ in the M-H step seems to not take advantage of how accurately $\bm{R}^{-1}$ is estimated and used in the proposal distribution.


\section{Real data applications}
\label{sec:real}

We illustrate the features of the proposed BVSGCR model by applying it to two real data sets, Ataxia-Telangiectasia disorder and individuals suffering from Temporal Lobe Epilepsy, which are typical examples of data routinely collected in clinical research where a combination of discrete and/or continuous outcome variables are used to assess patients' prognosis and disease progression. In the analysis of both real data sets our aim is twofold: (i) the identification of important associations between the outcome variables and the predictors that are either \lq\lq response-specific\rq\rq\ or \lq\lq shared\rq\rq, i.e., predictors that are linked with several responses at the same time and (ii) the estimation the correlation pattern between the responses in order to shed light on their conditional dependence not explained by set of predictors considered. Finally, we also assessed the out-of-sample prediction accuracy of the BVSGCR model by employing the method proposed by \cite{vehtari2017practical} in order to conduct approximate leave-one-out (LOO) cross-validation. In particular, we compared the predictive performance of the proposed model with the performance of single-response regression models by using the R-package \texttt{loo} \citep{loopack}.

\subsection{MCMC details}

The two real data sets have missing values on both the responses and the predictors. Missing values (completely at random or at random) are a frequent occurrence in clinical data since the propensity for a data point to be missing is either completely random or linked to some characteristics of the observed data. To overcome this problem, missing values in the outcome variables were imputed by modifying the designed MCMC presented in Algorithm \ref{algo:MCMC} as suggested by \cite{zhang2015bayesian} (see Supp. Mat. Section S.6). Missing values in the predictors were imputed using the median of the observed values for each variable. To produce the results presented in this section, we ran the MCMC algorithms for $70,000$ iterations. We considered the first $20,000$ as burn-in and then we stored the output every $50$th iteration in order to obtain $1,000$ (thinned) samples from the posterior distributions of interest.

\subsection{Ataxia-Telangiectasia disorder}
\label{sec:ataxia}

\subsubsection{Data and BVSGCR model}

We applied our model on a data set containing the measurements of 46 individuals suffering from Ataxia-Telangiectasia (A-T) disorder. A-T is a rare neurodegenerative disorder induced by mutations in the \emph{ATM} gene. 
Our data set is a subset of a larger multicentric cohort of $57$ patients presented in \cite{schon2018genotype}. 

The data set includes nine neurological responses (\% missing values in brackets) and $13$ predictors. In particular, four responses are continuous variables named as: Scale for Assessment and Rating of Ataxia (SARA) score (26\%), Ataxia-Telangiectasia Neurological Examination Scale Toolkit (A-T NEST) score (17\%), Age at first Wheelchair use (0\%) and Alpha-Fetoprotein (AFP) levels (28\%). Two responses are binary variables indicating the presence of Malignancy (0\%) and the presence of Peripheral Neuropathy (11\%) which is a term for a group of conditions in which the peripheral nervous system is damaged. Finally, three responses are ordinal variables which measure the overall Severity of the disorder (2\%), its Progression (2\%) and Eye Movements of the patients (2\%). The set of predictors includes genetic (Genetic Group, Missense Mutation, i.e., single base mutation responsible for the production of a different amino acid from the usual one, number of Mild Mutations, ATM Protein levels and Chromosomal Radiosensitivity, i.e. whether X-ray exposure induces chromosomal aberrations in individuals with A-T) and immunological (immunoglobulin IgM, IgG2, IgG, IgA and IgE and immune CD4 and CD8 T-cell counts and CD19 B-cell counts) characteristics of the patients. In addition, we used an intercept term (with a diffuse normal prior centered in zero) and three confounders (age, gender and age of onset) always included in the regression model. Both confounders and predictors were standardized and the continuous variables quantile-transformed before the analysis.

We modeled the responses by specifying the BVSGCR model as follows. For each $i=1,\ldots,n$ we set: for $k=1,\ldots,4$, $F_{k}(\cdot;\bm{x}_{ik},\bm{\beta}_{k},\theta_{k})$ to be the cdf of the Gaussian distribution with mean $\bm{x}_{ik}^{T}\bm{\beta}_{k}$ and variance $\theta_{k}$; for $k=5,6$, $F_{k}(\cdot;\bm{x}_{ik},\bm{\beta}_{k},\theta_{k})$ to be
the cdf of the Bernoulli distribution with probability of success $\Phi(\bm{x}_{ik}^{T}\bm{\beta}_{k})$ and $\theta_{k}=0$; for $k=7,8,9$, $F_{k}(\cdot;\bm{x}_{ik},\bm{\beta}_{k},\bm{\theta}_{k}) = \Phi(\theta_{kc}-\bm{x}_{ik}^{T}\bm{\beta}_{k})$, where $\bm{\theta}_{k}=(\theta_{k0},\ldots,\theta_{k,C_{k}-1})$ denotes the cut-points of the ordinal responses with $C_{7}=C_{9}=3$ and $C_{8}=4$ categories, respectively. We used the prior distributions presented in Section \ref{sec:priors} and in Supp. Mat. Section S.1.
Finally, we set $\E(\bm{\gamma}_{k}) = 3$ and $\Var(\bm{\gamma}_{k}) = 2$ for all $k$. These values imply \emph{a priori} a range of associations for each response between $0$ and $8$.

\subsubsection{Results}

Figure \ref{fig:MPPI_Ataxia} displays the estimated MPPI for each predictor-response pair. Despite the small sample size, strong associations are detected in SARA score, A-T NEST score, AFP levels, Peripheral Neuropathy and Eye Movements and some evidence of association in Malignancy, Severity and Progression. Amongst the genetic predictors, Missense Mutation seems to play an important role in predicting the disease status and its progression (Eye Movements, Severity and Progression) as well as Malignancy and Peripheral Neuropathy \citep{schon2018genotype}. The number of Mild Mutations appears to influence A-T NEST score and AFP levels. 
Regarding the role of the immune system, it is well documented that patients suffering A-T have often a weakened defence mechanism. We confirm this clinical finding and, in contrast to the genetic risk factors, immunological predictors seem to be more \lq\lq response-specific\rq\rq. 

Given the small sample size and (potentially important) unmeasured covariates, we do not expect that the genetic and immunological predictors are able to explain entirely the variability of the responses and their covariation. Therefore, it is important to model any source of extra variability that may induce false positives associations. Figure \ref{fig:EPPI_Ataxia} shows the conditional dependence structure of the responses estimated by the BVSGCR algorithm with a decomposable model specification. Disease status and its progression (Severity and Progression) are closely linked with A-T NEST score and Age at first Wheelchair use, the latter also strongly related. Interestingly, Severity and Progression seem to capture different aspects of the disease since they are almost conditionally independent once the effect of Missense Mutation is accounted for (see Figure \ref{fig:MPPI_Ataxia}). 
A-T NEST score is also important in predicting the level of SARA score and Eye Movements. Finally, SARA score seems to be a good proxy for Peripheral Neuropathy.

We have also checked whether the assumption of decomposability is supported by the data. When a non-decomposable graphical model is specified, we found that the number of edges with a non-zero posterior probability of inclusion is higher than in the decomposable case, although there is a good agreement between the two instances of the BVSGCR algorithm regarding the most important edges, see Supp. Mat. Figure S.13. Interestingly, the posterior mass is equally split between the two graphical models' specifications, with a small advantage for the non-decomposable case (53\%). The full results are reported in Supp. Section S.8.7.

\begin{figure}[H]
\centering
\includegraphics[height=0.44\textheight]{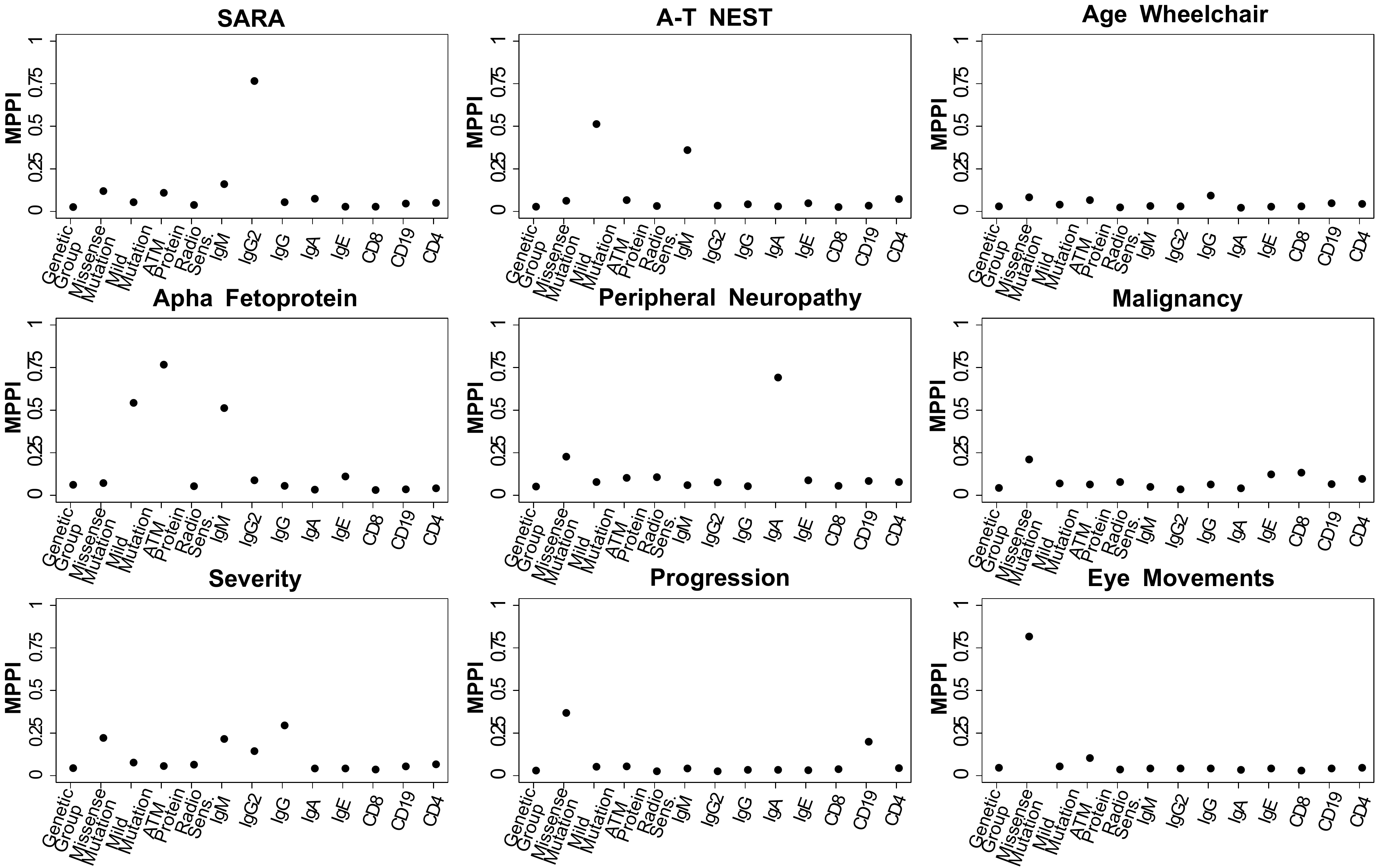}
\caption{Detection of important predictors for the neurological responses in the A-T data set. Marginal posterior probabilities of inclusion (MPPI) measures the strength of the predictor-response association.}
\label{fig:MPPI_Ataxia}

\includegraphics[height=0.365\textheight]{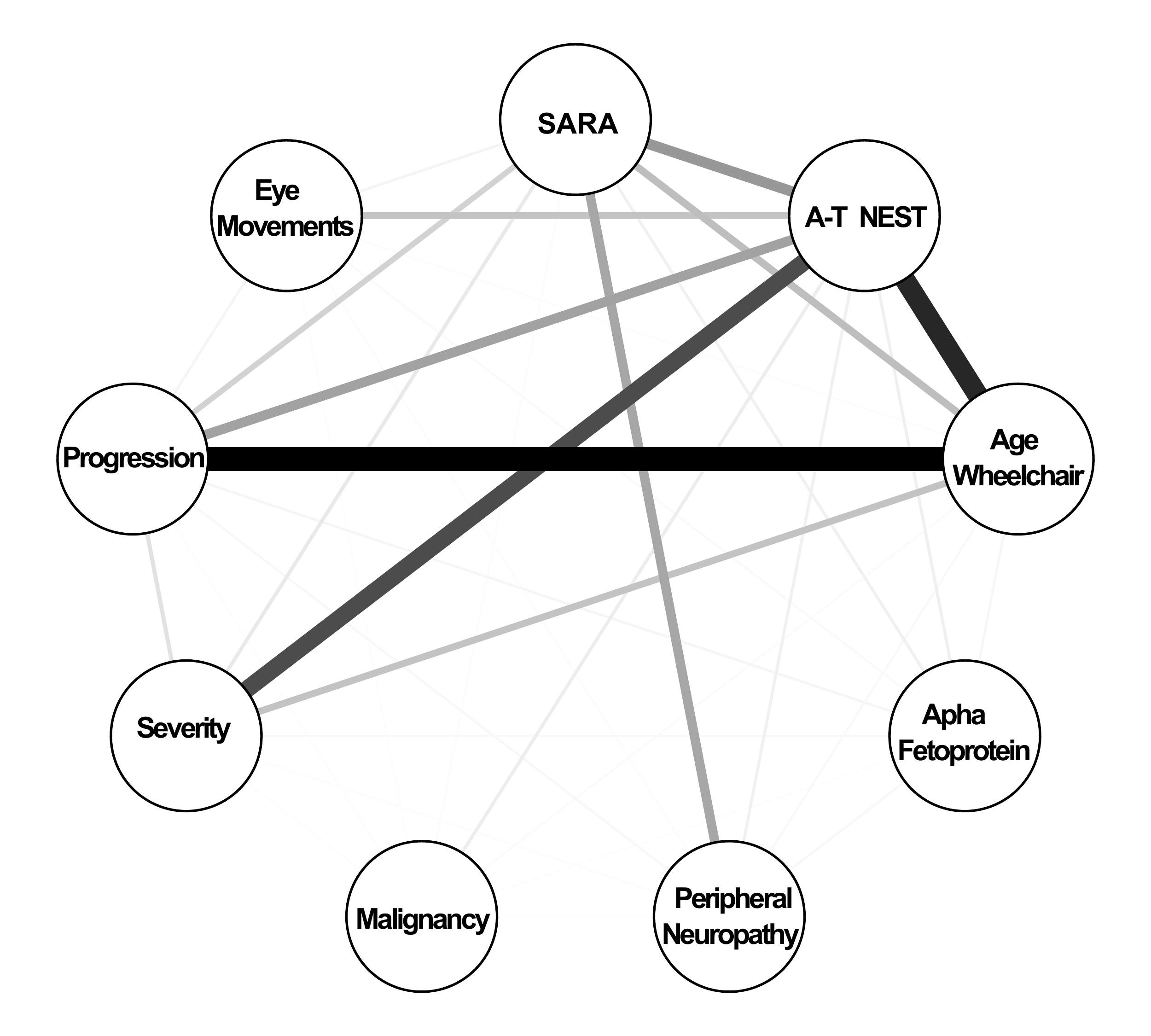}
\caption{Graph implied by the non-zero pattern of $\bm{R}^{-1}$ in the A-T data set and estimated by using the edge posterior probabilities of inclusion (EPPI) obtained by specifying a decomposable graphical model. Gray scale and edge thickness specify different levels of the EPPI (black thick line indicates large EPPI values).}
\label{fig:EPPI_Ataxia}
\end{figure}

Table \ref{tab:ataxia_loo} presents, for each response, the estimated difference in the expected log-pointwise predictive density \citep{vehtari2017practical} between the BVSGCR model and single-response regression models. To ensure a fair comparison, we used the same prior specification for the latent binary vector and implemented the same search algorithm, see Supp. Mat. Section S.2. It is clear that the proposed model delivers more accurate predictions than those obtained by any single-response regression models. As expected from the simulation study, for the binary responses (Peripheral Neuropathy and Malignancy) the difference is less remarkable, with the lowest ELPPD difference in Malignancy which is uncorrelated with the other responses and only mildly associated with Missense Mutation. Finally, Supp. Mat. Figure S.8 presents the associations detected by single-response regression models in the A-T data set.

\begin{table}[t]
\caption{Difference in the expected log-pointwise predictive density (ELPPD) between the BVSGCR model and independent single-response regression models. A positive difference indicates that the proposed model has better predictive performance (standard errors in brackets).}
\centering
\begin{tabular}{c c c}
\hline
 Type &Response & Difference in ELPPD \\
\hline
Gaussian& SARA  & $23.3$ $(6.4)$ \\
Gaussian& A-T NEST & $23.0$ $(6.7)$\\
Gaussian& Age Wheelchair  & $25.9$ $(7.1)$ \\
Gaussian& Alpha-Fetoprotein & $19.4$ $(5.4)$ \\
Binary& Peripheral Neuropathy & $4.1$ $(1.8)$ \\
Binary& Malignancy  & $0.1$ $(0.6)$ \\
Ordinal (3 categories)& Severity  & $17.1$ $(4.5)$ \\
Ordinal (4 categories)& Progression  & $35.6$ $(7.6)$ \\
Ordinal (3 categories)& Eye Movements& $15.1$ $(2.7)$ \\
\hline
\end{tabular}
\label{tab:ataxia_loo}
\end{table}

\subsection{Patients with Temporal Lobe Epilepsy}
\label{sec:epilepsy}

\subsubsection{Data and BVSGCR model}

We also applied the BVSGCR model to a second data set consisting of $m = 5$ responses and $p_k = 162$, $k=1,\ldots,m$, common predictors. We investigated the relationship between human cognition and epilepsy based on recent data collected by \cite{johnson2016systems}. The authors used $n = 122$ fresh-frozen whole-hippocampus samples, surgically resected from patients with Temporal Lobe Epilepsy (TLE) in order to determine whether genes belonging to their inferred gene-regulatory networks are related with human memory abilities and the number of seizures measured on the same individuals. 
More precisely, the responses (\% of missing values in brackets) comprise the average number of self-reported daily Seizures for each patient (10\% as we excluded some extremely large observations likely due to errors in self-reported number of Seizures), the memory category in which the patients have been assigned after the assessment by a neurologist (14\%) and the results (Learning (15\%), Post-Interference (15\%) and Delayed Recall (15\%)) of the Verbal Learning Test \citep{thiel2016effects} that quantifies the human cognition abilities. 
We also considered five confounding predictors: sex, age of manifestation of epilepsy, age at neurological assessment, anti-epileptic drugs load, handedness and laterality (brain lobe) of TLE. The $162$ correlated genes 
were obtained from a network analysis described in \cite{johnson2016systems}. Both confounders and gene expression predictors were standardized to have unit variance.

We modeled the number of self-reported seizures with a negative binomial distribution and we used an ordinal variable for the memory categories in which the patients have been assigned by the neurologist. Finally, we assumed that the number of correct words that each patient recalls in each one of the three tasks of the Verbal Learning Test is distributed as a binomial random variable with $15$ trials. 
To set the hyper-parameters of the prior that controls the level of sparsity, we followed the same procedure used in the simulation study with $\E(\bm{\gamma}_{k}) = 5$ and $\Var(\bm{\gamma}_{k}) = 9$ for all $k$. The prior distributions on $\theta_{k}$ for the negative binomial and the ordered categorical responses are presented in Section \ref{sec:priors}. 

\subsubsection{Results}

Figure \ref{fig:MPPI_Epilepsy} displays the estimated MPPI of the associations between the $162$ correlated genes and cognition abilities, the number of seizures and memory classification. The most striking finding is the ubiquitous role of \emph{RBFOX1} gene in Learning, Post-Interference and Delayed Recall as well as in Memory Category. It has been shown that mutations in this gene lead to neurodevelopmental disorder and it has also recently implied in cognitive functions \citep{Davies2018}. Regarding the Learning task, animal model studies have revealed critical functions for \emph{GABRB1} gene 
for maintenance of functioning circuits in the adult brain \citep{Gehman2011}. The genetic regulation of Delayed Recall is more complex and related to the difficult memory task that individuals were asked to perform. 

In contrast to cognition abilities, the associations with the number of Seizures are less strong. This phenomenon can be explained by the quality of the self-reported data: before the analysis, we removed $10$ measurements that appeared to be outliers. 
Despite that, the association with \emph{WNT3} gene seems interesting since deregulation in WNT signalling has a fundamental role in the origin of neurological diseases \citep{Oliva2013}. 

We conclude the description of the association results by comparing the outcome of the proposed BVSGCR model with single-response regression models, see Supp. Mat. Figure S.9. To conduct BVS for the ordinal response, we used the method proposed by \cite{holmes2002accounting} and for negative binomial and binomial responses we utilized the auxiliary mixture sampling method of \cite{fruhwirth2009improved} implemented in the R-package \texttt{pogit} \citep{dvorzak2016sparse}. For the latter, we matched the moments of the prior on $\bm{\gamma}_{k}$ with the hyper-parameters of the BVSGCR sparsity prior. From the comparisons, we noticed that for the binomial responses the number of associations identified by the single-response regression model is either too large (Learning and Delayed Recall) or almost nil (Post-Interference). This may depend on the Gibbs sampling search algorithm implemented in the R-package \texttt{pogit} that does not perform well when a large number of correlated predictors are considered \citep{bottolo2010evolutionary}. In contrast, our proposal distribution for $\bm{\gamma}_{k}$ allows the quick detection of relevant predictors that explain a large fraction of the responses' variability, see Supp. Mat. Figure S.11. 

Figure \ref{fig:EPPI_Epilepsy} presents the conditional independence graph for the group of responses considered when a decomposable graphical model is assumed. Similarly to the A-T disorder, we do not expect to capture the whole responses' variability and their covariation given the set of predictors considered and (potentially important) unmeasured covariates. Interestingly, the responses of the Verbal Learning Test are all connected, with Memory Category linked only with Delayed Recall, suggesting that the neurologist's patients classification strongly reflects the Delayed Recall score. Finally, the number of Seizures is conditionally independent of the memory tasks of the Verbal Learning Test. This can be explained either by the self-reported quality of the data or by the fact that we removed the effect of age of onset which is known to be negatively correlated with both Seizures and cognition abilities.

\begin{figure}[H]
\centering
\includegraphics[height=0.4\textheight]{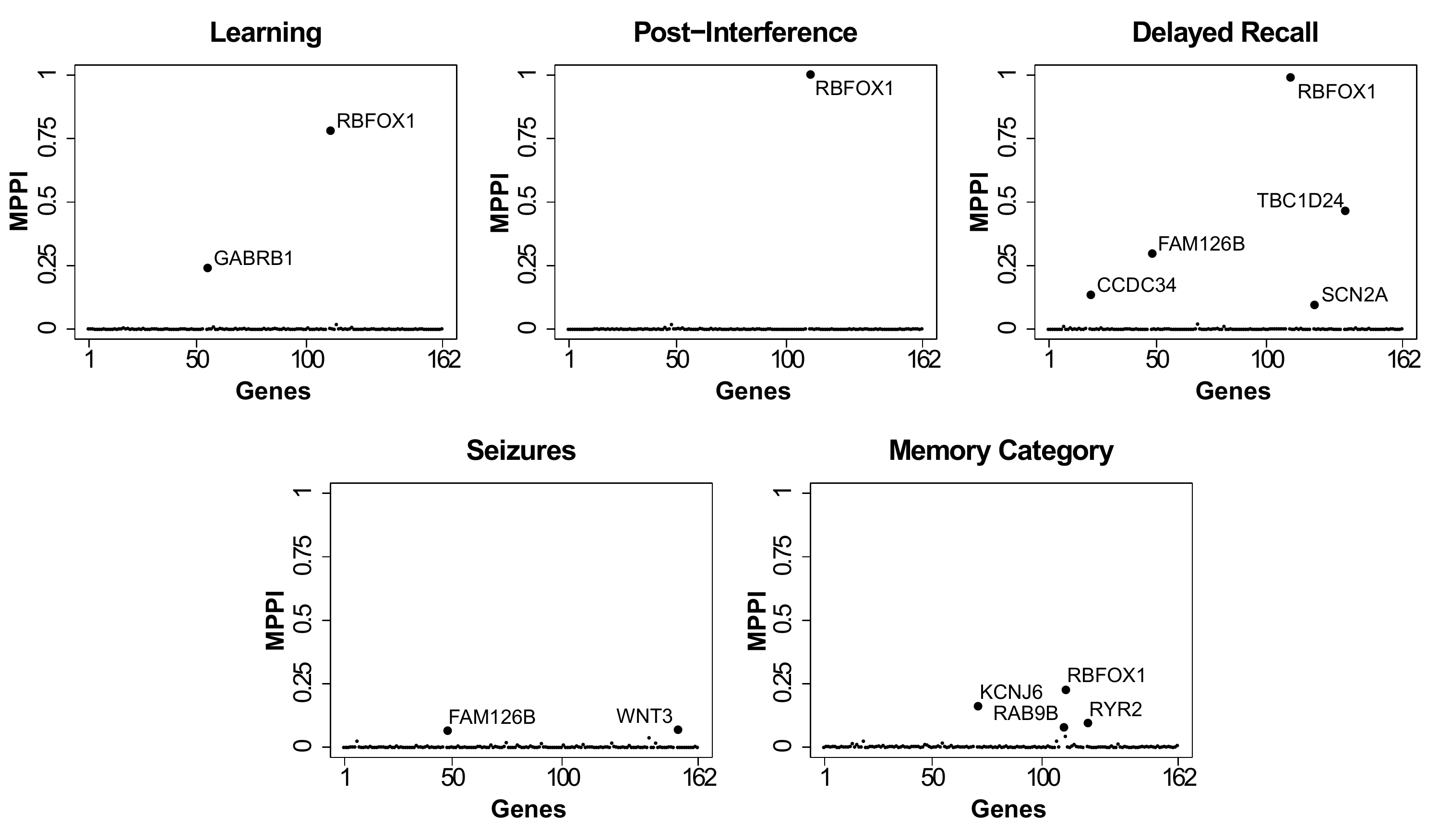}
\caption{Detection of important genes that predict human cognition abilities, number of Seizure and memory classification in $122$ individuals with TLE. Marginal posterior probabilities of inclusion (MPPI) measures the strength of the predictor-response association. For each response, only associations with MPPI $> 0.05$ are highlighted.}
\label{fig:MPPI_Epilepsy}

\includegraphics[height=0.365\textheight]{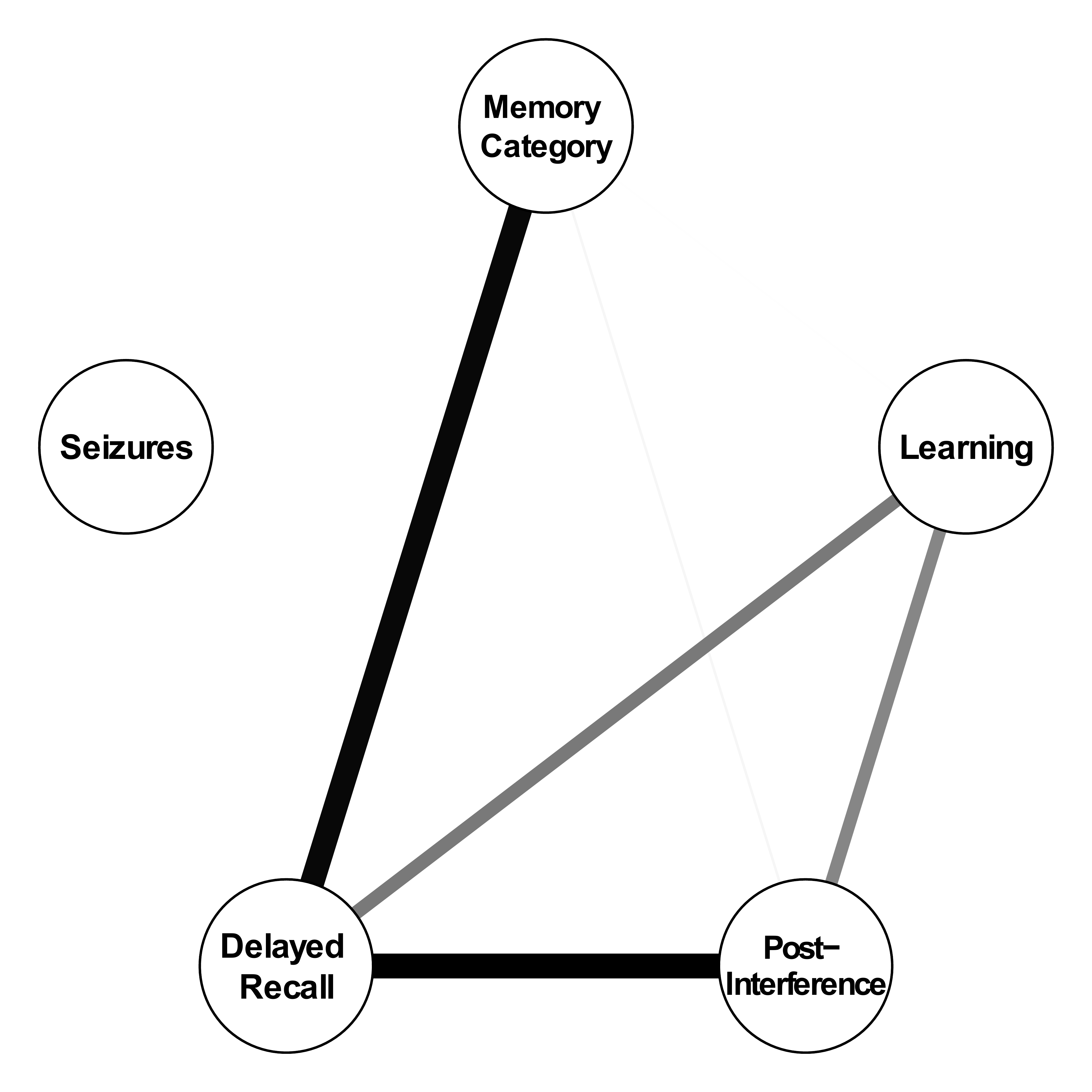}
\caption{Graph implied by the non-zero pattern of $\bm{R}^{-1}$ in the TLE data set and estimated by using the edge posterior probabilities of inclusion (EPPI) obtained by specifying a decomposable graphical model. Gray scale and edge thickness specify different levels of the EPPI (black thick line indicates large EPPI values).}
\label{fig:EPPI_Epilepsy}
\end{figure}

We also checked if the assumption of decomposability of the graphical model is realistic. The support for non-decomposable graphs is overwhelming with 96\% of the posterior mass concentrated on it. There is a general agreement regarding the detection of important edges between the two instances of the BVSGCR model, although for non-decomposable graphs, neurologist's memory classification of the patients seems to be based, more reasonably, on the whole set of results of the Verbal Learning Test, see Supp. Mat. Figure S.15.

Finally, Table \ref{tab:epil_loo} presents, for each response, the estimated difference in the expected log-pointwise predictive density \citep{vehtari2017practical} between the BVSGCR model and single-response regression models. They are all largely positive, indicating that the proposed model has better predictive performance, apart from the number of Seizures which is weakly associated with the set of genes and conditionally independent from the other responses. In any case, the small difference and the large standard deviation make it difficult to draw any clear conclusion about the best predictive model for this trait.

\begin{table}[t]
\caption{Difference in the expected log-pointwise predictive density (ELPPD) between the BVSGCR model and independent single-response regression models. A positive difference indicates that the proposed model has better predictive performance (standard errors in brackets).}
\centering
\begin{tabular}{c c c}
\hline
 Type & Response & Difference in ELPPD \\
\hline
Binomial & Learning & $51.1$ $(9.6)$\\
Binomial& Post-Interference & $49.4$ $(10.4)$ \\
Binomial& Delayed Recall & $93.1$ $(14.5)$ \\
Neg. Binomial& Seizures & -$8.8$ $(10.4)$ \\
Ordinal (5 categories) & Memory Category  & $59.3$ $(12.9)$ \\
\hline
\end{tabular}
\label{tab:epil_loo}
\end{table}

\section{Discussion}
\label{sec:discussion}

In this paper, we have presented a novel approach for BVS when a combination of discrete and/or continuous responses are jointly considered. The proposed method allows the exploration of the model space consisting of all possible subsets of predictors while estimating the conditional dependence structure among the responses and \emph{vice versa}.

We have shown that for some continuous and discrete outcomes, including Gaussian, binary and ordinal responses, the regression coefficients can be \lq\lq implicitly marginalised\rq\rq\ in the acceptance probability of the M-H step for the joint update of $(\bm{\beta}_{k},\bm{\gamma}_{k})$. This allows the reduction of the posterior correlation between these two quantities and thus improves the mixing of the designed sampler. The \lq\lq implicit marginalisation\rq\rq\ of the non-zero regression coefficients holds also when an unordered categorical variable is considered. In contrast to binary or ordinal responses, where only one latent variable is required, in the unordered categorical case, the state-space is expanded by $C-1$ latent variables where $C$ is the number of categories. Combined with an effective proposal distribution for the latent binary vector, based on the marginal screening of important predictors for each response, our approach allows an efficient exploration of the ultra-high predictors model space. 

We tested the proposed method on simulated and real data sets and compared it with widely-used sparse Bayesian single-response linear and non-linear regression models. In all examples considered, BVSGCR outperformed existing BVS algorithms in terms of selection of important predictors and/or estimation of the non-zero regression coefficients, except for the binary case. In the simulation study, we have also demonstrated that covariance selection is key when the sample size is small and the number of responses is large.

We conclude with some final remarks regarding directions for future research. As we have demonstrated in the simulation study, in the case of multiple-response count data, the proposed BVSGCR method performs better than single-response regression models because it exploits the Gaussian copula model. However, in the same simulated example, we have shown that the proposal distribution that takes into account the correlation between responses performs no better than the proposal distribution that does not use this information. When the \lq\lq implicit marginalisation\rq\rq\ is not possible, the M-H step doesn't seem to take advantage of how accurately the residual correlations between the responses are estimated and this information used in the proposal distribution.
Thus, it is paramount to specify the cdf of the marginal distribution so that it allows the marginalization of the regression coefficients. For count data, this may be accomplished by using the generalized ordered-response Probit (GORP) model presented in \cite{Castro2012}  which can be expressed as a function of the Gaussian cdf. 
An interesting venue for future work will be the assessment of the similarities of the GORP model with Generalized Linear Models to fully exploits the benefits of the marginalization of the regression coefficients and the estimation of the residual correlations between the responses in the analysis of multivariate count data.

In summary, our new BVSGCR algorithm is tailored to jointly analyze correlated responses of diverse types with missing observations and a large set of predictors. Besides the application to clinically relevant data, the proposed method can be also used in the analysis of other problems for which sparse regression algorithms for combinations of discrete and/or continuous responses are required.


\bigskip
\begin{center}
{\large\bf SUPPLEMENTAL MATERIALS}
\end{center}

\begin{description}

\item[Supplementary material:] It provides additional material regarding the proposal density and the \lq\lq implicit marginalisation\rq\rq\ of the regression coefficients as well as further results of the simulation study and the analysis of the real data sets.

\item[R-package \texttt{BVS4GCR}:] The package is freely available on \url{https://github.com/lb664/BVS4GCR/}. It includes examples that explain how to generate the simulated data and run the algorithm.

\end{description}

\spacingset{1}
\bibliographystyle{chicago}
\bibliography{BVSGCR_bib_Final}

\end{document}


\def\spacingset#1{\renewcommand{\baselinestretch}%
{#1}\small\normalsize} \spacingset{1}


\spacingset{1.45}

\if0\blind
{
\title{Supplementary Material \\
\textbf{Bayesian Variable Selection for Gaussian \\
copula regression models}}
\author{Angelos Alexopoulos and Leonardo Bottolo}
  \maketitle
} \fi

This document supplements the material presented in the manuscript \lq\lq Bayesian variable selection for Gaussian copula regression models\rq\rq. In Section \ref{sec:thetaprior} we specify the prior distributions for the response-specific parameters $\bm{\Theta}$. In Sections \ref{sec:stephens}, \ref{sec:propdistr} and \ref{sec:marginalisation}, we present the proposal distribution for the latent binary matrix $\bm{\Gamma}$, the regression coefficients $\bm{B}$ and their joint update in the Metropolis-Hastings (M-H) step. The M-H step for the
adjacency matrix $\bm{G}$ 
is detailed in Section \ref{sec:acceptG} for both decomposable and non-decomposable graphs. 
Section \ref{sec:impute} describes the strategy that we utilized in order to impute missing observations. Finally, Section \ref{sec:rocs} presents additional results regarding the simulation study that we performed to test the proposed Bayesian variable selection for Gaussian copula regression (BVSGCR) model and the analysis of the real data sets.

\section{Prior distribution for the response-specific parameters}
\label{sec:thetaprior}

Let $m$ the number of response variables in the Gaussian copula regression model defined in eq. (1) of the main paper. The marginal cumulative distribution function (cdf) $F_{k}(\cdot)$ that is used for modeling the $k$th response may depend on the vector of parameters $\bm{\theta}_{k}$. By assuming that each $F_{k}(\cdot)$, $k=1,\ldots,m$, is a member of the exponential family, then $\bm{\theta}_{k}$ is the vector of response-specific parameters of the $k$th response in the multivariate Generalized Linear Model \citep{song2009joint}. Specifically:

\begin{itemize}
\item If $F_{k}(\cdot)$ is the cdf of the Gaussian distribution, we have that $\theta_{k}$ is the variance of the distribution, i.e., $\theta_{k}=\delta_{k}^{2}$, and the prior density for this parameters is presented in Section 2.3.2 of the main paper where we set $\delta_{k}^{2}|r^{kk} {\sim} \mathrm{IGam} \left( (m+1)/2, r^{kk}/2 \right)$ with $r^{kk}=(\bm{R}^{-1})_{kk}$. 
\item If $\bm{y}_{k}$ is an ordered categorical response with $C_{k}$ levels, then $F_{k}(\cdot)$ depends on the vector of cut-points $\bm{\theta}_{k}$ \citep{mccullagh1980regression}. Following \cite{schliep2015data}, we set for identifiability conditions that $\eta_{k0} = \theta_{k0} = -\infty$, $\eta_{k1} = \theta_{k1} = 0$, $\eta_{kC_{k}} = \theta_{kC_{k}} = \infty$ and $\eta_{kc} = \log(\theta_{kc} - \theta_{k,c-1})$ for $c=2,\ldots,C_{k}-1$. Finally, we assume that $\eta _{kc}\overset{\text{iid}}{\sim }\mathrm{N}(0,10^{2})$ for $c=2,\ldots,C_{k}-1$.
\item If $\bm{y}_{k}$ is modeled using the negative binomial distribution, then $\theta_{k}$ is a scalar known as the number of failures until the experiment is stopped $(\theta_{k}>0)$. Following \cite{dvorzak2016sparse}, we assume that $\theta_{k} \sim \mathrm{Gam}(2,1)$.
\end{itemize}

\section{Proposal distribution for the latent binary matrix}
\label{sec:stephens}
Here, we present the distribution that we used in the BVSGCR model to propose a new latent binary vector $\bm{\gamma}_{k}^{\ast}$, $k=1,\ldots,m$, at each iteration of the MCMC algorithm. The proposal distribution is based on the idea developed by \cite{guan2011bayesian} for single-response sparse linear regression models. The aim is to explore efficiently the model space consisting of all the possible subsets of predictors by either adding predictors in the regression model with a non-uniform probability or deleting them with a uniform probability in order to ensure the reversibility of Markov chain.

We extended the idea of \cite{guan2011bayesian} and generalized for non-linear regression models. Let $\bm{X}_{k}$ be the $n \times p_{k}$ matrix of available predictors for the $k$th response. Before running the MCMC algorithm, we rank the predictors according to the marginal $p$-value obtained by regressing each column of $\bm{X}_{k}$ on $\bm{y}_{k}$. Although based on a marginal screening, this allows quick detection of relevant predictors that explain a large fraction of each response's variability. Then, for each $k=1,\ldots,m,$
the proposal distribution consists of three moves:
\begin{itemize}
\item [(i)] With probability $0.45$, we sample an integer $\rho_{k}$ from a mixture of a truncated distribution with a uniform distribution on $1,..., p_{k}-p_{\bm{\gamma}_{k}}$ with $p_{\bm{\gamma}_{k}}= |\bm{\gamma}_{k}|$ and we propose to add a predictor currently not included in the model with rank $\rho_{k}$.
\item [(ii)] With probability $0.45$ we propose to delete uniformly at random a predictor among those currently in the model.
\item [(iii)] With probability $0.1$, we propose to add a predictor not in the model and to remove another one currently in the model based on a uniform distribution.
\end{itemize}

\noindent Regarding (i), we found useful setting the mean of the truncated geometric distribution equal to the \emph{a priori} expected number of important predictors associated with each response $(E(\bm{\gamma}_{k}))$ and the weights of mixture equal to $(0.7, 0.3)$ in order to favor the geometric component of the rank-based proposal.

\section{Proposal distribution for the regression coefficients}
\label{sec:propdistr}

From eq. (13) of the main paper, the Gaussian latent variable $Z_{ik}$ is normal distributed with mean $\mu_{ik}=\bm{x}_{i,\bm{\gamma}_{k}}^{T}\bm{\beta}_{\bm{\gamma}_{k}}$ and variance either $\sigma^2_{k}=\delta_{k}^{2}$ (if continuous) or $\sigma^2_{k}=1$ (if discrete). Since any multivariate distribution can be written as the product of its marginal densities with the density of the associated copula, and by taking into account eq. (2), we have that the distribution of the Gaussian latent variables $\widetilde{\bm{Z}}$ is
\begin{equation}
p(\widetilde{\bm{Z}}|\bm{B},\bm{\Gamma},\bm{D},\bm{R})=|\bm{R}|^{-n/2}\prod_{i=1}^{n}\exp \left\{\frac{1}{2}\tilde{\bm{z}}_{i}^{T}(\bm{I}_{m}-\bm{R}^{-1})\tilde{\bm{z}}_{i}\right\}
\prod_{k=1}^{m}\mathcal{N}(z_{ik};\bm{x}_{i,\bm{\gamma}_{k}}^{T}\bm{\beta}_{
\bm{\gamma}_{k}},\sigma _{k}^{2}).
\label{eq:proposal_lik}
\end{equation}
By setting $\widetilde{Z}_{ik}=(Z_{ik}-\bm{x}_{i,\bm{\gamma}_{k}}^{T}\bm{\beta}_{\bm{\gamma}
_{k}})/\sigma _{k}$, eq. \eqref{eq:proposal_lik} implies that
\begin{align}
p(\bm{z}_{k}|& \bm{\beta}_{k},\bm{\gamma}_{k},\widetilde{\bm{Z}}_{-k},\bm{D},\bm{R})\propto
\notag \\
& \propto \exp \left\{ \frac{1}{2}(1-r^{kk})\sum_{i=1}^{n}\tilde{z}
_{ik}^{2}-\sum_{i=1}^{n}\sum_{\ell =1,\ell \neq k}^{m}r^{k\ell }\tilde{z}
_{ik}\tilde{z}_{i\ell }-\frac{1}{2\sigma _{k}^{2}}\sum_{i=1}^{n}(z_{ik}-
\bm{x}_{i,\bm{\gamma}_{k}}^{T}\bm{\beta}_{\bm{\gamma}_{k}})^{2}\right\}
\notag \\
& =\exp \left\{ -\frac{r^{kk}}{2\sigma_{k}^{2}}\sum_{i=1}^{n}(z_{ik}-\bm{x}
_{i,\bm{\gamma}_{k}}^{T}\bm{\beta}_{\bm{\gamma}_{k}})^{2}-\frac{1}{\sigma
_{k}}\sum_{i=1}^{n}\sum_{\ell =1,\ell \neq k}^{m}r^{k\ell }(z_{ik}-\bm{x}_{i,\bm{\gamma}_{k}}^{T}\bm{\beta}_{\bm{\gamma}_{k}})\tilde{z}_{i\ell }\right\}
\notag \\
& =\exp \left\{ -\frac{r^{kk}}{2\sigma_{k}^{2}}(\bm{z}_{k}-\bm{X}_{\bm{\gamma}
_{k}}\bm{\beta}_{\bm{\gamma}_{k}})^{T}(\bm{z}_{k}-\bm{X}_{\bm{\gamma}_{k}}
\bm{\beta}_{\bm{\gamma}_{k}})-\frac{1}{\sigma _{k}}(\bm{z}_{k}-\bm{X}_{\bm{\gamma}
_{k}}\bm{\beta}_{\bm{\gamma}_{k}})^{T}\tilde{\bm{\zeta}}_{k}\right\} ,
\label{eq:cond_lik}
\end{align}
where $r^{k\ell} = (\bm{R}^{-1})_{k\ell}$, $\bm{X}_{\bm{\gamma}_{k}}$ denotes the $n \times |\bm{\gamma}_{k}|$ matrix in which the columns are selected according to the latent binary vector $\bm{\gamma}_{k}$ and $\tilde{\bm{\zeta}}_{k} = (\tilde{\zeta}_{1k},\ldots,\tilde{\zeta}_{nk})$ with $\tilde{\zeta}_{ik} = \sum_{\ell=1, \ell \neq k}^m r^{k\ell}\tilde{z}_{i\ell}$.

\remark{The Gaussian latent variable $Z_{ik}$ conditionally on $\widetilde{\bm{Z}}_{i,-k}$ is normal distributed with mean $\mu _{ik|-k}=\bm{x}_{i,\bm{\gamma}_{k}}^{T}\bm{\beta}_{\bm{\gamma}_{k}}+\sigma _{k}\bm{R}_{k,-k}\bm{R}_{-k,-k}^{-1}\tilde{\bm{z}}_{i,-k}$ and variance $\sigma_{k|-k}^{2}=\sigma_{k}^{2}(1-\bm{R}_{k,-k}\bm{R}_{-k,-k}^{-1}\bm{R}_{-k,k})$. Since  $(\bm{R}^{-1})_{kk}^{-1}=1-\bm{R}_{k,-k}\bm{R}_{-k,-k}^{-1}\bm{R}_{-k,k}$ and$ (\bm{R}^{-1})_{kk}^{-1}(\bm{R}^{-1})_{k,-k}=-\bm{R}_{k,-k}\bm{R}_{-k,-k}^{-1}$ (\cite {Harville1997}, Corollary 8.5.12), then $\sigma _{k|-k}^{2}=\sigma
_{k}^{2}/r^{kk}$ with $r^{kk}=(\bm{R}^{-1})_{kk}$ and $\mu _{ik|-k}=\bm{x}_{i,\bm{\gamma}_{k}}^{T}\bm{\beta}_{\bm{\gamma}_{k}}-(\sigma
_{k}/r^{kk})\tilde{\zeta} _{ik}=\bm{x}_{i,\bm{\gamma}_{k}}^{T}\bm{\beta}_{
\bm{\gamma}_{k}}-(\sigma _{k|-k}^{2}/\sigma _{k})\tilde{\zeta} _{ik}$.
\label{rem:conddistrI}}

\remark{The Gaussian latent variable $\widetilde{Z}_{ik}$ conditionally on $\widetilde{\bm{Z}}_{i,-k}$ is normal distributed with variance $\tilde{\sigma} _{k|-k}^{2}=1/r^{kk}$ and mean $\tilde{\mu} _{ik|-k}=-(1/r^{kk})\tilde{\zeta} _{ik}=-\tilde{\sigma} _{k|-k}^{2}\tilde{\zeta} _{ik}$. Since $\tilde{\sigma}_{k|-k}^{2}=\sigma _{k|-k}^{2}/\sigma _{k}^{2}$, then $\tilde{\mu}_{ik|-k}=-(\sigma _{k|-k}^{2}/\sigma^{2} _{k})\tilde{\zeta} _{ik}$. \label{rem:conddistrII}}


\remark{In the case of a Gaussian response, since $\tilde{z}_{ik} = \Phi^{-1}\{F_{k}(y_{ik};\bm{x}_{i,\bm{\gamma}_{k}},\bm{\beta}_{\bm{\gamma}_{k}}, \theta_{k})\}$ and $F_{k}(\cdot) = \Phi(\cdot)$, we have that $y_{ik}=z_{ik}$ and eq. \eqref{eq:cond_lik} is a special case of the likelihood in eq. (11) for Gaussian data with $\sigma^{2} _{k}=\theta_{k}=\delta_{k}^{2}$. \label{rem:normal}}

\bigskip

\noindent Exploiting the fact that $\bm{\beta}_{k}$ is quadratic in eq. \eqref{eq:cond_lik}, we have
\begin{align*}
p(\bm{\beta}_{k}|& \bm{\gamma}_{k},\bm{z}_{k},\widetilde{\bm{Z}}_{-k},\bm{D},\bm{R})\propto  \\
& \propto \exp \left\{ -\frac{r^{kk}}{2\sigma _{k}^{2}}\bm{\beta}_{
\bm{\gamma}_{k}}^{T}\bm{X}_{\bm{\gamma}_{k}}^{T}\bm{X}_{\bm{\gamma}_{k}}\bm{\beta}_{
\bm{\gamma}_{k}}+\frac{r^{kk}}{\sigma _{k}^{2}}\bm{z}_{k}^{T}X_{
\bm{\gamma}_{k}}\bm{\beta}_{\bm{\gamma}_{k}}+\frac{1}{\sigma _{k}}\tilde{\bm{\zeta}}_{k}^{T}\bm{X}_{\bm{\gamma}_{k}}\bm{\beta}_{\bm{\gamma}_{k}}\right\} p(\bm{\beta}_{k}|\bm{\gamma}_{k}) \\
& \propto \exp \left\{ -\frac{r^{kk}}{2\sigma _{k}^{2}}\bm{\beta}_{
\bm{\gamma}_{k}}^{T}\bm{X}_{\bm{\gamma}_{k}}^{T}\bm{X}_{\bm{\gamma}_{k}}\bm{\beta}_{
\bm{\gamma}_{k}}+\left(\frac{r^{kk}}{\sigma _{k}^{2}} \bm{z}_{k}^{T}+\frac{1}{\sigma _{k}}\tilde{\bm{\zeta}}_{k}^{T}\right) \bm{X}_{\bm{\gamma}_{k}}\bm{\beta}_{\bm{\gamma}_{k}}\right\} p(\bm{\beta}_{k}|\bm{\gamma}_{k}).
\end{align*}
Finally, using the prior distribution for the regression coefficients in eq. (4) and Remark \ref{rem:conddistrI}, we get
\begin{align}
p(\bm{\beta}_{k}|& \bm{\gamma}_{k},\bm{z}_{k},\widetilde{\bm{Z}}
_{-k},\bm{D},\bm{R})\propto   \notag \\
& \propto \exp \left\{ -\frac{1}{2}\bm{\beta}_{\bm{\gamma}_{k}}^{T}\left(
\frac{r^{kk}}{\sigma _{k}^{2}}\bm{X}_{\bm{\gamma}_{k}}^{T}\bm{X}_{\bm{\gamma}
_{k}}+v^{-1}\bm{I}_{\left\vert \bm{\gamma}_{k}\right\vert }\right) \bm{\beta}_{\bm{\gamma}_{k}}+\left( \frac{r^{kk}}{\sigma _{k}^{2}}\bm{z}_{k}^{T}+\frac{1}{\sigma _{k}}\tilde{\bm{\zeta}}_{k}^{T}\right) \bm{X}_{\bm{\gamma}_{k}}\bm{\beta}_{\bm{\gamma}_{k}}\right\}   \notag \\
& \propto \exp \left\{ -\frac{1}{2}\bm{\beta}_{\bm{\gamma}_{k}}^{T}\bm{V}_{
\bm{\gamma}_{k}}^{-1}\bm{\beta}_{\bm{\gamma}_{k}}+\bm{\beta}_{\bm{\gamma}
_{k}}^{T}\bm{V}_{\bm{\gamma}_{k}}^{-1}\bm{V}_{\bm{\gamma}_{k}}\bm{X}_{\bm{\gamma}
_{k}}^{T}\left( \frac{r^{kk}}{\sigma _{k}^{2}}\bm{z}_{k}+\frac{1}{\sigma _{k}}\tilde{\bm{\zeta}}_{k}\right) \right\} ,  \notag \\
& \propto \exp \left\{ -\frac{1}{2}\bm{\beta}_{\bm{\gamma}_{k}}^{T}\bm{V}_{
\bm{\gamma}_{k}}^{-1}\bm{\beta}_{\bm{\gamma}_{k}}+\bm{\beta}_{\bm{\gamma}
_{k}}^{T}\bm{V}_{\bm{\gamma}_{k}}^{-1}\bm{V}_{\bm{\gamma}_{k}}\bm{X}_{\bm{\gamma}
_{k}}^{T}\sigma _{k|-k}^{-2}\left( \bm{z}_{k}+\frac{\sigma _{k|-k}^{2}}{
\sigma _{k}}\tilde{\bm{\zeta}}_{k}\right) \right\},
\label{eq:SBpropI}
\end{align}
where $\bm{V}_{\bm{\gamma}_{k}}^{-1}=\sigma
_{k|-k}^{-2}\bm{X}_{\bm{\gamma}_{k}}^{T}\bm{X}_{\bm{\gamma}_{k}}+v^{-1}\bm{I}_{\left\vert
\bm{\gamma}_{k}\right\vert }$. Given a candidate value $\bm{\gamma}_{k}^{\ast}$ for latent binary vector $\bm{\gamma}_{k}$, we propose the non-zero regression coefficients $\bm{\beta}_{k}^{\ast}$ from the distribution with density
\begin{align*}
q(\bm{\beta}_{k}^{\ast }|\bm{\gamma}_{k}^{\ast },\bm{z}_{k},\widetilde{\bm{Z}}_{-k},\bm{D},\bm{R})& \propto p(\bm{z}_{k}|\bm{\beta}_{k}^{\ast },\bm{\gamma}_{k}^{\ast },\widetilde{\bm{Z}}_{-k},\bm{D},\bm{R})p(\bm{\beta}_{k}^{\ast }|\bm{\gamma}_{k}^{\ast }) \\
& \propto \mathcal{N}_{|\bm{\gamma}_{k}^{\ast }|}(\bm{\beta}_{k}^{\ast };\bm{V}_{\bm{\gamma}_{k}^{\ast }}\bm{X}_{\bm{\gamma}_{k}^{\ast }}^{T}\sigma _{k|-k}^{-2}\{\bm{z}_{k}+\sigma _{k|-k}^{2}/\sigma _{k}\tilde{\bm{\zeta}} _{k}\},\bm{V}_{\bm{\gamma}_{k}^{\ast }}).
\end{align*}

\remark{By Remark \ref{rem:conddistrI}, $\E(\bm{Z}_{k}|\widetilde{\bm{Z}}_{-k})=\bm{X}_{\gamma }\bm{\beta} _{\gamma }-\sigma _{k|-k}^{2}/\sigma
_{k}\tilde{\bm{\zeta}} _{k}$. Then, $\E(\bm{Z}_{k}|\widetilde{\bm{Z}}_{-k}+\sigma _{k|-k}^{2}/\sigma _{k}\tilde{\bm{\zeta}} _{k})=\E(\bm{Z}_{k})=\bm{X}_{\bm{\gamma}
}\bm{\beta} _{\gamma }$. Thus, in the proposal density, we recent the realizations of $\bm{Z}_{k}|\widetilde{\bm{Z}}_{-k}$ such that their mean does not depend on $\widetilde{\bm{Z}}_{-k}$. By Remark \ref{rem:conddistrII}, we can also write the proposal density in a more compact form as
\begin{equation}
q(\bm{\beta}_{k}^{\ast }|\bm{\gamma}_{k}^{\ast },\bm{z}_{k},\widetilde{\bm{Z}}
_{-k},\bm{D},\bm{R})\propto \mathcal{N}_{|\bm{\gamma}_{k}^{\ast }|}(\bm{\beta}
_{k}^{\ast };\bm{V}_{\bm{\gamma}_{k}^{\ast }}\bm{X}_{\bm{\gamma}_{k}^{\ast
}}^{T}\sigma_{k|-k}^{-2}\{\bm{z}_{k}-\sigma_{k}\tilde{\bm{\mu}}_{k|-k}\},\bm{V}_{
\bm{\gamma}_{k}^{\ast }}).
\label{eq:SBpropII}
\end{equation}}

\remark{If in eq. (4) of the main paper $v\rightarrow \infty $, then $\bm{V}_{\bm{\gamma}_{k}}=\sigma_{k|-k}^{2}(\bm{X}_{\bm{\gamma}_{k}}^{T}\bm{X}_{\bm{\gamma}
_{k}})^{-1}$ and
\begin{eqnarray}
q(\bm{\beta}_{k}^{\ast }|\bm{\gamma}_{k}^{\ast},\bm{z}_{k},\widetilde{\bm{Z}}_{-k},\bm{D},\bm{R}) &\propto &\mathcal{N}_{|\bm{\gamma}_{k}^{\ast }|}(\bm{\beta}_{k}^{\ast };(\bm{X}_{\bm{\gamma}_{k}^{\ast}}^{T}\bm{X}_{\bm{\gamma}_{k}^{\ast}})^{-1}
\bm{X}_{\bm{\gamma}_{k}^{\ast }}^{T}\{\bm{z}_{k}-\sigma_{k}\tilde{\bm{\mu}}
_{k|-k}\}),\bm{V}_{\bm{\gamma}_{k}^{\ast }})   \notag \\
&\propto &\mathcal{N}_{|\bm{\gamma}_{k}^{\ast }|}(\bm{\beta}_{k}^{\ast };\hat{
\bm{\beta}}_{\bm{z}_{k}}(\bm{\gamma}_{k}^{\ast})-\sigma_{k}\hat{\bm{\beta}}
_{\tilde{\bm{\mu}}_{k|-k}}(\bm{\gamma}_{k}^{\ast}),\bm{V}_{\bm{\gamma}_{k}^{\ast }}),
\label{eq:SBpropbeta}
\end{eqnarray}
where $\hat{\bm{\beta}}_{\bm{z}_{k}}(\bm{\gamma}_{k})=(\bm{X}_{\bm{\gamma}_{k}}^{T}\bm{X}_{
\bm{\gamma}_{k}})^{-1}\bm{X}_{\bm{\gamma}_{k}}^{T}\bm{z}_{k}$ and $\hat{\bm{\beta}}_{\tilde{\bm{\mu}}_{k|-k}}(\bm{\gamma}_{k})=(\bm{X}_{\bm{\gamma}
_{k}}^{T}\bm{X}_{\bm{\gamma}_{k}})^{-1}\bm{X}_{\bm{\gamma}_{k}}^{T}\tilde{\bm{\mu}}_{k|-k}$. Thus, assuming $n>|\bm{\gamma}_{k}^{\ast }|$, the mean of the proposal distribution is centered in $\hat{\bm{\beta}}_{\bm{z}_{k}}(
\bm{\gamma}_{k}^{\ast })$ and adjusted by the rescaled MLEs of the coefficients of the regression of the conditional mean $\tilde{\bm{\mu}}_{k|-k}$ on $\bm{X}_{\bm{\gamma}_{k}^{\ast}}$, where the factor $\sigma_{k}$ brings the MLEs of the regression coefficients $\hat{\bm{\beta}}_{\bm{z}_{k}}(\bm{\gamma}_{k})$ and $\hat{\bm{\beta}}_{\tilde{\bm{\mu}}_{k|-k}}(\bm{\gamma}_{k})$ on the same scale. If $\bm{R}=\bm{I}_{m}$, eq. \eqref{eq:SBpropbeta} simplifies to the expression provided in \cite{albert1993bayesian}.

\section{Implicit marginalization of the regression coefficients in the Metropolis-Hastings step}
\label{sec:marginalisation}

The acceptance probability of the Metropolis-Hastings step for the target in eq. (10) is
\begin{equation*}
\alpha =1\wedge \frac{p(\bm{y}_{k}|\bm{\beta}_{k}^{\ast },\bm{\gamma}
_{k}^{\ast},\bm{\theta}_{k},\widetilde{\bm{Z}}_{-k},\bm{D},\bm{R})p(\bm{\beta}_{k}^{\ast }|\bm{\gamma}_{k}^{\ast })p(\bm{\gamma}_{k}^{\ast })q(\bm{\beta}_{k}|
\bm{\gamma}_{k},\bm{z}_{k},\widetilde{\bm{Z}}_{-k},\bm{D},\bm{R})q(\bm{\gamma}
_{k}|\bm{\gamma}_{k}^{\ast})}{p(\bm{y}_{k}|\bm{\beta}_{k},\bm{\gamma}_{k},
\bm{\theta}_{k},\widetilde{\bm{Z}}_{-k},\bm{D},\bm{R})p(\bm{\beta}_{k}|\bm{\gamma}_{k})p(
\bm{\gamma}_{k})q(\bm{\beta}_{k}^{\ast }|\bm{\gamma}_{k}^{\ast },\bm{z}_{k},\widetilde{\bm{Z}}_{-k},\bm{D},\bm{R})q(\bm{\gamma}_{k}^{\ast }|
\bm{\gamma}_{k})},
\end{equation*}
where $q(\bm{\beta}_{k}^{\ast }|\bm{\gamma}_{k}^{\ast},\bm{z}_{k},\widetilde{\bm{Z}}_{-k},\bm{D},\bm{R})$ is defined in eq. (14) of the main paper.
\begin{itemize}
\item For a Gaussian response, using Remark \ref{rem:normal}, we have that $\bm{y}_{k}=\bm{z}_{k}$. Given a candidate value $\bm{\gamma}_{k}^{\ast}$ for latent binary vector $\bm{\gamma}_{k}$, to obtain eq. \eqref{eq:SBpropII}, we multiply and divide eq. \eqref{eq:SBpropI} by
    $\left\vert \bm{V}_{\bm{\gamma}^{\ast} _{k}}\right\vert ^{-1/2}\exp\{-(1/2)\bm{m}_{\bm{\gamma}^{\ast}_{k}}^{T}\bm{V}_{    \bm{\gamma}^{\ast}_{k}}\bm{m}_{\bm{\gamma}_{k}}\}$ with $\bm{m}_{\bm{\gamma}^{\ast}_{k}}=\bm{V}_{\bm{\gamma}^{\ast}_{k}}
    \bm{X}_{\bm{\gamma}^{\ast}_{k}}^{T}\sigma_{k|-k}^{-2}(\bm{z}_{k}-
    \sigma_{k}\tilde{\bm{\mu}}_{k|-k})$. Then, the acceptance probability of the M-H step becomes
   \begin{equation}
   \alpha =1\wedge \frac{v^{|\bm{\gamma}_{k}|/2}\left\vert \bm{V}_{\bm{\gamma}_{k}}\right\vert ^{-1/2}\exp \left( -\frac{1}{2}\bm{m}_{\bm{\gamma}   _{k}}^{T}\bm{V}_{\bm{\gamma}_{k}}^{-1}\bm{m}_{\bm{\gamma}_{k}}\right) p(
   \bm{\gamma}_{k}^{\ast })q(\bm{\gamma}_{k}|\bm{\gamma}_{k}^{\ast })}{v^{|\bm{\gamma}_{k}^{\ast }|/2}\left\vert \bm{V}_{\bm{\gamma}_{k}^{\ast }}\right\vert
   ^{-1/2}\exp \left( -\frac{1}{2}\bm{m}_{\bm{\gamma}_{k}^{\ast }}^{T}\bm{V}_{\bm{\gamma}_{k}^{\ast }}^{-1}\bm{m}_{\bm{\gamma}_{k}^{\ast }}\right) p(\bm{\gamma}_{k})q(\bm{\gamma}_{k}^{\ast }|\bm{\gamma}_{k})}.
   \label{eq:implicitmarg}
   \end{equation}
\item For a binary/ordinal response, we choose the Probit regression model which is specified by setting $\Pr (Y_{ik}=c|\bm{\beta}_{k},\bm{\gamma}_{k},\theta _{kc})=\Phi (\theta _{kc}-\bm{x}_{i,\bm{\gamma}_{k}}^{T}\bm{\beta}_{\bm{\gamma}_{k}})$, where $\theta _{kc}$, $c=1,\ldots ,C_{k}-1$, are cut-points with $\theta _{k0}=-\infty $ and \ $\theta _{kC_{k}}=+\infty $.
    This representation is justified by the data augmentation approach which assumes that $Y_{ik}=c$ if $Z_{ik}\in (\theta _{k,c-1},\theta _{kc}]$ and $Z_{ik}\sim \mathrm{N}(\bm{x}_{i,\bm{\gamma}_{k}}^{T}\bm{\beta}_{\bm{\gamma}_{k}},1)$ \citep{albert1993bayesian}. For a binary response, setting $C_{k}=2$ with $\theta _{k}=0 $ and defining $Y_{ik}=1$ if $Z_{ik}> 0$ and $Y_{ik}=0$ otherwise, then $\Pr (Y_{ik}=0|\bm{\beta}_{k},\bm{\gamma}_{k},\theta _{k})=\Pr
    (Z_{ik}\leq0)=\Pr(\widetilde{Z}_{ik}\leq-\bm{x}_{i,\bm{\gamma}_{k}}^{T}\bm{\beta}_{
    \bm{\gamma}_{k}})=\Phi (-\bm{x}_{i,\bm{\gamma}_{k}}^{T}\bm{\beta}_{
    \bm{\gamma}_{k}})$ which implies the transformation in eq. (3) of the main paper, i.e.,  $F_{k}(y_{ik}|\bm{x}_{ik},\bm{\beta}_{k},\bm{\gamma}_{k},\theta _{k})= \mathbb{I}\{y_{ik}=0\}\Phi (-\bm{x}_{i,\bm{\gamma}_{k}}^{T}\bm{\beta}_{\bm{\gamma}_{k}})+\mathbb{I}\{y_{ik}=1\}\Phi (\bm{x}_{i,\bm{\gamma}_{k}}^{T}\bm{\beta}_{\bm{\gamma}_{k}})$. Thus, the discrete marginal distribution is replaced by the density
    \begin{eqnarray*}
    p(y_{ik},z_{ik};\bm{\beta}_{k},\bm{\gamma}_{k},\theta _{k}) & = &(\mathbb{I}\{y_{ik}=0\}\mathbb{I}\{Z_{ik}\leq 0\}+\mathbb{I}\{y_{ik}=1\}\mathbb{I}\{Z_{ik}>0\})\times   \notag \\
    &&\mathcal{N}(z_{ik};\bm{x}_{i,\bm{\gamma}_{k}}^{T}\bm{\beta}_{\bm{\gamma}
    _{k}},1)
    \end{eqnarray*}
    in eq. \eqref{eq:proposal_lik} from which is it straightforward to derive $p(\bm{\beta}_{k}|\bm{y}_{k}, \bm{\gamma}_{k},\bm{z}_{k},\widetilde{\bm{Z}}_{-k},\bm{D},\bm{R})$. 
    Then, the \lq\lq implicit marginalization\rq\rq\ of the regression coefficients follows. 
    Similar arguments can be used for an ordinal response since $p(y_{ik}, z_{ik};\bm{\beta}_{k},\bm{\gamma}_{k},\bm{\theta}_{k})=(\sum
    \nolimits_{c=1}^{C_{k}}\mathbb{I}\{y_{ik}=c\}\mathbb{I\{}\theta
    _{k,c-1}<Z_{ik}\leq \theta _{kc}\})\mathcal{N}(z_{ik};\bm{x}_{i,\bm{\gamma}_{k}}^{T}
    \bm{\beta}_{\bm{\gamma}_{k}},1)$. Thus, for both binary/ordinal responses, the acceptance probability has the form shown in eq.
    \eqref{eq:implicitmarg}.
\item For an unordered categorical response, we follow \cite{albert1993bayesian}. More precisely, we assume that for each one of the $i=1,\ldots,n$ individuals and $c=1,\ldots,C_{k}$ categories ($C_{k} > 2$), there is an unobserved (latent) random variable $Z_{ikc} = \bm{x}^{T}_{i,\bm{\gamma}_{k},c}\bm{\beta}_{\bm{\gamma}_{k}} +\epsilon_{ikc}$, where $\bm{x}_{i,\bm{\gamma}_{k},c}$ is a $p_{\bm{\gamma}_{k}}$-dimensional vector of predictors associated with the $c$th category, $\bm{\beta}_{\bm{\gamma}_{k}}$ are the corresponding non-zero regression coefficients that are not category-specific \citep{Fruehwirth-Schnatter2012} and $\bm{\epsilon}_{ik} = (\epsilon_{ik1},\ldots,\epsilon_{ikC_{k}}) \sim \N_{C_{k}}(0,\bm{\Lambda})$ with $\bm{\Lambda}$ such that at least $\bm{\Lambda}_{11} =1$. Then, the category $c$ is observed, i.e, $y_{ik} =c$, if $z_{ikc} >z_{ikc'}$ for all $c\neq c'$. The density of the latent variables given the observed data can be written as
    \begin{equation}
    p(\bm{Z}_{k}|\bm{y}_{k},\bm{\beta}_{k},\bm{\gamma}_{k},\bm{\Lambda} )\propto |\bm{\Lambda}|^{-1/2}\exp \left\{ -\frac{1}{2}\sum_{i=1}^{n}(\bm{z}_{ik}-\bm{X}_{i,\bm{\gamma}_{k}}^{T} \bm{\beta}_{\bm{\gamma}_{k}})^{T}\bm{\Lambda} ^{-1}(\bm{z}_{ik}-\bm{X}_{i,\bm{\gamma}_{k}}^{T}\bm{\beta}_{\bm{\gamma}_{k}})\right\},
    \label{unordered}
    \end{equation}
    where $\bm{Z}_{k}$ is the $n \times C_{k}$ matrix of unobserved random variables, $\bm{y}_{k}$ is the vector of observed categories and $\bm{X}_{i,\bm{\gamma}_{k}}$ is a $p_{\bm{\gamma}_{k}} \times C_{k}$ matrix of predictors for the unordered categories. For identifiability reasons, the $C_{k}$ level is usually taken as reference level and thus $C_{k}-1$ latent variables are required.

    Without loss of the generality, we assume that the unordered categorical response $\bm{y}_m$ is the last observation, with $\bm{Z}_{m},\bm{Z}_{m+1},\ldots,\bm{Z}_{m+C_{m}-2}$ the $C_{m}-1$ continuous latent variables used to model $\bm{y}_m$. Based on the centered parametrization of the Gaussian latent variables in eq. (13), for $k > m-1$, we define $Z_{ikc} = \bm{x}^{T}_{i,\bm{\gamma}_{k},c}\bm{\beta}_{\bm{\gamma}_{k}} + \delta_{k} \widetilde{Z}_{ikc}$ with $\delta_{m}=1$. Since for $k > m-1$ we use the Gaussian copula to model the unordered categorical response for which a linear regression model has been assumed, and  using eq. \eqref{unordered}, it follows that the \lq\lq implicit marginalisation\rq\rq\ property holds as in the case of Gaussian responses. For multivariate unordered categorical responses, see the extension of eq. \eqref{unordered} in
    \cite{zhang2015bayesian}.
\end{itemize}

\section{Computing the  acceptance probability in eq. (18)}}
\label{sec:acceptG}

To sample the adjacency matrix $\bm{G}$, we target the distribution with density $p(\bm{G}|\widetilde{\bm{W}}) \propto p(\widetilde{\bm{W}}|\bm{G})p(\bm{G})$ with $p(\bm{G})$ the induced marginal prior on the adjacency matrix and
\begin{equation*}
p(\widetilde{\bm{W}}|\bm{G})=\int_{M(\mathcal{G})}p(\widetilde{\bm{W}}|
\bm{\Sigma},\bm{G})p(\bm{\Sigma}|\bm{G})\,d\bm{\Sigma},
\end{equation*}
where $M(\mathcal{G})$ denotes the set of all possible symmetric positive definite matrices consistent with $\mathcal{G}$, $\widetilde{\bm{W}}=\widetilde{\bm{Z}}\bm{D}$ is the $(n\times m)$-dimensional matrix of the scaled Gaussian latent variables defined in eq. (7) of the main paper, $p(\widetilde{\bm{W}}|\bm{\Sigma},\bm{G})$ is the Markov Gaussian density and $p(\bm{\Sigma}|\bm{G})$ is the density of the hyper-inverse Wishart distribution.

\subsection{Decomposable graphs}

On the space of the scaled Gaussian latent variables $\widetilde{\bm{W}}$, if $\bm{\Sigma}$ is a covariance matrix consistent with a decomposable graph $\mathcal{G}$, then $p(\widetilde{\bm{W}}|\bm{\Sigma},\bm{G})$ is Markov with respect to any decomposable graphs $\mathcal{G}$ \citep{dawid1993hyper}. Based on the results presented in \cite{talhouk2012efficient}, i.e., $\bm{\Sigma} |\bm{G} \sim \H\I\W_{\mathcal{G}}(2,\bm{I}_{m})$, it is straightforward to show that
\begin{equation}
p(\widetilde{\bm{W}}|\bm{G})=(2\pi)^{-mn/2}\dfrac{H(\bm{G},2,\bm{I}_{m})}
{H(\bm{G},2+n,\bm{I}_{m}+\widetilde{\bm{W}}^{T}\widetilde{\bm{W}})},
\label{eq:Wmarg}
\end{equation}
where the constant of normalization $H(\cdot)$ we can be calculate explicitly \citep{Roverato2002}.

\subsection{Non-decomposable graphs}

Let $\bm{K} = \bm{\Sigma}^{-1} = (\bm{DRD})^{-1}$, where $\bm{D}$ is diagonal and $\delta_{k} \overset{\text{iid}}{\sim }\mathrm{IGam}((m+1)/2,r^{kk}/2)$, $k=1,\ldots,m$ and assume that $\bm{K}$ follows a G-Wishart distribution
defined as
\begin{equation}
p(\bm{K}|\bm{G})=\frac{1}{I_G (b,\bm{\Lambda})} |\bm{K}|^{(b-2)/2} \exp \left\{ -\frac{1}{2} \tr(\bm{\Lambda} \bm{K}) \right\} \mathbb{I}\{\bm{K} \in \mathbb{P}_{G}\},
\label{eq:G-Wishart_dens}
\end{equation}
where $b$ are the degrees of freedom, $\bm{\Lambda}$ is a symmetric positive definite matrix, $I_G (b,\bm{\Lambda})$ is the normalizing constant with respect to $\bm{G} \in \mathcal{G}$ and $\mathbb{I}\{\cdot\}$ is the indicator function.

While for decomposable graphs, the normalizing constant in eq. \eqref{eq:G-Wishart_dens} has an explicit form, for non-decomposable graphs, it does not. In that case it can be calculated by using the Monte Carlo method \citep{Atay-Kayis2005}, the Laplace approximation \citep{Lenkoski2011} or the more recent attempts by \cite{Wang2012}, \cite{Lenkoski2013} and \cite{Mohammadi2015}.

Assuming without loss of generality that $\bm{\Lambda} =\bm{I}_{m}$, it can be shown that
\begin{equation}
p (\bm{\Sigma} |b,\bm{I}_{m})=H(b,\bm{I}_{m})\det (I_{SS}(\bm{I}_{m}))^{1/2}\frac{\det (I_{SS}(\bm{\Sigma}
))^{-1}}{\det (\bm{\Sigma} )^{(b-2)/2}}\exp \left\{ -\frac{1}{2}\tr(\bm{\Sigma}
^{-1})\right\},
\label{eq:HIWnodecomp}
\end{equation}
where $H(b,\bm{I}_{m})$ is the constant of normalization and $I_{SS}(\bm{I}_{m})$ is the Isserlis matrix of $\bm{I}_{m}$. The density in eq. \eqref{eq:HIWnodecomp} is the generalization of the hyper-inverse Wishart for non-decomposable graphs and \cite{Roverato2002} referred to the corresponding distributions as hyper-inverse Wishart as well. He also proved that given $\bm{G} \in \mathcal{G}$, the density in eq. \eqref{eq:HIWnodecomp} can be written as
\begin{equation}
p(\bm{\Sigma}|\bm{G}, b, \bm{I}_{m}) = \frac{\prod_{P \in \mathcal{P}} p(\bm{\Sigma}^P|b,\bm{I}_{m})}{\prod_{S \in \mathcal{S}} p(\bm{\Sigma}^S|b,\bm{I}_{m})},
\label{eq:SigPrior}
\end{equation}
where $\mathcal{P}$ and $\mathcal{S}$ are the sets of prime components and separators of $\mathcal{\bm{G}}$. Note that, since $\mathcal{\bm{G}}$ is non-decomposable, at least one prime component is incomplete.

Similarly to \cite{talhouk2012efficient}, and based on eq. \eqref{eq:SigPrior}, the prior $\bm{R}|\bm{G}$ can be factorized as
\begin{equation*}
p(\bm{R}|\bm{G}) = \frac{\prod_{P \in \mathcal{P}} p(\bm{R}^P)}{\prod_{S \in \mathcal{S}} p(\bm{R}^S)}.
\end{equation*}
For each complete component $P \in \mathcal{P}$ and separator $S \in \mathcal{S}$, the density is marginally uniform
\begin{equation*}
p(\bm{R}^U) \propto \det(\bm{R}^U)^{\tfrac{|U|(|U|-1)}{2}}\Big( \prod_{k \in U} \det(\bm{R}_{kk}^U) \Big)^{-\tfrac{|U|+1}{2}},
\end{equation*}
where $|U|$ is the cardinality of $U$, $\bm{R}_{kk}^U$ is the principal sub-matrix of $\bm{R}^U$ and the use of an upper script
$U$ has the same interpretation as in eq. \eqref{eq:SigPrior}. For the incomplete prime components $U$, the prior $p(\bm{R}^U)$ is obtained by applying the transformation $\bm{\Sigma} = \bm{DRD}$ on the density in eq. \eqref{eq:HIWnodecomp} and integrating out $\bm{D}$. However, due to the presence of the determinant of the Isserlis matrix, this integration cannot be conducted analytically.

\section{Missing observations}
\label{sec:impute}


Imputation of missing observations under the Bayesian framework can be conducted by treating any missing data as unknowns. Thus, instead of targeting the posterior distribution with density $p(\bm{B},\bm{\Gamma},\bm{\Theta},\widetilde{\bm{Z}},\bm{D},\bm{R},\bm{G}|\bm{Y})$,  the designed MCMC algorithm has to target the posterior $p(\bm{B},\bm{\Gamma},\bm{\Theta},\widetilde{\bm{Z}},\bm{D},\bm{R},\bm{G},\bm{Y}_{mis}|
\bm{Y}_{obs})$, where we have assumed that the mixed data are made of the observed data $\bm{Y}_{obs}$ and a portion of missing observations denoted by $\bm{Y}_{mis}$. We also denote by $\bm{Y}_{mis}^{c}$ the missing observations from the continuous response variables and with $\bm{Y}_{mis}^{d}$ the missing discrete observations, that is $\bm{Y}_{mis}=(\bm{Y}_{mis}^{c},\bm{Y}_{mis}^{d})$. To draw sample from the target posterior we modify Algorithm 1 as follows.

If observations are missing exclusively from the discrete response variables the update of the latent variables $\widetilde{\bm{Z}}$ is conducted as described by Algorithm 1. In the presence of missing values in the continuous responses, we need to sample the latent variables that correspond to the missing values $\bm{Y}_{mis}^{c}$ from a conditional univariate normal distribution without truncation. When $\bm{Y}_{mis}=(\bm{Y}_{mis}^{c},\bm{Y}_{mis}^{d})$, $\bm{Y}_{mis}^{d}$ have to be drawn conditionally on the Gaussian latent variables $\widetilde{\bm{Z}}$ and this results in a reducible Markov chain since each component $\bm{Y}_{mis}^{d}$ will not make any transition from the state determined by the initial values of $\widetilde{\bm{Z}}$. To overcome this problem, we follow \cite{zhang2015bayesian} and we integrate $\bm{Y}_{mis}^{d}$ from the target posterior distribution. This implies that an additional step has to be added in Algorithm 1. More precisely, the latent variables that correspond to $\bm{Y}_{mis}^{d}$ are drawn from Gaussian distributions without truncation and the remaining steps of the MCMC algorithm are the same ones as described in Algorithm 1 of the main paper. The resulting Markov chain will be irreducible and the MCMC will target the desired stationary distribution.

\section{Interval score for the regression coefficients}
\label{sec:mis}

To evaluate the posterior credible intervals of the non-zero regression coefficients in the simulated examples we used a modified version of the interval score presented in \cite{gneiting2007strictly}. It was introduced to assess the forecast ability of a model, but we use here to assess the accuracy of the regression coefficients' posterior credible intervals. Specifically, if $\beta$ is the value of the non-zero simulated regression coefficient and $\beta_{\ell}$ and $\beta_{u}$ denote the $\alpha/2$ and $1-\alpha/2$ quantiles of the $(1-\alpha)\times 100\%$ posterior credible interval for $\beta$ then the interval score is given by the formula
\begin{equation*}
(\beta_{u}-\beta_{\ell}) + \frac{2}{\alpha}(\beta_{\ell}-\beta) \mathbb{I}_{\beta < \beta_{\ell}} + \frac{2}{\alpha}(\beta-\beta_{u}) \mathbb{I}_{\beta > \beta_{u}}.
\end{equation*}

\section{Additional results from the simulation study}
\label{sec:rocs}

\subsection{Estimated graph}

\begin{figure}[h]
\centering
\includegraphics[height=0.5\textheight]{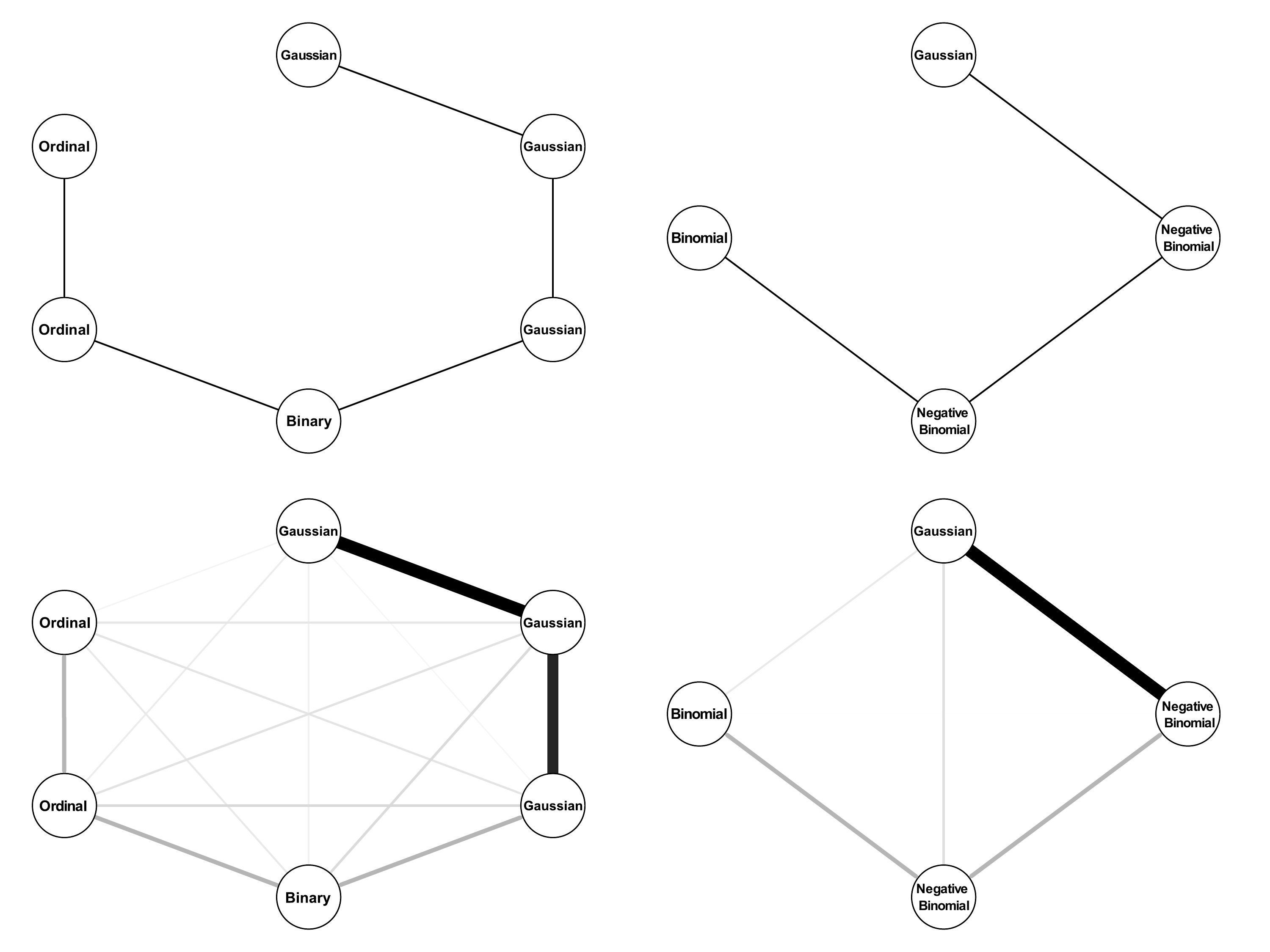}
\caption{Top: Simulated graphs used in Section 4.2 (left) and in Section 4.3 (right) of the main paper. Bottom: The corresponding graphs implied by the non-zero pattern of $\bm{R}^{-1}$ and estimated by using the edge posterior probabilities of inclusion (EPPI) obtained by specifying a decomposable graphical model and averaged over $20$ replicates in Scenario II (left) and IV (right), respectively. Gray scale and edge thickness specify different levels of the EPPI (black thick line indicates large EPPI values).}
\label{fig:EPPI_Mixed}
\end{figure}

Figure \ref{fig:EPPI_Mixed} presents the simulated and the graph implied by the non-zero pattern of $\bm{R}^{-1}$ and estimated by using the edge posterior probabilities of inclusion (EPPI) averaged across $20$ replicates of the simulation study presented in Section 4 of the main paper. For each replicates, EPPI were obtained by specifying a decomposable graphical model. 
For these particular simulated examples, the graph reconstruction that we achieve based on the EPPI converges to a slightly denser graph. 

\clearpage
\spacingset{1}

\subsection{ROC curves for the BVSGCR model and for independent single-response regression models}

\vspace{-0.5cm}

\begin{figure}[H]
\centering
\includegraphics[height=0.385\textheight]{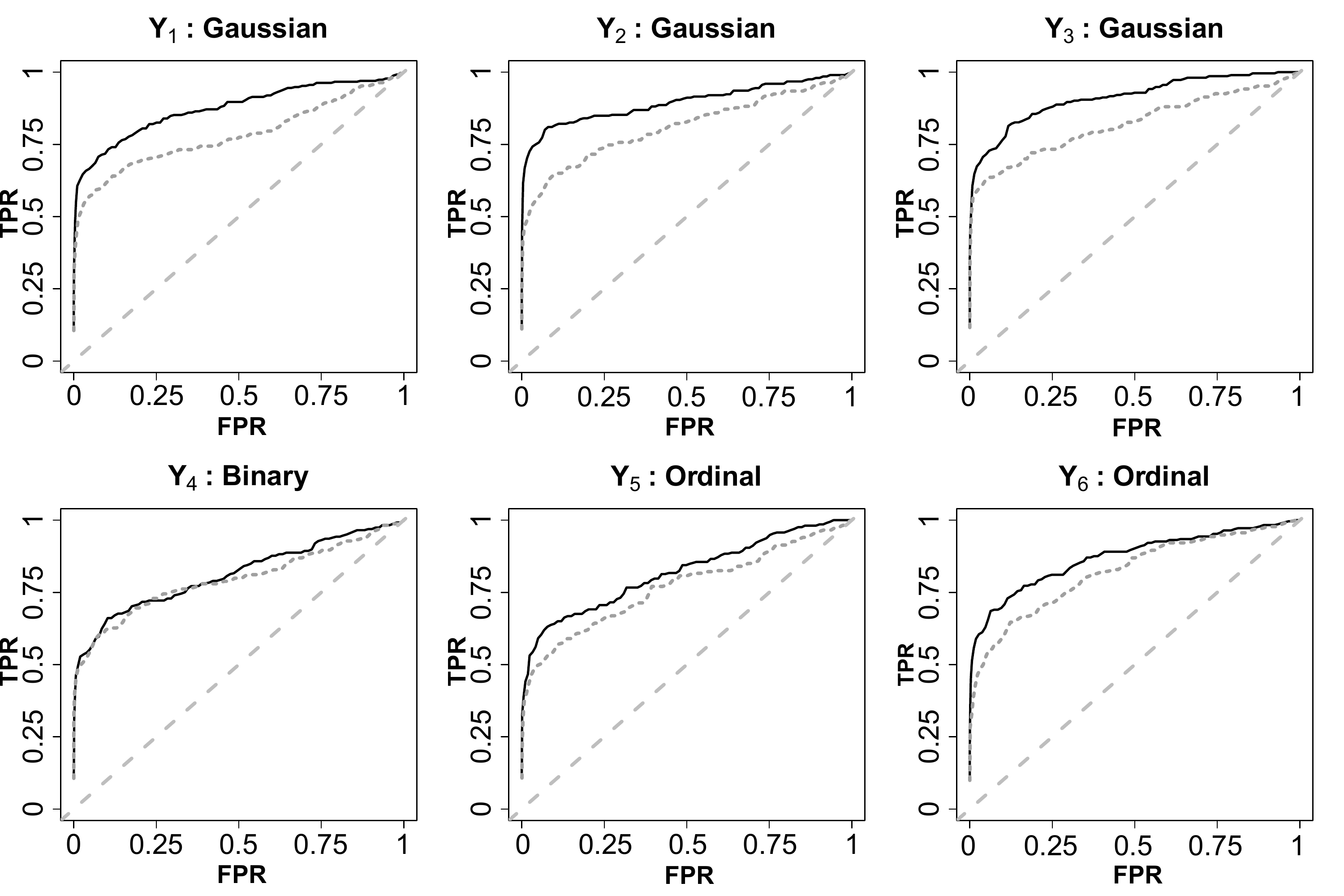}
\caption{Average (over $20$ replicates) ROC curves for the BVSGCR model (black solid lines) and for independent single-response regression models gray dotted lines) for each response in the simulated Scenario II described in Section 4.2 of the main paper.}
\label{fig:ROC_Mixed_n100}

\includegraphics[height=0.385\textheight]{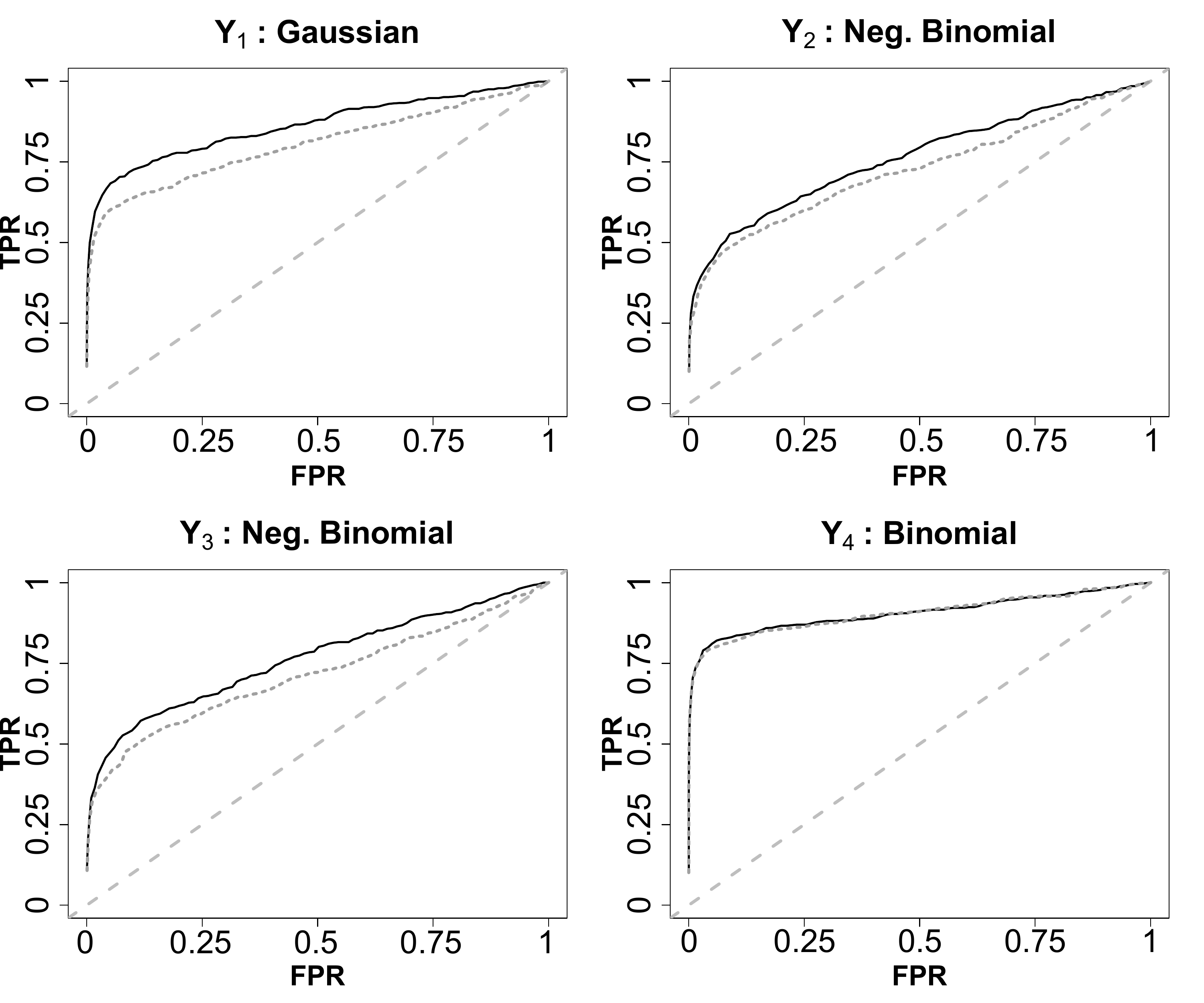}
\caption{Average (over $20$ replicates) ROC curves for the BVSGCR model (black solid lines) and for independent single-response regression models (gray dotted lines) for each response in the simulated Scenario IV described in Section 4.3 of the main paper.}
\label{fig:ROC_Count_n100}
\end{figure}

\subsection{ROC curves for the BVSGCR model with different proposal densities for the regression coefficients
\label{sec:naive}}

\vspace{-0.5cm}

\begin{figure}[H]
\centering
\includegraphics[height=0.385\textheight]{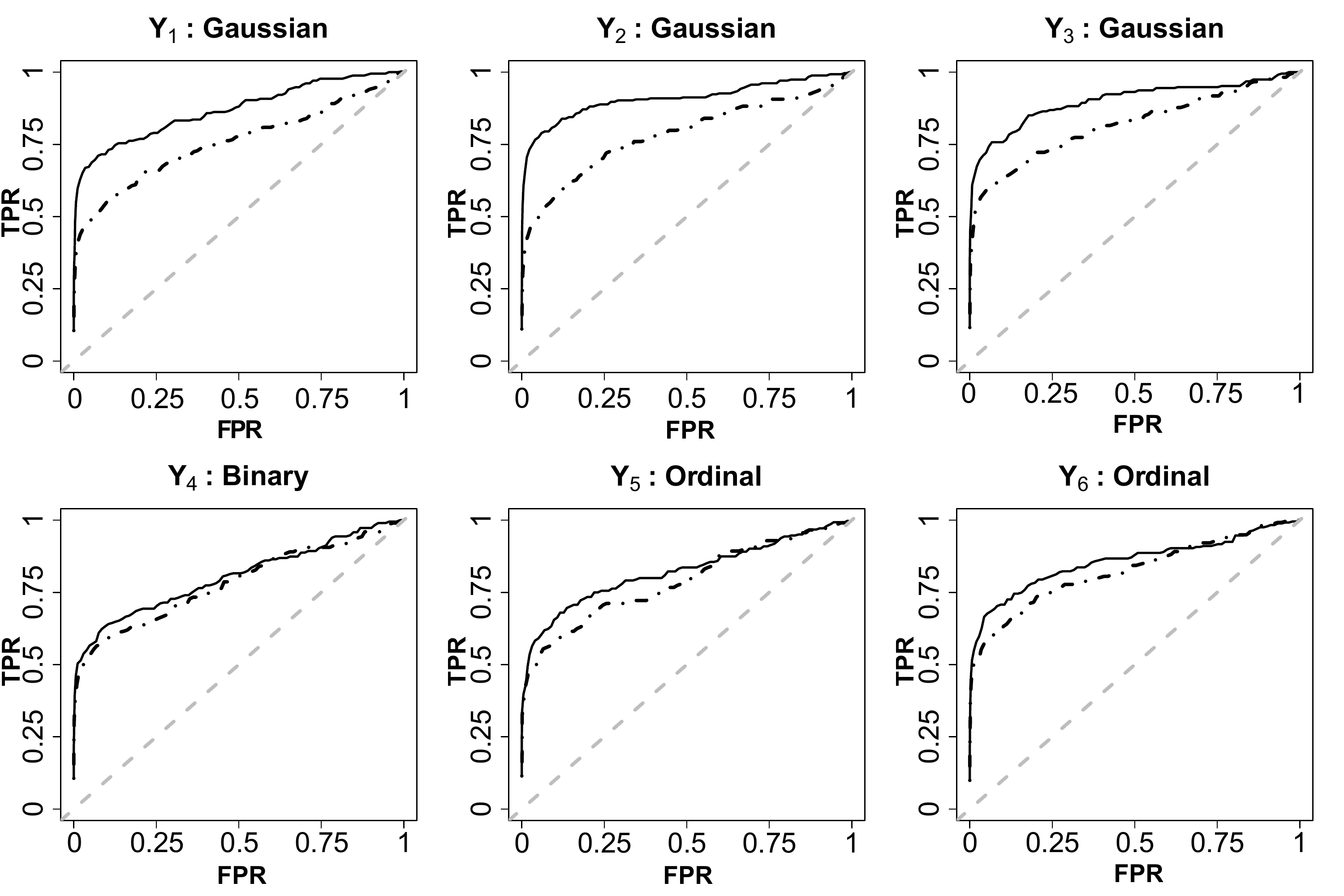}
\caption{Average (over $20$ replicates) ROC curves for the BVSGCR model with the proposal density in eq. (14) (black solid lines) and the simpler version in eq. (15) (black dashed-dotted lines) of the main paper for each response in the simulated Scenario II.}
\label{fig:ROC_Mixed_n100_naive}

\includegraphics[height=0.385\textheight]{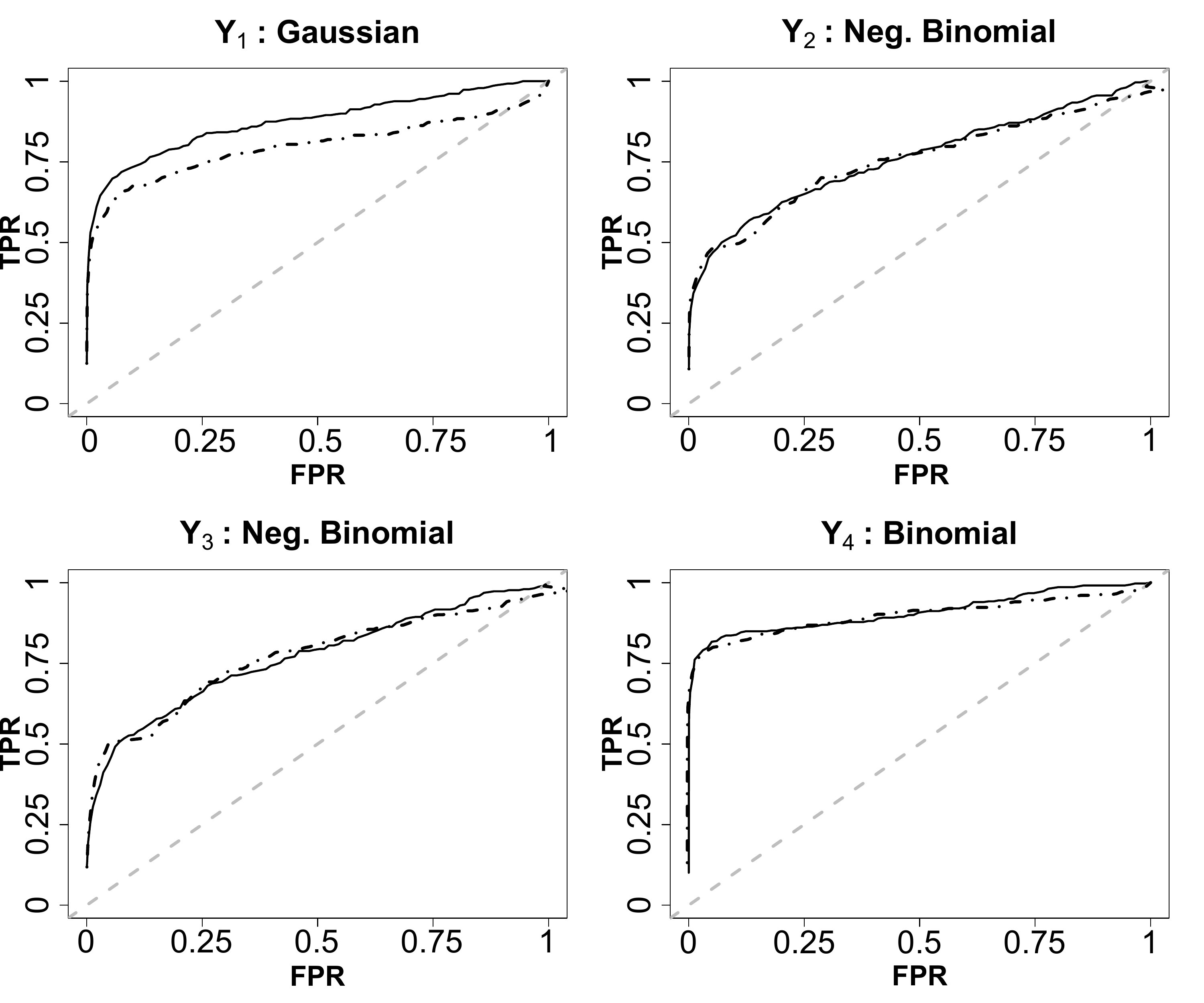}
\caption{Average (over $20$ replicates) ROC curves for the BVSGCR model with the proposal density in eq. (14) (black solid lines) and the simpler version in eq. (15) (black dashed-dotted lines) of the main paper for each response in the simulated Scenario IV.}
\label{fig:ROC_Count_n100_naive}
\end{figure}

\subsection{ROC curves for the BVSGCR model with sparse and full inverse correlation matrix}
\label{sec:compfull}

\vspace{-0.5cm}

\begin{figure}[H]
\centering
\includegraphics[height=0.385\textheight]{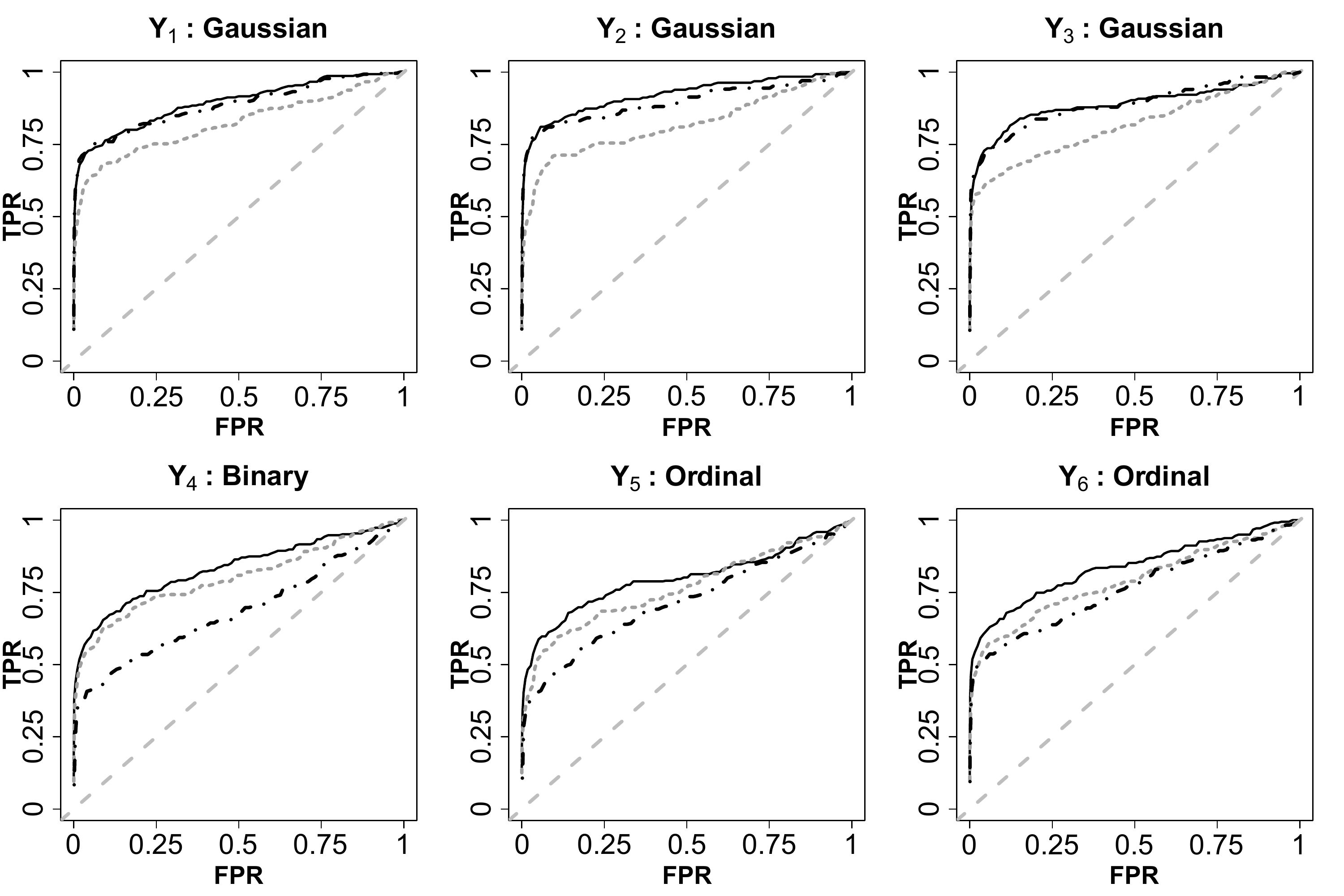}
\caption{Average (over $20$ replicates) ROC curves for the BVSGCR model with sparse (black solid lines) and full (black dashed-dotted lines) inverse correlation matrix $\bm{R}^{-1}$, and for single-response regression models (gray dotted lines) for each response in the simulated Scenario II.}
\label{fig:ROC_Mixed_n100_full}

\includegraphics[height=0.385\textheight]{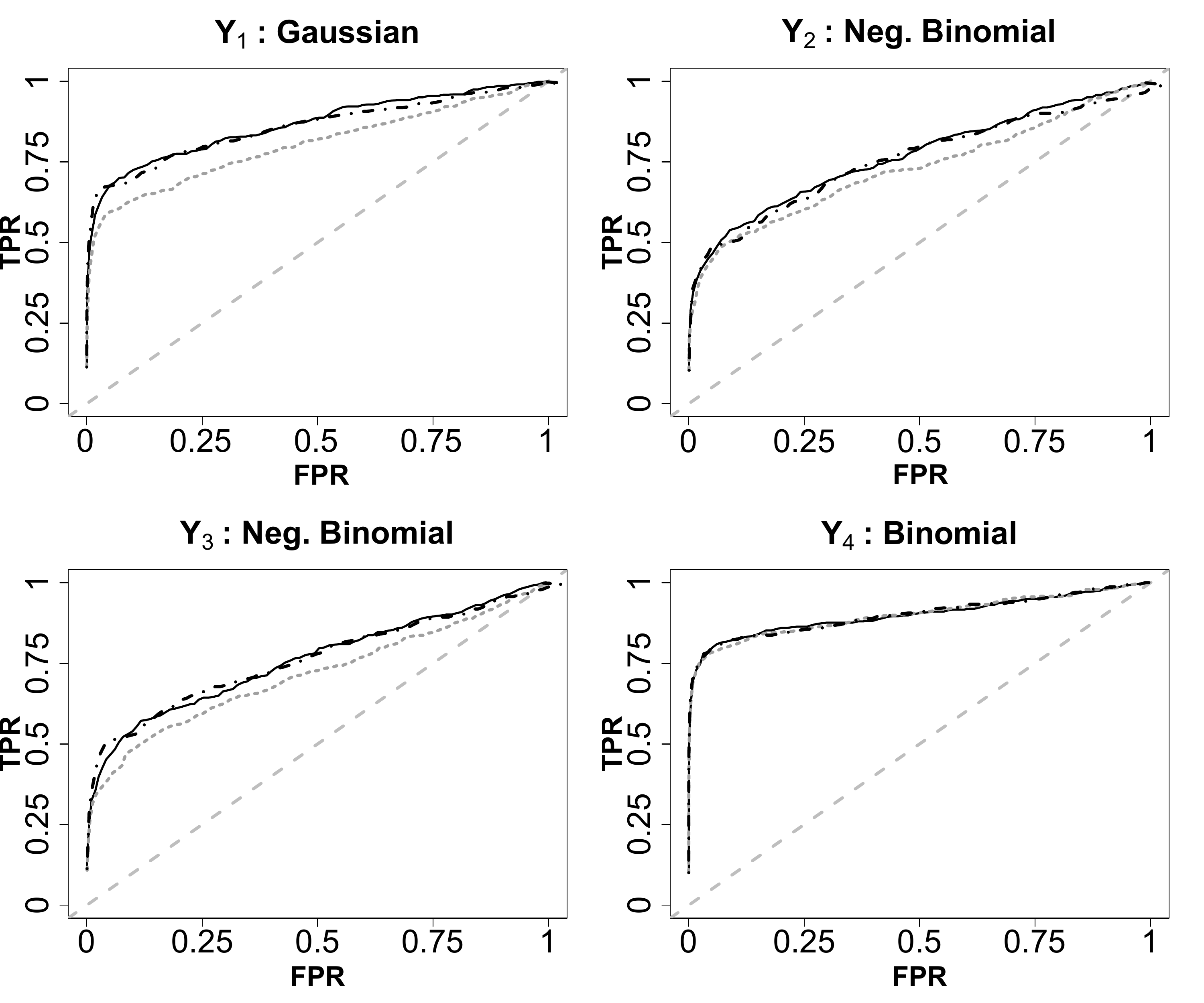}
\caption{Average (over $20$ replicates) ROC curves for the BVSGCR model with sparse (black solid lines) and full (black dashed-dotted lines) inverse correlation matrix $\bm{R}^{-1}$, and for single-response regression models (gray dotted lines) for each response in the simulated Scenario IV.}
\label{fig:ROC_Count_n100_full}
\end{figure}

\subsection{Real data sets: single-response BVS}
\label{sec:compsingle}

\vspace{-0.5cm}

\begin{figure}[H]
\centering
\includegraphics[height=0.385\textheight]{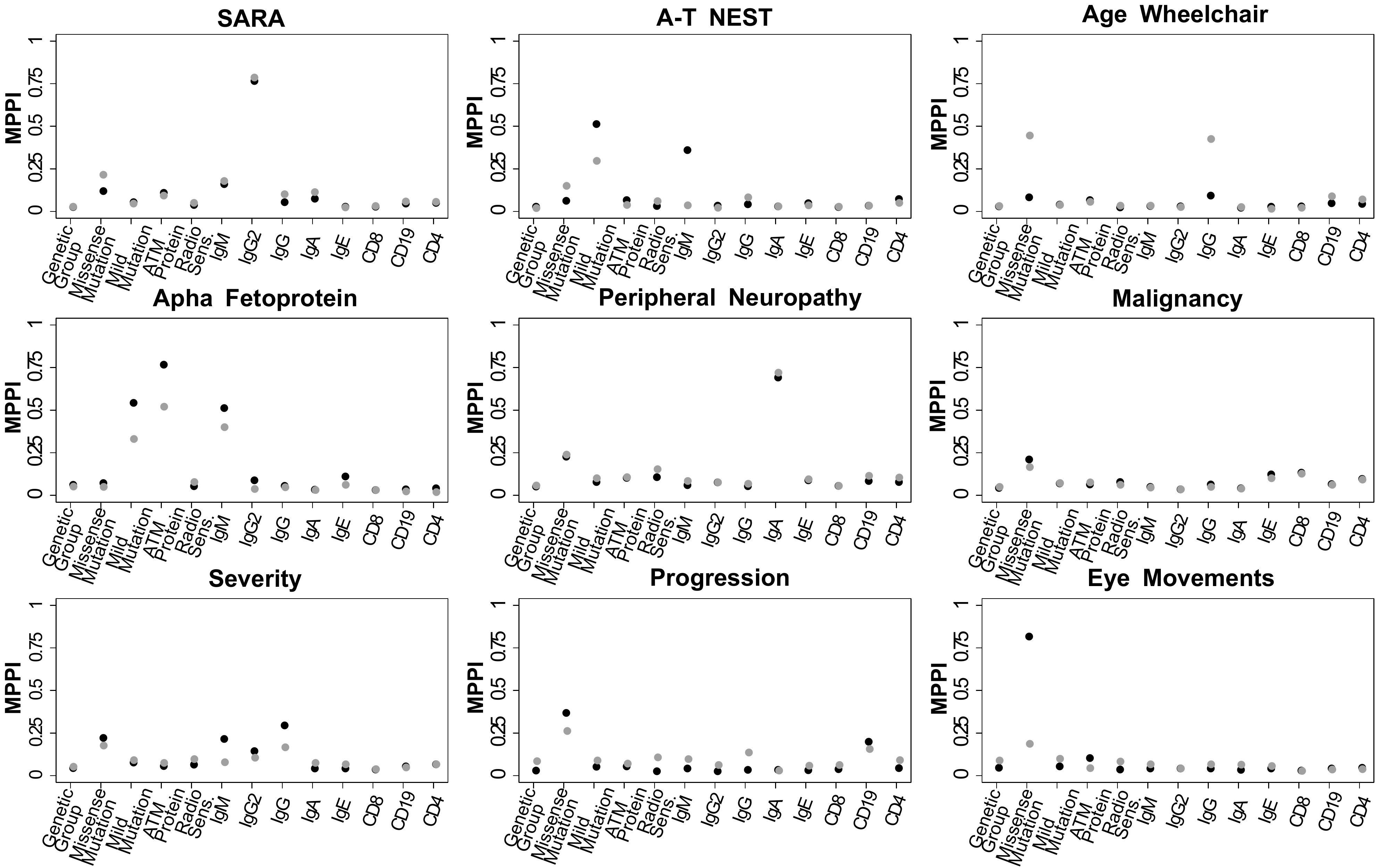}
\caption{Detection of important predictors for the neurological responses in the A-T data set using single-response regression models (gray dots) with superimposed associations obtained by the BVSGCR model (black dots). Marginal posterior probabilities of inclusion (MPPI) measures the strength of the predictor-response association.}
\label{fig:MPPI_Ataxia_single}

\includegraphics[height=0.385\textheight]{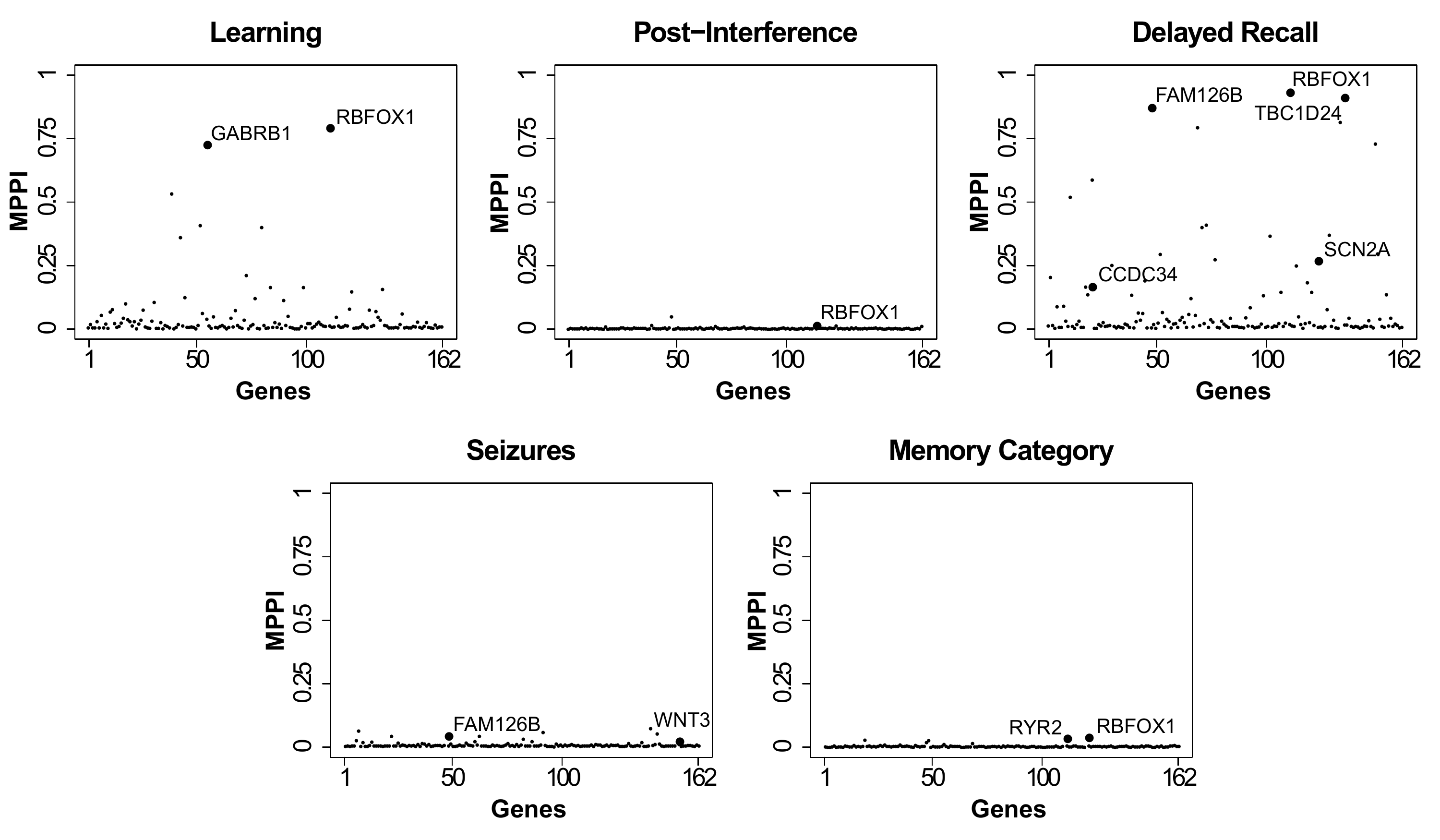}
\caption{Detection of important genes that predict human cognition abilities, number of seizure and neurological memory classification using single-response regression models. Marginal posterior probabilities of inclusion (MPPI) measures the strength of the predictor-response association. Genes detected by the BVSGCR model with MPPI $> 0.05$ are highlighted.}
\label{fig:MPPI_Epilepsy_single}
\end{figure}

\subsection{Real data sets: GLM marginal screening}
\label{sec:compfull}

\vspace{-0.5cm}

\begin{figure}[H]
\centering
\includegraphics[height=0.385\textheight]{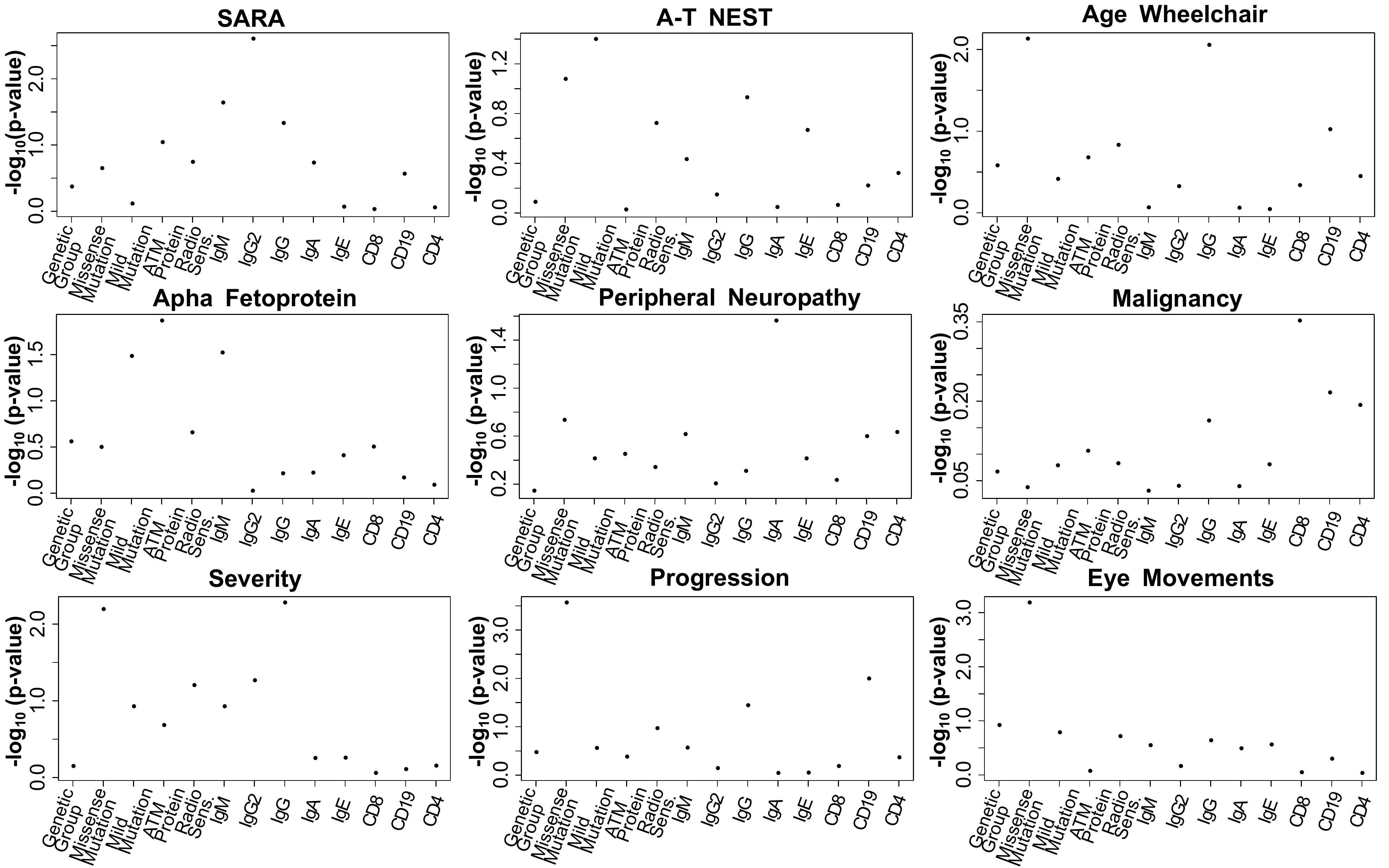}
\caption{Marginal GLM screening of the neurological responses in the A-T data set used in the proposal distribution of the binary latent matrix $\bm{\Gamma$} described in Section \ref{sec:stephens}.}
\label{fig:Proposal_Ataxia}

\includegraphics[height=0.385\textheight]{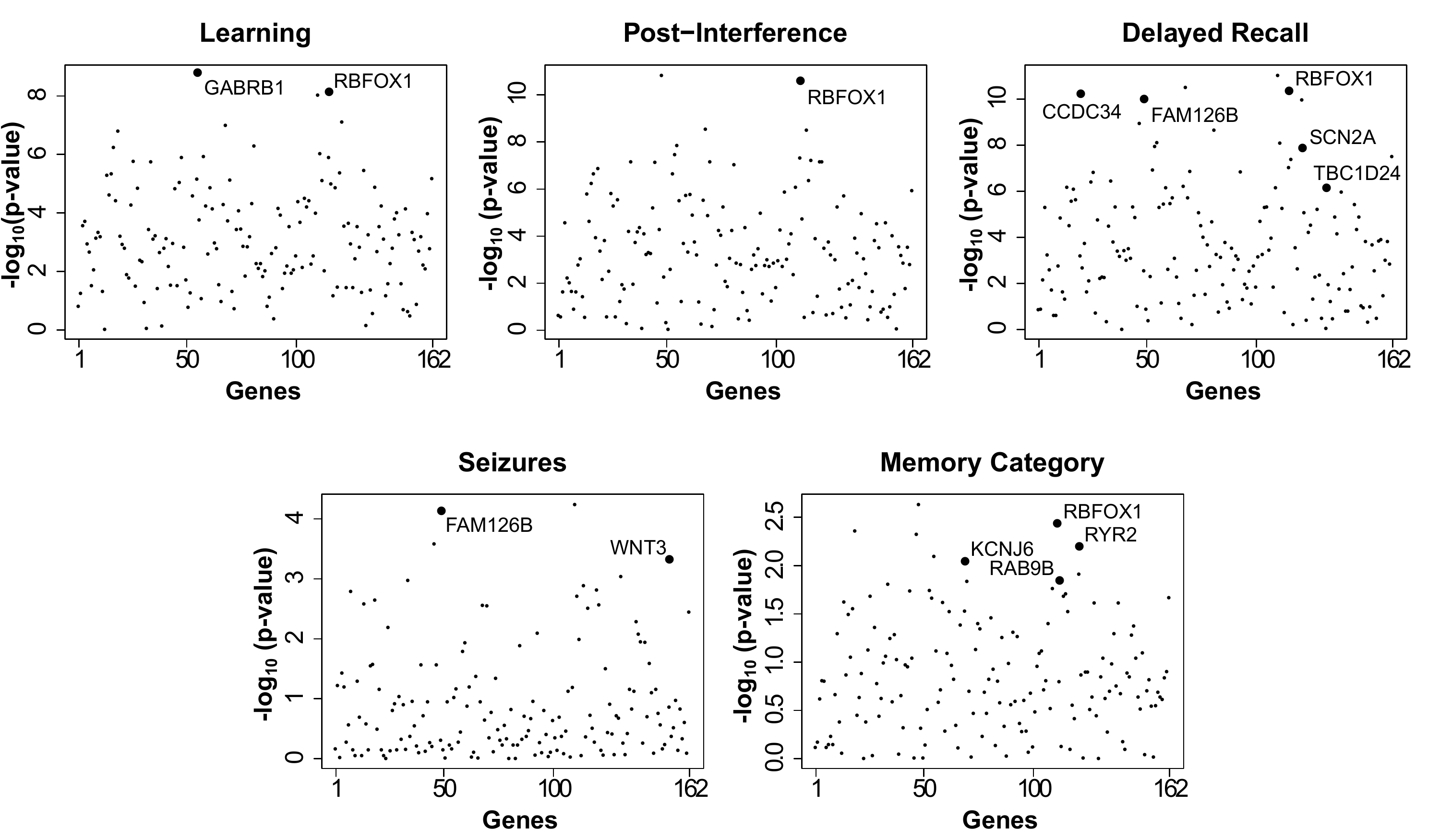}
\caption{Marginal GLM screening of human cognition abilities, number of seizure and neurological memory classification used in the proposal distribution of the binary latent matrix $\bm{\Gamma$} described in Section \ref{sec:stephens}. Genes detected by the BVSGCR model with MPPI $> 0.05$ are highlighted.}
\label{fig:Proposal_Epilepsy}
\end{figure}

\subsection{Ataxia-Telangiectasia disorder: BVSGCR model with non-decomposable graphs specification}
\label{sec:Ataxiacompnndec}

\vspace{-0.5cm}

\begin{figure}[H]
\centering
\includegraphics[height=0.385\textheight]{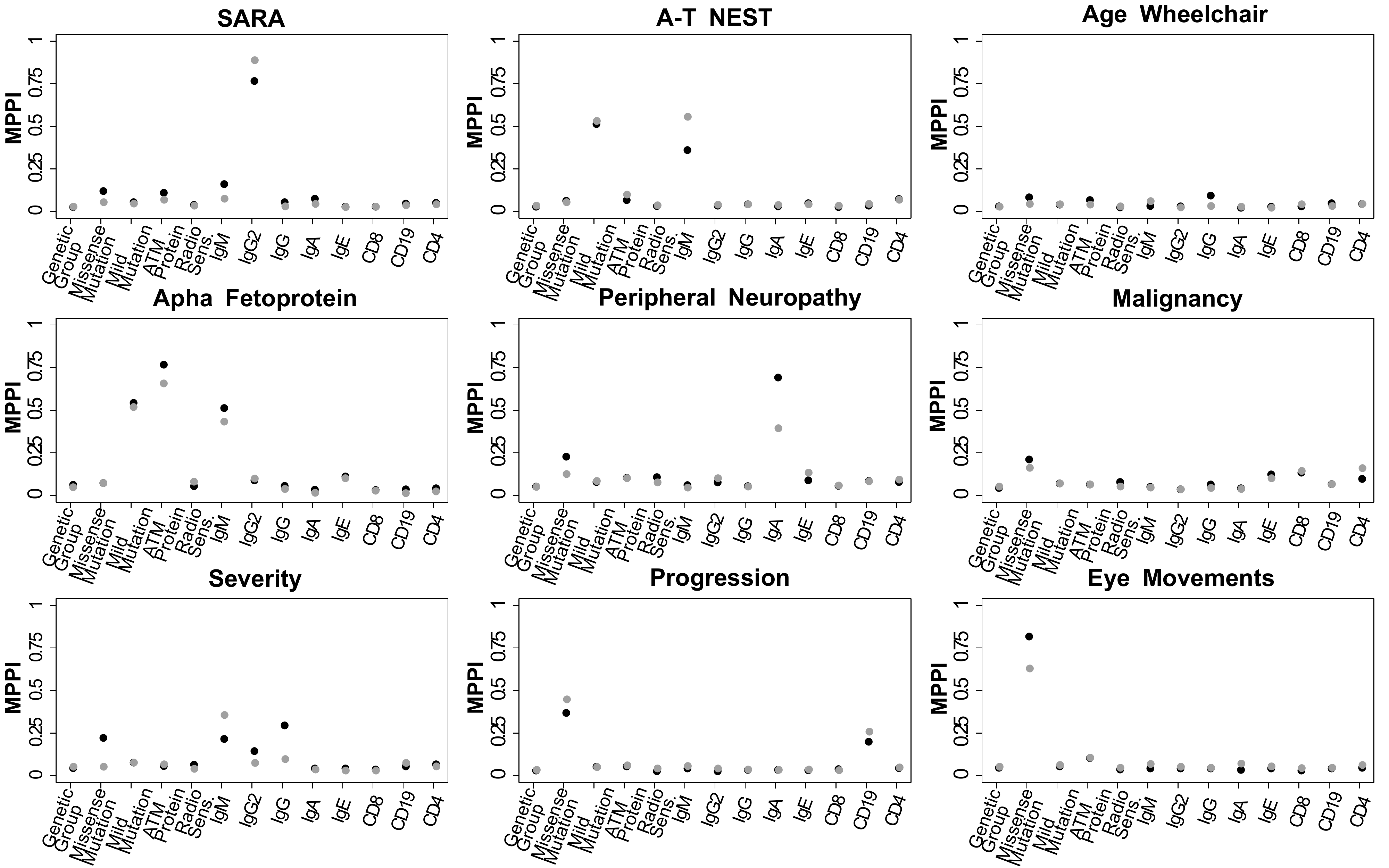}
\caption{Detection of important predictors for the neurological responses in the A-T data set using the BVSGCR algorithm
with a non-decomposable (gray dots) and decomposable (black dots) graphical model specification. MPPI measures the strength of the predictor-response association.}
\label{fig:MPPI_Ataxia_nndec}

\includegraphics[height=0.35\textheight]{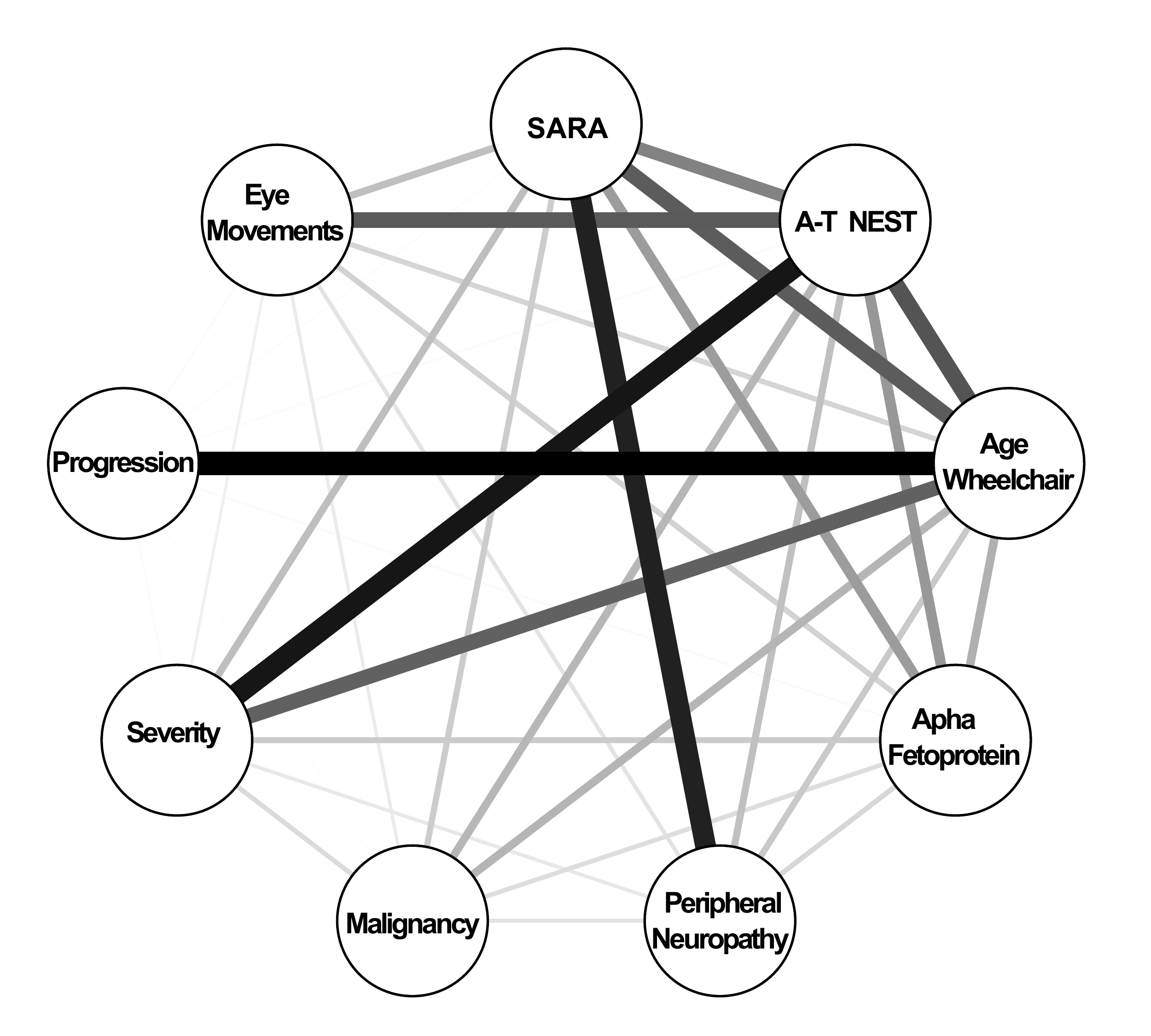}
\caption{Graph implied by the non-zero pattern of $\bm{R}^{-1}$ in the A-T data set and estimated by using the edge posterior probabilities of inclusion (EPPI) obtained by specifying a non-decomposable graphical model. Gray scale and edge thickness specify different levels of the EPPI (black thick line indicates large EPPI values).}
\label{fig:EPPI_Ataxia_nndec}
\end{figure}

\subsection{Patients with Temporal Lobe Epilepsy: BVSGCR model with non-decomposable graphs specification}
\label{sec:LTEcompnndec}

\vspace{-0.5cm}

\begin{figure}[H]
\centering
\includegraphics[height=0.385\textheight]{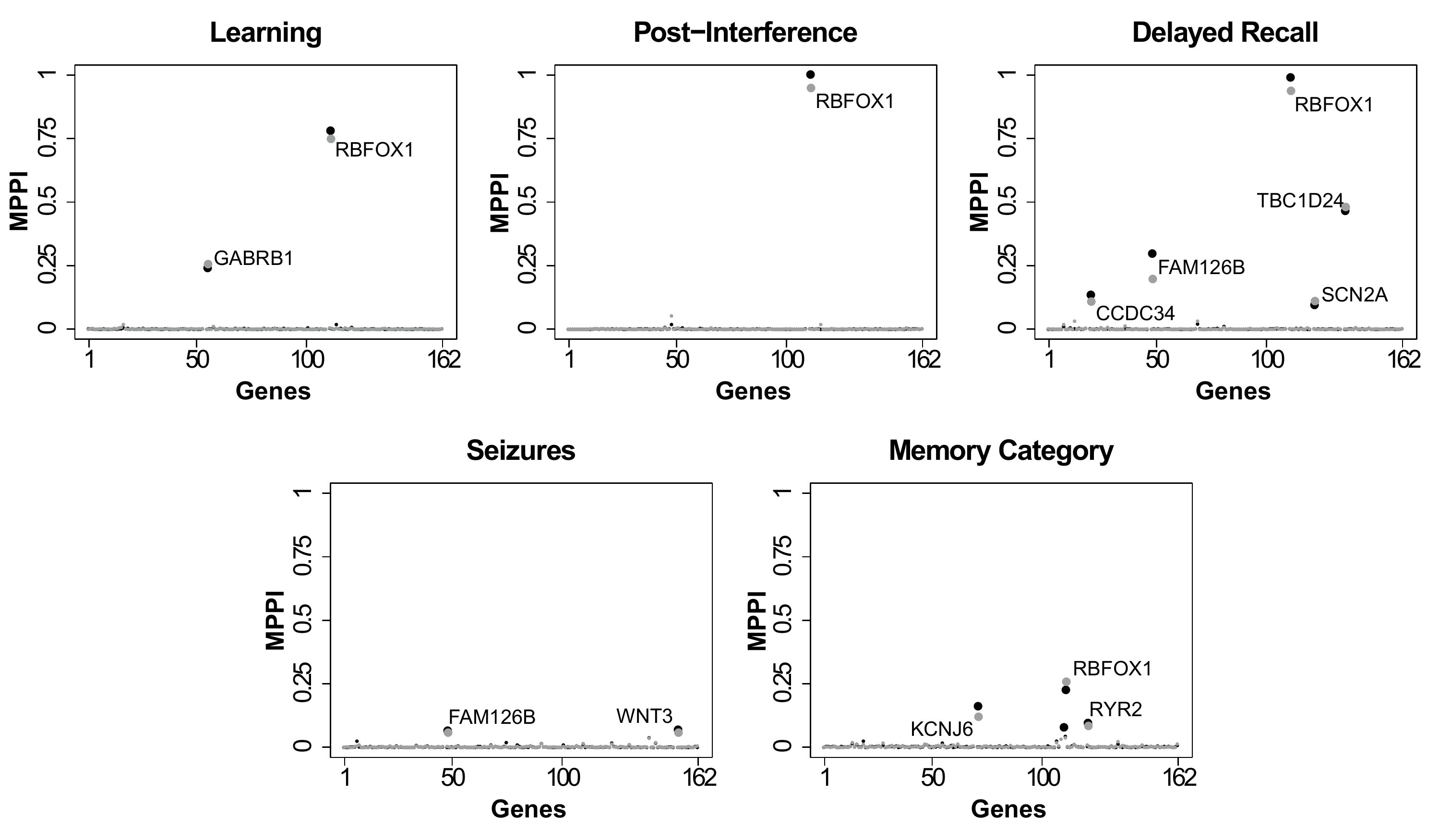}
\caption{Detection of important genes that predict human cognition abilities, number of seizure and memory classification in TLE data set using the BVSGCR algorithm with a non-decomposable (gray dots) and decomposable (black dots) graphical model specification. MPPI measures the strength of the predictor-response association. For each response, only associations with MPPI $> 0.05$ are highlighted.}
\label{fig:MPPI_Epilepsy_nndec}

\includegraphics[height=0.35\textheight]{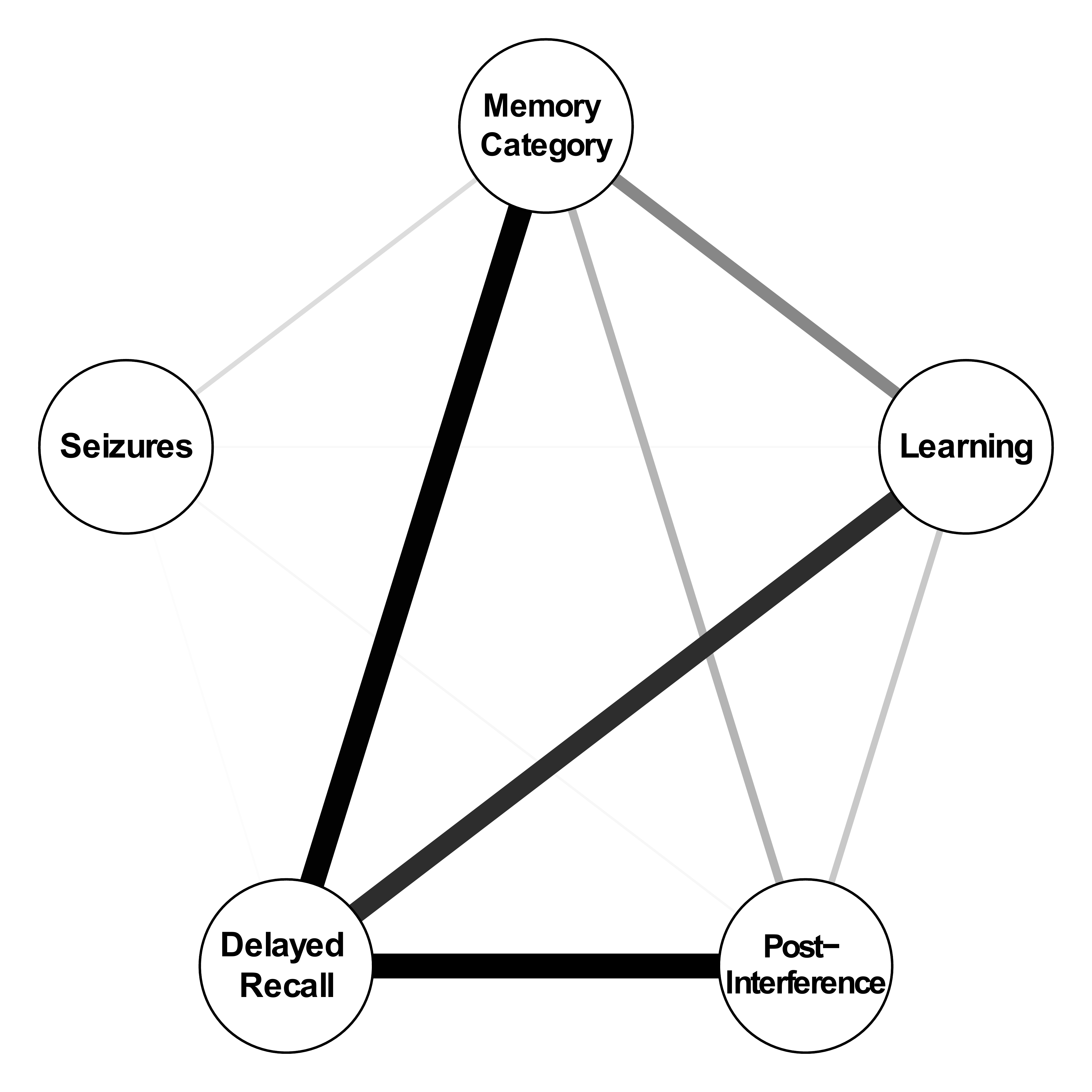}
\caption{Graph implied by the non-zero pattern of $\bm{R}^{-1}$ in the TLE data set and estimated by using the edge posterior probabilities of inclusion (EPPI) obtained by specifying a non-decomposable graphical model. Gray scale and edge thickness specify different levels of the EPPI (black thick line indicates large EPPI values).}
\label{fig:EPPI_Epilepsy_nndec}
\end{figure}

\clearpage
\spacingset{1}
\bibliographystyle{chicago}
\bibliography{BVSGCR_bib_Final}